\documentclass[useAMS,usenatbib]{mn2e}
\pdfoutput=1
\usepackage{graphicx}
\usepackage{subfigure}
\usepackage[english]{babel}
\usepackage{amssymb}
\usepackage{xspace}
\usepackage{amsmath}
\usepackage[draft]{hyperref}
\usepackage{color}

\newcommand\ionm[2]{#1$\,${\small\rmfamily{#2}}}%
\newcommand{\hi}{\ionm{H}{I}\xspace}
\def\hii{\ifmmode {\mbox H{\scshape ii}}\else H{\scshape ii}\fi\xspace}
\def\h2{\ifmmode {\mbox H$_2$}\else H$_2$\fi\xspace}
\def\frach2star{$f_{\rm H2*}$\xspace}

\def\barh2frac{$\frac{\rm{M}_{\rm H2}}{\rm{M}_{\rm gas} + \rm{M}_*}$}
\def\cosmos{\ifmmode {\mbox {\it COSMOS}}\else {\it COSMOS} \fi}

\title [The dust content of galaxies from $z=0$ to
$z=9$]{The dust content of galaxies from $z=0$ to $z=9$}
\author[G. Popping, R. S. Somerville, and M. Galametz]{Gerg\"o Popping$^{1}$\thanks{E-mail:
    gpopping@eso.org}, Rachel S. Somerville$^{2, 3}$, and Maud Galametz$^{1}$\\
$^{1}$European Southern Observatory, Karl-Schwarzschild-Strasse 2,
85748, Garching, Germany\\
$^{2}$Department of Physics and Astronomy, Rutgers University, 136 Frelinghuysen Road, Piscataway, NJ 08854, USA\\
$^{3}$Center for Computational Astrophysics, 160 5th Ave NY NY\\
}
\begin{document}

\maketitle

\begin{abstract}
We study the dust content of galaxies from $z=0$ to
$z=9$ in semi-analytic models of galaxy formation that include new recipes to
track the production and destruction of dust. We include condensation
of dust in stellar ejecta, the growth of dust in the interstellar
medium (ISM), the destruction of dust by supernovae and in the hot halo, and dusty winds and inflows. The rate of dust
growth in the ISM depends on the metallicity and density of molecular
clouds. Our fiducial model reproduces the
relation between dust mass and stellar mass from $z=0$ to $z=7$, the number density of galaxies with dust masses less than
$10^{8.3}\,\rm{M}_\odot$, and the cosmic density of dust at $z=0$. The
model accounts for the double power-law trend between dust-to-gas
(DTG) ratio and gas-phase metallicity of local galaxies and the
relation between DTG ratio and stellar mass. The
dominant mode of dust formation is dust growth in the ISM, except for
galaxies with $M_*<10^7\,\rm{M}_\odot$, where condensation of dust in
supernova ejecta dominates. The dust-to-metal ratio of galaxies
depends on the gas-phase metallicity, unlike what is
typically assumed in cosmological simulations. Model variants including higher condensation
efficiencies, a fixed timescale for dust growth in the ISM, or no growth
at all reproduce some of the observed constraints, but fail to simultaneously
reproduce the shape of dust scaling relations and the dust mass of
high-redshift galaxies.
\end{abstract}


\section{Introduction}
Dust is a key ingredient in interstellar medium (ISM) and galaxy
physics. For example, dust influences interstellar chemistry via
surface reactions and acts as a catalyst for the formation of
molecules \citep{Hollenbach1971,Mathis1990,Li2001,Draine2003}. Dust
depletes metals from the gas phase ISM
\citep{Calzetti1994,Calzetti2000,Netzer2007,Spoon2007,Melbourne2012}.
Dust grains absorb stellar radiation in the ultraviolet (UV) and
re-emit this radiation in the infrared
\citep[IR,][]{Spitzer1978,Draine1984,Mathis1990,Tielens2005}. Dust
contributes significantly to the metals in the circumgalactic medium
(CGM) and can be an additional cooling channel for gas
\citep[e.g.][]{Ostriker1973,Menard2010,Peeples2014,Peek2015}.

Interstellar dust is produced in the ejecta of asymptotic giant branch
(AGB) stars and supernovae
\citep[SNe,][]{Gehrz1989,Todini2001,Nozawa2003,Ferrarotti2006,Nozawa2007,Zhukovska2008,Nanni2013}. After
the initial formation, dust growth can occur in the dense ISM via
accretion of metals onto dust particles
\citep{Draine1990,Dominik1997,Dwek1998,Draine2009,Hirashita2011,Zhukovska2014}. The
exact contribution to the dust mass of a galaxy by the different dust
formation channels is still unknown, although several authors have suggested that
dust growth via accretion in the ISM plays an important role
\citep[e.g.,][]{Dwek2007,Zhukovska2014,Michalowski2015,Schneider2016}. Dust
can be destroyed via thermal sputtering, collisions with other dust
grains, and SN shocks
\citep{Dwek1980,Draine1979,McKee1989,Jones1996}. Besides the
  aforementioned processes, AGN can
  act as an additional channel for the formation of dust
  \citep{Elvis2002}. This dust, however, is likely to dominate only
  in the very central regions of the galaxies, and should not have a
  major impact on the total dust content of galaxies.

The dust content of galaxies at low and high redshifts has intensely
been studied over the past decades. Such studies provide additional constraints for galaxy
formation models and the baryonic physics that regulates the dust and
gas content of galaxies. These observational constraints include for instance the relation between dust mass and stellar mass
\citep{Corbelli2012,Santini2014}, the gas fraction of
galaxies and dust mass \citep{Cortese2012}, dust mass and star-formation rate
\citep[SFR;][]{daCunha2010,Casey2012,Santini2014}, and the dust mass function of
galaxies \citep{Dunne2003,Vlahakis2005,Dunne2011,Eales2009,Clemens2013}.
Two particularly interesting scaling relations are the ratio between
dust mass and gas mass in the ISM (dust-to-gas ratio; DTG), or the
ratio between dust mass and the total mass in metals (dust-to-metal
ratio; DTM) as a function of metallicity or galaxy stellar mass
\citep{Issa1990,Lisenfeld1998,Hirashita2002,James2002,Hunt2005,Draine2007,Engelbracht2008,Galametz2011,Magrini2011,Remy-ruyer2014}. \citet{Remy-ruyer2014}
demonstrated that the DTG ratio in galaxies cannot be described
by a single
power-law as a function of metallicity, but is better represented by a double power-law with
a break around a metallicity of  0.1 Z$_\odot$ \citep{Edmunds2001}.

Absorption line studies using gamma ray burst and Damped Lyman-alpha
absorbers have suggested that the
DTM ratio in galaxies at redshifts $z=0.1$ to $z=6.3$ is
surprisingly similar to the DTM ratio in the local group
\citep{DeCia2013,Zafar2013,Sparre2014,DeCia2016,Wiseman2016}.The DTM
ratios measured in these studies drop at metallicities lower than 0.05
Z$_\odot$. These results demonstrate that high-redshift absorbers can
already be significantly enriched with dust but also that the dust production efficiency can vary significantly between
different environments.

Far-infrared (FIR) and submillimeter observations have shown that even
at the highest redshifts ($z > 4$) galaxies can have significant
reservoirs of dust \citep[$10^7\,\rm{M}_\odot$ or even
  greater,][]{Bertoldi2003,Hughes1997,Valiante2009,Venemans2012,Casey2014,Riechers2014}. \citet{Watson2015}
found a galaxy at $z=7.5\pm0.2$ with a dust mass of
$4\times10^7\,\rm{M}_\odot$ and a DTG ratio that is half of the Milky
Way value. Although these dusty examples may not be representative of
typical high-redshift galaxies, they set strong constraints on our
understanding of dust formation and growth in galaxies in the early
Universe. The Atacama Large sub/Millimeter Array (ALMA) and the James
Webb Space Telescope (JWST) are expected to further revolutionise our
understanding of dust physics in the low- and high-redshift
Universe. It is therefore becoming important to develop
  cosmological galaxy formation models that include dust physics, in
  order to provide a theoretical context for the observations with
  these instruments.

Despite the observational prospects and theoretical importance,
cosmological models of galaxy formation typically do not include
self-consistent tracking of the production and destruction of dust nor
dust chemistry. Traditionally, a linear scaling between dust and metal
abundance is assumed
\citep[e.g.,][]{Silva1998,Granato2000,Baugh2005,Lacey2008,Lacey2010,Fontanot2011,Niemi2012,Somerville2012,Hayward2013,Cowley2016}. A
few groups have started to include self-consistent tracking of dust in
hydrodynamic simulations \citep{Bekki2013,Bekki2015,McKinnon2015,Aoyama2017}, but
these studies used zoom-simulations of individual objects and didn't
focus on trends between global galaxy properties and dust mass
covering a large range of parameter space and cosmic time. Recently,
\citet{McKinnon2016} used a hydrodynamic model to make predictions for
the dust content of galaxies in cosmological volumes, focusing on the
redshift regime $z<2.5$. \citet{Dayal2011} and \citet{Mancini2016}
tracked the dust content of galaxies in cosmological simulations to
look at the dust absorption properties of galaxies at
redshifts $z>5$.

Most implementations of dust chemistry in galaxy formation have been
made using specialised models
\citep[e.g.][]{Dwek1998,Hirashita2002,Inoue2003,Morgan2003,Calura2008,Zhukovska2008,Valiante2009,Asano2013,Calura2014,Zhukovska2014,Feldmann2015}. %
These models have been essential for developing our understanding of
the relevance of the individual channels of dust formation to the dust
content of galaxies.  However, these models are often idealised to
reproduce specific objects and are not placed within a cosmological
context. Furthermore, they generally do not include all physical
processes thought to be relevant for galaxy formation.

Semi-analytic models (SAMs) offer a good alternative approach for
self-consistently tracking the production and destruction of dust in
galaxies within the framework of a $\Lambda$ cold dark matter
cosmology. Simplified but physically motivated recipes are used to
track physical processes such as the cooling of hot gas into galaxies,
star formation, the energy input from supernovae and active galactic
nuclei into the ISM, the sizes of galaxy discs, and the enrichment of
the ISM by supernovae ejecta and stellar winds \citep[see][for a
  recent review]{SomervilleDave2015}. The low computational cost of
SAMs makes them a powerful tool to model a broad range of galaxy
masses probing large volumes, provide predictions for future studies,
and explore different recipes for physical processes in galaxies.

In this paper, we include tracking of dust production and destruction
in the most recent version of the Santa Cruz semi-analytic model
\citep{Popping2014sam,Somerville2015}. We explore how the dust content
of galaxies and our Universe evolves over time and how this is
affected by different implementations of the processes that produce
dust. We extend the \citet{Arrigoni2010} galactic chemical evolution
(GCE) model to include the condensation of dust in stellar ejecta, the
growth of dust in the dense ISM, the destruction of dust through
thermal sputtering by supernovae (SNe), dusty winds from star-forming
regions, dust destruction in the hot halos, and the infall of dust
from the CGM. 
In this work we only focus on the evolution of dust masses and the
different dust formation channels, leaving the rest of the underlying
galaxy properties unchanged from the models published in
\citet{Popping2014sam} and \citet{Somerville2015}. In a future work we
will extend this model by including a self-consistent treatment of the
impact of dust on the galaxy formation physics (i.e., cooling through
dust channels, \h2 formation recipes based on the dust abundance, and
dust absorption based on the estimated dust abundance).

This paper is structured as follows. In Section \ref{sec:model} we
present the galaxy formation model and GCE used in this work. We
present the newly implemented dust related processes in Section
\ref{sec:dust_model}. We briefly summarise how observational estimates
of dust masses in galaxies are typically obtained, and also discuss
the uncertainties on these estimates in Section
\ref{sec:observing_dust}. In Section \ref{sec:results} we present our
predictions for the dust scaling relations in galaxies and how these
evolve with cosmic time. We discuss our finding in Section
\ref{sec:discussion} and summarise our work in Section
\ref{sec:summary}. Throughout this paper we adopt a flat $\Lambda$CDM
cosmology with $\Omega_0=0.28$, $\Omega_\Lambda = 0.72$,
$h=H_0/(100\,\rm{km}\,\rm{s}^{-1}\,\rm{Mpc}^{-1}) = 0.7$,
$\sigma_8=0.812$, and a cosmic baryon fraction of $f_b=0.1658$
\citep{Komatsu2009}.

\section{Galaxy formation model}
\label{sec:model}
In this section we present the galaxy formation model within which we
include the tracking of dust production and destruction. We provide a
general introduction to the semi-analytic model employed in this work
and will focus in more detail on the elements of the code that are
relevant for the tracking of dust (Section \ref{sec:sam}). We then
discuss the GCE model (Section \ref{sec:GCE}), relevant for the
condensation of dust in stellar ejecta. We adopt a flat $\Lambda$CDM
cosmology with $\Omega_0=0.28$, $\Omega_\Lambda = 0.72$,
$h=H_0/(100\,\rm{km}\,\rm{s}^{-1}\,\rm{Mpc}^{-1}) = 0.7$,
$\sigma_8=0.812$, and a cosmic baryon fraction of $f_b=0.1658$
\citep{Komatsu2009}. Unless stated otherwise we leave the free
parameters associated with the galaxy-formation model fixed to the
values given in \citet{Somerville2015}.

\subsection{Semi-analytic model framework}
\label{sec:sam}
The galaxy formation model was originally presented in
\citet{Somerville1999} and \citet{Somerville2001}. Significant updates
to this model are described in \citet[][S08]{Somerville2008},
\citet{Somerville2012}, \citet[PST14]{Popping2014sam},
\citet{Porter2014}, and \citet[SPT15]{Somerville2015}. The model
tracks the hierarchical clustering of dark matter haloes, shock
heating and radiative cooling of gas, SN feedback, star formation,
active galactic nuclei (AGN) feedback (by quasars and radio jets),
metal enrichment of the interstellar and intracluster medium, mergers
of galaxies, starbursts, and the evolution of stellar populations. The
PST14 and SPT15 models include new recipes that track the abundance of
ionised, atomic, and molecular hydrogen and a molecule-based
star-formation recipe. These models have been fairly successful in
reproducing the local properties of galaxies such as the stellar mass
function, gas fractions, gas mass function, SFRs, and stellar
metallicities, as well as the evolution of the galaxy sizes, quenched
fractions, stellar mass functions, and luminosity functions
\citep[PST14,
SPT15]{Somerville2008,Somerville2012,Porter2014,Popping2014PDR,Brennan2015,Popping2016}.

Fundamentally, semi-analytic models track the flows of material
between different reservoirs. In our models, all galaxies form within
a dark matter halo. The reservoirs for gas include the ``hot'' gas
that is assumed to be in a quasi-hydrostatic spherical configuration
throughout the virial radius of the halo, the ``cold'' gas in the
galaxy, assumed to be in a thin disk, and ``ejected'' gas which is gas
that has been heated and ejected from the halo by stellar winds. We
can schematically think of the cold disk gas as corresponding to the
ISM, and the hot halo gas as corresponding to the circumgalactic,
intra-group, or intra-cluster medium. The interpretation
  of the ``ejected'' gas is less clear, but it is thought to correspond
  either to the circumgalactic or intergalactic medium or a
  combination of the two. Gas moves between these reservoirs as
follows. As dark matter halos grow in mass, pristine gas is accreted
from the intergalactic medium into the hot halo. In addition, a simple
cooling model is used to estimate the rate at which gas accretes from
the hot halo into the cold gas reservoir, where it becomes available
to form stars. Gas is removed from the cold gas reservoir as it
becomes locked up in stars, and also by stellar and AGN-driven
winds. Part of the gas that is ejected by stellar winds is returned to
the hot halo, and the rest is deposited in the ``ejected''
reservoir. The fraction of gas that escapes the hot halo
  is determined by the virial velocity of the progenitor galaxy (see
  S08 for more detail). Gas ``re-accretes'' from the ejected
reservoir back into the hot halo according to a parameterized
timescale (again see S08 for details). In the
present work, we add new ``dust'' reservoirs corresponding to all of
these gas reservoirs. We track the production and destruction of dust
within the relevant reservoirs, as well as the movement of dust
between reservoirs, as will be described a bit later.

The galaxy that initially forms at the center of each halo is called
the ``central'' galaxy. When dark matter halos merge, the central
galaxies in the smaller halos become ``satellite'' galaxies and orbit
within the larger halo until their orbit decays and they merge with
the central galaxy, or until they are tidally destroyed.

For this work, we construct the merging histories (or
  merger trees) of dark matter haloes based on the extended
  Press–Schechter (EPS) formalism using the method described in
  \citet{SomervilleKolatt1999}, with improvements described in
  S08. These merger trees record the growth of dark matter haloes via
  merging and accretion, with each ‘branch’ representing a merger of
  two or more haloes. Each branch is followed back in time to a
  minimum progenitor mass $M_{\rm res}$, which we refer to as the mass
  resolution of our simulation. \citet{Lu2014} and \citet{Porter2014}
  showed that our SAMs give nearly identical results when run on the
  EPS merger trees or on merger trees extracted from dissipationless
  N-body simulations. We prefer EPS merger trees here because they
  allow us to attain extremely high resolution. In this paper, haloes
  are resolved down to a resolution of $M_{\rm res} =
  10^{10}\,\rm{M}_\odot$ for all root haloes, where $M_{\rm root}$ is
  the mass of the root halo and represents the halo mass at the
  output redshift. We furthermore impose a minimum
  resolution of $M_{\rm res} = 0.01 M_{\rm root}$ (see Appendix A of
  SPT15 for tests supporting these choices). The simulations were run
  on a grid of haloes with root halo masses ranging from $5\times 10^8$
  to $5 \times 10^{14}\,\rm{M}_\odot$ at each redshift of interest, with 100 random realizations
  created at each halo mass. Each individual halo has a different
  merger history with a stochastic element, which gives us an ensemble
  of modelled halos at fixed halo mass, allowing us to explore the
  scatter between halos.

Here we briefly summarise the recipes employed to compute the size of
galaxy discs and to track the molecular hydrogen abundance. These play
an important role in modelling the growth of dust by accretion in the
ISM (see Section \ref{sec:dust_accretion} and Section
\ref{sec:free_parameters}). We point the reader to
\citet{Somerville2008}, \citet{Somerville2012}, PST14, and SPT15 for a
more detailed description of the model.

The sizes of the galaxy discs are important as they set the surface
densities for our \h2 partitioning recipe and growth rate of dust by
accretion in the ISM. When gas cools onto a galaxy, we assume it
initially collapses to form a rotationally supported disc. The scale
radius of the disc is computed based on the initial angular momentum
of the gas and the halo profile, assuming that angular momentum is
conserved and that the self-gravity of the collapsing baryons causes
contraction of the matter in the inner part of the halo
\citep{Blumenthal1986,Flores1993,Mo1998}. This approach successfully
reproduces the evolution of the size-stellar mass relation of
disc-dominated galaxies from $z\sim2$ to $z=0$
\citep{Somerville2008size}, the sizes of \hi discs in the local
Universe and the observed sizes of CO discs in local and high-redshift
galaxies (PST14).

To compute the \h2 fraction of the cold gas we use an approach based
on the work of \citet{Gnedin2011}. These authors performed
high-resolution `zoom-in' cosmological simulations with the Adaptive
Refinement Tree (ART) code \citep{Kravtsov99}, including gravity,
hydrodynamics, non-equilibrium chemistry, and simplified 3D
on-the-fly radiative transfer \citep{Gnedin2011}.  The authors present a
fitting formula for the \h2 fraction of cold gas based on the
DTG ratio relative to solar, $D_{\rm MW}$, the ionising background radiation field, $U_{\rm MW}$, and the surface density of the cold gas (see PST14, SPT15).
We assume that the local UV background scales with the SFR relative to
the Milky Way value, $U_{\rm MW} = SFR/SFR_{\rm MW}$, where we choose
$SFR_{\rm MW} = 1.0\,\rm{M}_\odot\,\rm{yr}^{-1}$
\citep{Murray2010,Robitaille2010}.
As in PST14 and SPT15, we assume that the DTG ratio is
proportional to the metallicity of the gas in solar units $D_{\rm MW}
= Z_{\rm gas}/Z_\odot$ (where in this case the metallicity is given by
all the available metals in the ISM). In a future paper we
will make our models self-consistent by instead using the modeled dust
mass to estimate the molecular hydrogen fraction. However, initially
we prefer to leave the underlying galaxy formation model unchanged and
explore how successful our simple model is at reproducing fundamental
observations of dust content.

We considered other recipes
for the partitioning of \hi and \h2 in PST14 and SPT15. We found that metallicity based recipes that do not
include a dependence on the UV background predict less efficient
formation of \h2, less star formation, and less metal enrichment at early
times in low-mass haloes ($M_{\rm h}<
10^{10.5}\,\rm{M}_\odot$). PST14 also considered a pressure-based
recipe \citep{Blitz2006}, but found that the pressure-based version
of the model is less successful in reproducing the \hi density of our
Universe at $z>0$.

The SF in the SAM is modelled based on an empirical relationship
between the surface density of molecular hydrogen and the surface
density of star-formation
\citep{Bigiel2008,Genzel2010,Bigiel2012}. Observations of high-density
environments (especially in starbursts and high-redshift objects) have indicated that above
some critical surface density, the relation between molecular hydrogen
surface density and SFR surface density steepens
\citep{Sharon2013,Hodge2015}. To account for this steepening we use the following expression to model star formation 
\begin{equation}
\label{eqn:bigiel2}
\Sigma_{\rm SFR} = A_{\rm SF} \, \Sigma_{\rm H_2}/(10 M_\odot {\rm pc}^{-2}) \left(1+
\frac{\Sigma_{H_2}}{\Sigma_{\rm H_2, crit}}\right)^{N_{\rm SF}},
\end{equation}
where $\Sigma_{\rm H_2}$ is the surface density of molecular hydrogen
and with $A_{\rm SF}=5.98\times 10^{-3}\, M_\odot {\rm yr}^{-1} {\rm
  kpc}^{-2}$, $\Sigma_{\rm H_2, crit} = 70 M_\odot$ pc$^{-2}$, and
$N_{\rm SF}=1$.The free parameters are chosen based on
  the observations presented in \citet{Bigiel2008} and \citet[see
    PST14 and STP15 for more details]{Hodge2015}.

Following PST14 and SPT15, we adopt a metallicity floor of
$Z=10^{-3}\,$Z$_\odot$ and a floor for the fraction of molecular
hydrogen of $f_{\rm mol}=10^{-4}$.  These floors represent the
enrichment of the ISM by `Pop III' stars and the formation of
molecular hydrogen through other channels than on dust grains
\citep{Haiman1996,Bromm2004}. We showed in SPT15 that our model results are not
  sensitive to the precise values of these parameters.

\subsection{Galactic Chemical Evolution}
\label{sec:GCE}
We use the Galactic Chemical Evolution (GCE) model presented in
\citet{Arrigoni2010} to track the abundance of individual
elements. \citet{Arrigoni2010} extended the Santa Cruz semi-analytic
model to include the detailed multi-element metal enrichment by type
Ia and type II supernovae and long-lived stars. With this extension
our model tracks the abundances of 19 individual elements, as well as
the rate of SNIa and SNII. We refer the reader to \citet{Arrigoni2010}
for a detailed description of the GCE and its ingredients. In this
paper we will discuss several updates to the Arrigoni model adopted
here, including modified stellar yields and the delay time
distribution formulation for SNIae.  We assume a Chabrier
\citep{Chabrier2003} initial stellar mass function with a slope of
$x=-1.15$ in the mass range 0.1--100 $\rm{M}_\odot$. This yields
good agreement with the observed mass metallicity and alpha-to-iron
ratio of galaxies \citep{Fontanot2016}.
We adopt the solar abundances from \citet{Asplund2009}.

\subsubsection{Stellar yields}
Stellar yields are critical ingredients in a chemical evolution model.
The stellar yield describes the amount of material of a given element
that a star produces (per unit mass of stars formed) and ejects into
the ISM. The \citet{Arrigoni2010} GCE adopts different nucleosynthesis
prescriptions for stars in different mass ranges.

We adopt the yields from \citet{Karakas2007} for low- and
intermediate-mass stars ($0.8 < M/M_\odot < 8$). These stars produce
He, C, N, and heavy s-process elements, which they eject during the
formation of a planetary nebulae.

We adopt the yields from \citet{Woosley1995} for massive stars ($M > 8
M_\odot$). These stars mainly produce $\alpha$-elements (O, Na, Ne,
Mg, Si, S, Ca), some Fe-peak elements, light s-process elements and
r-process elements. Our SNII yields account for the
  contribution of radioactive $^{56}$Ni to the Fe yield in stars with
  $M > 12 M_\odot$ \citep{Arrigoni2012}.

We assume that SNIa are C--O white dwarfs in binary systems, exploding
by C-deflagration after reaching the Chandrasekhar mass via accretion
of material from the companion star. SNIa mainly produce Fe and
Fe-peak elements. We adopt the yields from \citet{Iwamoto1999}, which
calculates the SNIa yields using a delayed detonation. We furthermore
assume that the primary stars also enrich the medium as a normal AGB
prior to the SN event. We assume a SNIa binary fraction of 0.025.

We use metallicity-dependent yields for low- and intermediate-mass
stars, tabulated for metallicities of $Z = 0.0002,\,0.004,\,0.02$, and
interpolate between these values. If the metallicity falls below or
above the limiting values, we use yields corresponding to the minimum
or maximum metallicity, respectively. The SNIa yields are only given
for solar metallicity.

\subsubsection{Delay time distribution for SN Ia}
The delay time distribution (DTD) describes the SN rate after a burst of
star-formation. Here we adopt the DTD from \citet{Maoz2012}, where
\begin{equation}
  \rm{SNR}_{\rm Ia}(t) = 4\times10^{13}\rm{yr}^{-1}m_*\bigg(\frac{t}{1\,\rm{Gyr}}\bigg)^{-1},
\end{equation}
is the rate of SNIa produced by a stellar population with mass $m_*$
that was formed at $t=0$. \citet{Walcher2016} showed that a GCE
adopting a power law DTD such as in
\citet{Maoz2012} reproduces the age and $\alpha$-element abundances of
early-type galaxies better than the more classical single or double
Gaussian shaped DTD.

\begin{table*}
 \caption{Summary of the dust-related model parameters in our
   fiducial model\label{tab:free_parameters}}
 \begin{tabular}{cccc}
\hline
 Parameter & Description & Section defined & value \\
\hline
$\delta^{\rm AGB}_{j}$ & Condensation efficiency in AGB
                                       ejecta &
                                                \ref{sec:condensation}
                                          & 0.2 \\
$\delta^{\rm SNIa}_{j}$ & Condensation efficiency in SNIa
                                       ejecta &
                                                \ref{sec:condensation}
                                          & 0.15 \\
$\delta^{\rm SNII}_{j}$ & Condensation efficiency in SNII
                                       ejecta &
                                                \ref{sec:condensation}
                                          & 0.15 \\
$\tau_{\rm acc,0}$& Time scale of dust growth &
                                                \ref{sec:dust_accretion}  & 15 Myr \\
$f_{\rm SN}$ & Supernova efficiency & \ref{sec:destruction} & 0.36\\
$M_{\rm cleared,carbonaceous}$ & Gas mass cleared of carbonaceous dust by 
                               one SN event & \ref{sec:destruction} &
                                                                      $600\,\rm{M}_\odot$\\
$M_{\rm cleared,sillicates}$ & Gas mass cleared of silicate dust by 
                               one SN event & \ref{sec:destruction} & $980\,\rm{M}_\odot$\\
\hline
\hline
 \end{tabular}
 \end{table*}

\section{Dust evolution}
\label{sec:dust_model}
There are a number of physical processes that contribute to the formation,
destruction, and removal of dust from galaxies. The net
rate of change in the mass of dust of the $j$th element in the ISM of a
galaxy is given by
\begin{multline}
\label{eq:dust_balance}
\dot{M}_{j,\rm{dust}} = \dot{M}_{j,\rm{dust}}^{\rm produced} +
\dot{M}_{j,\rm{dust}}^{\rm growth} \\ - \dot{M}_{j,\rm{dust}}^{\rm destruct} -
 \dot{M}_{j,\rm{dust}}^{\rm SF} + \dot{M}_{j,\rm{dust}}^{\rm infall} -
 \dot{M}_{j,\rm{dust}}^{\rm out},
\end{multline}
where $\dot{M}_{j,\rm{dust}}^{\rm produced}$ is the condensation rate
of dust in the ejecta of long-lived stars and SNe,
$\dot{M}_{j,\rm{dust}}^{\rm growth}$ the growth rate of dust in the
ISM, $\dot{M}_{j,\rm{dust}}^{\rm destruct}$ the destruction rate of
dust in the ISM due to SNe, $\dot{M}_{j,\rm{dust}}^{\rm SF}$ the rate
that dust is locked up in stars, $\dot{M}_{j,\rm{dust}}^{\rm infall}$
the rate of dust accreting onto the galaxy from the CGM, and
$\dot{M}_{j,\rm{dust}}^{\rm out}$ the rate of decrease in the dust
mass due to outflows. An equation similar to Equation
\ref{eq:dust_balance} can be constructed for the gas-phase
metallicity, where when a given amount of dust condenses, the same
amount of metals is removed from the cold gas reservoir, and when dust
is destroyed, the same amount of metals are added back to the gas
reservoirs.

The individual recipes for the dust-related processes are described
below in separate subsections. The model tracks the dust evolution of
the refractory elements C, Mg,
Si, S, Ca, Ti, Fe, and O. We summarise the physical parameters adopted in
the various recipes in Table \ref{tab:free_parameters} and describe
them in further detail in sub-section \ref{sec:free_parameters}.

\subsection{Dust production}
\label{sec:condensation}
Some of the metals returned to the ISM by stars and supernovae may
condense into dust. To model the condensation of dust we follow the
approach presented in \citet{Dwek1998}, with updated condensation
efficiencies based on recent theoretical and observational work. In
the following, $m^k_{j,\rm ej}$ is the mass of the $j$th element (C,
Mg, Si, S, Ca, Ti, Fe, or O) returned by the $k$th stellar process
(SNIa, SNII, or AGB stars), whereas $m^k_{j,\rm{dust}}$ marks the mass
of dust of the $j$th element from the $k$th type of process.

The amount of dust produced by AGB stars with a carbon-to-oxygen ratio
C/O$>$1 in their returned mass is described as
 \begin{equation}
m^{\rm AGB}_{j,\rm{dust}} = 
 \begin{cases}
 \delta^{\rm AGB}_{\rm C}(m^{\rm AGB}_{\rm C,ej} - 0.75m^{\rm
   AGB}_{\rm O,ej}) &\quad
 \text{if } j=\text{C}\\
&\\
 0 & \quad \text{else,}\\
   \end{cases}
 \end{equation}
where $\delta^{\rm AGB}_{j}$ is the condensation efficiency of
element $j$ for AGB stars. When the carbon-to-oxygen ratio of the AGB
mass return is less than 1 (C/O$<$1), the mass of dust produced can be
described as
\begin{equation}
m^{\rm AGB}_{j,\rm{dust}} = 
\begin{cases}
0 &\quad \text{if } j=\text{C,}\\
&\\
16\displaystyle\sum_{j={\rm{Mg, Si, S, Ca, Ti, Fe}}}\delta^{\rm
  AGB}_{j}m^{\rm AGB}_{j,\rm{ej}}/\mu_j  &\quad \text{if } j=\text{O,}\\
&\\
\delta^{\rm AGB}_{j} m^{\rm AGB}_{j,\rm{ej}}& \quad \text{otherwise,}\\
  \end{cases}
\end{equation}
where $\mu_j$ is the mass of element $j$ in atomic mass units.

The mass of dust produced via the ejecta of SNII is
\begin{equation}
m^{\rm SNII}_{j,\rm{dust}} = 
\begin{cases}
\delta^{\rm SNII}_{C} m^{\rm SNII}_{C,\rm{ej}} &\quad \text{if } j=\text{C,}\\
&\\
16\displaystyle\sum_{j={\rm{Mg, Si, S, Ca, Ti, Fe}}}\delta^{\rm
  SNII}_{j}m^{\rm SNII}_{j,\rm{ej}}/\mu_j  &\quad \text{if } j=\text{O,}\\
&\\
\delta^{\rm SNII}_{j} m^{\rm SNII}_{j,\rm{ej}}& \quad \text{else,}\\
  \end{cases}
\end{equation}
where $\delta^{\rm SNII}_{j}$ is the dust condensation efficiency of
element $j$ for SNII. The same approach is used for the dust
condensation in the ejecta of SNIa, where the condensation efficiency
for SNII $\delta^{\rm SNII}_{j}$  is replaced by the condensation
efficiency for SNIa $\delta^{\rm SNIa}_{j}$. 

The total mass of dust condensation in the ejecta of
AGB stars and SNae
is then given by
\begin{equation}
\dot{M}_{j,\rm{dust}}^{\rm produced} = \frac{d m^{\rm AGB}_{j,\rm{dust}}}{dt} +
\frac{d m^{\rm SNIa}_{j,\rm{dust}}}{dt} + \frac{d m^{\rm SNII}_{j,\rm{dust}}}{dt}. 
\end{equation}

\subsection{Growth of dust by accretion in the ISM}
\label{sec:dust_accretion}
Collisions between gas-phase elements and existing dust grains can
lead to the growth of the dust mass in galaxies
\citep{Draine1990,Dwek1998,Draine2009}. To model this process
we follow the prescription in \citet{Zhukovska2008} and
\citet{Zhukovska2014}. We also present the growth
recipe first described in \citet{Dwek1998} and later adopted by others
\citep[e.g.,][]{Calura2008,McKinnon2015,Feldmann2015} and further
explore this in the appendix.

\subsubsection{The Zhukovska growth model}
Our model makes the explicit assumption that dust can only grow in the
dense regions of the ISM. Within our model, these regions are
associated with molecular hydrogen. Not all the ISM resides in such
states. We can define an ``effective'' exchange time $\tau_{\rm
  exch,eff}$, over which all the ISM in a galaxy is cycled through
molecular clouds \citep{Zhukovska2014}. This exchange time is given by
\begin{equation}
\tau_{\rm exch,eff} = \tau_{\rm exch}\frac{1 - f_{\rm mol}}{f_{\rm mol}},
\end{equation}
where $\tau_{\rm exch} = 20\,\rm{Myr}$ \citep{Murray2010} is the
lifetime of molecular clouds and the timescale for exchange from the
dense to the diffuse ISM.
The quantity $f_{\rm mol}$ is the molecular
fraction of the cold gas, computed as described in section
\ref{sec:sam}.

The growth rate of element $j$ on dust grains $\dot{M}^{\rm
  growth}_{j,\rm{dust}}$ can be expressed as \citep{Zhukovska2014}: 
\begin{equation}
\dot{M}^{\rm growth}_{j,\rm{dust}}  =\frac{1}{\tau_{\rm
    exch,eff}}\bigg(f_{j,\rm{cond}} M_{j,\rm{metal}} - M_{j,\rm{dust}}\bigg),
\end{equation}
where $f_{j,\rm{cond}}$ is the mass fraction of metal species $j$ in
a molecular cloud condensed into dust at the end of the molecular
 cloud lifetime. The quantity $M_{\rm metal,j}$ is the total mass of element $j$
(the sum of the elements located in the cold gas and locked up in the
dust).

We adopt an approximation from
\citet{Zhukovska2008} to describe $f_{j,\rm{cond}}$:
\begin{equation}
\label{eq:equation10}
f_{j,\rm{cond}} = \bigg[\big(f_{j,0}(1 + \tau_{\rm exch}/\tau_{\rm
  acc})\big)^{-2} + 1\bigg]^{-1/2},
\end{equation}
where $f_{j,0}$ is the initial degree of condensation for dust species
$j$, and $\tau_{\rm acc}$ is the timescale for dust growth.
  It is important to note that in this framework $f_{j,\rm{cond}}$ can
  be approximately equal to or even smaller than $f_{j,0}$ when $\tau_{\rm exch}/\tau_{\rm
  acc}<<1$, effectively reducing the net dust growth rate to
  zero.

We adopt an expression for the timescale for dust growth that has been
used in many previous works
\citep{Hirashita:2000,Inoue2003,Asano2013,deBennassuti2014,Schneider2016} :
\begin{equation}
\label{eq:tau_acc}
\tau_{\rm acc} =  \tau_{\rm acc,0} \times \bigg(\frac{n_{\rm mol}}{100
  \,\rm{cm}^{-3}}\bigg)^{-1}\,\bigg(\frac{T_{\rm
    cl}}{50\,\rm{K}}\bigg)^{-1/2}\,\bigg(\frac{Z_j}{Z_{j,\odot}}\bigg)^{-1}.
\end{equation}
$\tau_{\rm acc,0}$ is the timescale of dust growth in Milky Way
molecular clouds and is treated in this work as a free parameter. The
quantity $T_{\rm cl}$ is the temperature in molecular clouds, which we
assume to be 50 K \citep{Wilson1997}. The variable $n_{\rm mol}$ is
the volume density of molecular clouds and $Z_j$ the gas-phase
abundance of species $j$ with respect to the solar abundance. We will
discuss our choice for $\tau_{\rm acc,0}$ in Section
\ref{sec:free_parameters}.

Equation \ref{eq:tau_acc} is derived from the expression for dust mass growth rate in
clouds \citep[e.g.,][]{Dwek1998,Hirashita2000,Inoue2003}:
\begin{equation}
\dot{M}^{\rm growth}_{j,\rm{dust}}  = f_{\rm mol} N \pi \langle a^2\rangle
\alpha \rho^{\rm gas}_{\rm Z}\langle v \rangle,
\end{equation}
where N is the number of dust grains, $\langle a^2\rangle$ is the 2nd moment of the grain size $a$, $\alpha$
is the mean sticking coefficient of metals, $\rho^{\rm gas}_{\rm Z}$
is the mass density of gaseous metals that are not contained in dust, and
$\langle v \rangle$ is the mean velocity of metals in the gas phase. This
derivation is presented in detail in \citet{Asano2013}, and
implicitly assumes spherical dust grains, a sticking coefficient $\alpha=1$ and that the solid matter in dust grains has a
fixed mass density of 3 g cm$^{-3}$. This approach furthermore assumes a
  fixed mean  grain radius of $a=0.1 \mu$m. The grain sizes for
  dust produced by SNe are expected to be larger than $0.01 \mu$m
  \citep{Bianchi2007,Nozawa2007} and the grain size distribution of
  dust produced by AGB stars is thought to peak near $0.1 \mu$m
  \citep{Groenewegen1997,Winters1997,Yasuda2012,Asano2013}.

\begin{figure}
\includegraphics[width = 0.95\hsize]{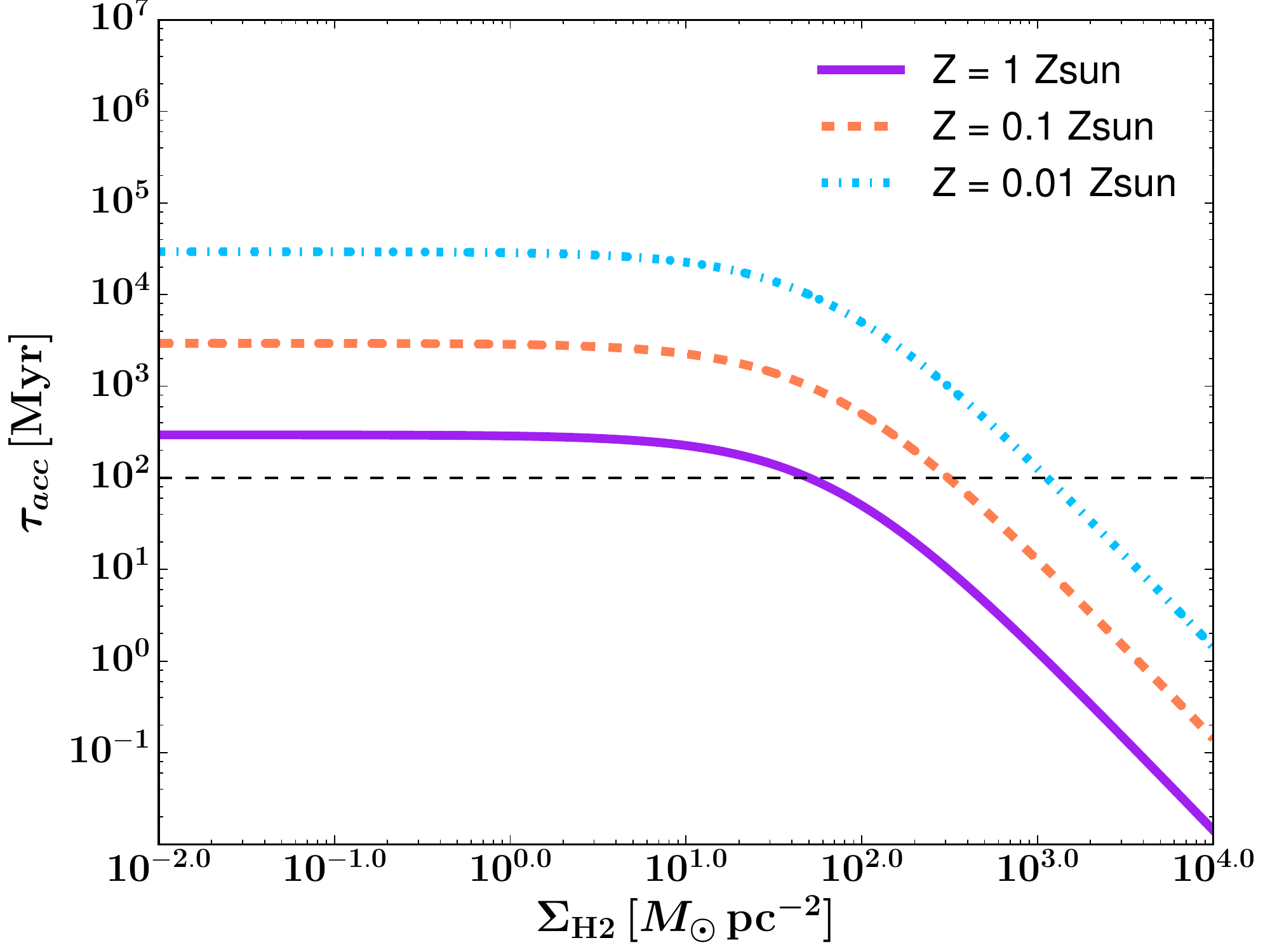}
\caption{The timescale for dust growth in the ISM as a function of gas
  surface density for different gas-phase metallicities (see Equation
  \ref{eq:tau_acc}). The horizontal dashed line marks the fixed
  timescale assumed in the `fix-tau' model variant. 
\label{fig:accretion_time}}
\end{figure}

Our SAMs do not provide volume
densities of individual clouds. To overcome this we express the SFR
surface density in terms of the volume density dependent free-fall
time of the molecular gas,
\begin{equation}
\label{eq:collapse}
\Sigma_{\rm SFR} = \epsilon \,\frac{\Sigma_{\rm H_2}}{t_{\rm
  ff}},
\end{equation}
where $\epsilon$ is the efficiency of
star-formation. Observations constrain this efficiency to $\sim$1\%
\citep{Krumholz2007,Krumholz2012,Krumholz2014}. The free-fall time is given by
\begin{equation}
\label{eq:tff1}
t_{\rm ff} = \sqrt{\frac{3\,\pi}{32\,G\,n_{\rm mol}}},
\end{equation}
with $G$ the gravitational constant.  In Equation
\ref{eqn:bigiel2} we presented a recipe that relates  SFR surface density to
 molecular hydrogen surface density. By combining Equation
\ref{eq:collapse} with Equation \ref{eqn:bigiel2} we can write a new
expression for the free-fall time
\begin{equation}
\label{eq:tff2}
t_{\rm ff} = \epsilon \,\bigg[A_{\rm SF} \, /(10 M_\odot {\rm pc}^{-2}) \left(1+
\frac{\Sigma_{H_2}}{\Sigma_{\rm H_2, crit}}\right)^{N_{\rm SF}}\bigg]^{-1}.
\end{equation}
We can now simply solve for $n_{\rm mol}$ by combining Equations
\ref{eq:tff1} and \ref{eq:tff2}. To illustrate this recipe we plot the accretion time scale
of dust as a function of molecular hydrogen surface density and
metallicity in Figure \ref{fig:accretion_time}. 

The gas surface density and molecular hydrogen fraction are important
physical parameters when calculating the growth rate of dust due to
metal accretion. These parameters are not constant throughout a galaxy,
but vary with radius. We assume that the cold gas is distributed in an
exponential disc with scale radius $r_{\rm gas}$ and a central gas
surface density of $M_{\rm cold}\,_{\rm gas}/(2\pi r_{\rm gas}^2)$,
where $M_{\rm cold}\,_{\rm gas}$ is the mass of all cold gas in the
disc. \citet{Bigiel:2012} find that this is a good approximation for
nearby spiral galaxies.
We divide the gas disc into radial annuli and compute the fraction of
molecular gas and the growth rate of dust in each annulus as described
above. The total growth-rate of dust in the galaxy at each time step
is then calculated using a fifth order Runge Kutta integration scheme.

\subsubsection{The Dwek growth model}
\label{Sec:Dwek_approach}
A more traditionally used approach for the growth of dust through
accretion in the dense ISM was presented in
\citet{Dwek1998}. Following this approach, the growth rate of dust is
given by
\begin{equation}
\dot{M}^{\rm growth}_{j,\rm{dust}}  = \bigg(1 - \frac{M_{j,\rm{dust}}
}{M_{j,\rm{metal}}}\bigg)\bigg(\frac{f_{\rm H2}M_{j,\rm{dust}}}{\tau_{\rm acc}} \bigg).
\end{equation}
The biggest difference between the \citet{Dwek1998} and the
\citet{Zhukovska2008} approach is that the \citet{Dwek1998} approach
does not account for the exchange time over which all of the ISM in a
galaxy is cycled through molecular clouds. This lowers the net
efficiency of dust growth in the ISM for the \citep{Dwek1998}  approach, especially in galaxies with low
molecular hydrogen fractions. Furthermore, as we discussed before, the Zhukovska
approach yields dust accretion rates close to zero in environments
where the metal accretion time is much larger than the cloud exchange
time. In the Dwek approach, dust can still grow efficiently in such
environments. 

\subsection{Dust destruction}
\label{sec:destruction}
There are a number of processes that can destroy dust in the
ISM. SN blast waves in particular efficiently destroy dust grains
through inertial and thermal sputtering
\citep[e.g.,][]{Dwek1980,McKee1989,Jones1994,Jones1996}. The time scale over
which dust grains in the ISM
are destroyed due to SN blast waves is given by \citep{Dwek1980,McKee1989}
\begin{equation}
\tau_{\rm destruct} = \frac{M_{\text{HII+HI}}}{M_{\rm cleared}f_{\rm SN}R_{\rm SN}},
\end{equation}
where $M_{\text{HII+HI}}$ is the mass of diffuse gas (ionised and
atomic) in the galaxy, $M_{\rm cleared}$ is the mass of gas cleared
from dust by one supernova event (which is different for carbonaceous
and silicate grains, Section \ref{sec:free_parameters}), $f_{\rm SN}$
accounts for the effects of correlated SNe (SNe exploding
  in existing super-bubbles created by previous SNe in the association)
and SNe out of the plane of the galaxy, and $R_{\rm SN}$ is the rate
of supernova type I and type II combined. We note that we assume here
that destruction in the dense cold ISM is inefficient, because shock
velocities are lower, and that destruction only works in the warm ISM
\citep[in our case \hi and \hii; ][]{Jones1996}.\footnote{Though see
  \citet{Temim2015} who find $\tau_{\rm destruct}$ is only weakly
  dependent on density.}.

The supernova rate is calculated as a part of the GCE model presented
in Section \ref{sec:GCE}. The destruction rate of the dust is then
given by
\begin{equation}
\dot{M}^{\rm destruct}_{j,\rm{dust}} = \frac{M_{j,\rm{dust}}}{\tau_{\rm destruct}}\,\rm{M}_\odot\,\rm{yr}^{-1}.
\end{equation}

\subsection{Star Formation, Infall, Outflow, and Mergers}
There are a number of physical processes that act on the dust in the
ISM that we have not yet discussed in detail. These are processes that
affect the ISM as a whole, and therefore also the dust within it. Our
model includes the following additional dust-related processes.
\begin{itemize}
\item When stars are formed out of the ISM the dust that is locked up
  in these stars is assumed to be destroyed and added to the metal
  content of the stars. The rate at which dust is locked up in stars is
  proportional to the SFR of the galaxy and equals
  $\dot{M}_{j,\rm{dust}}^{\rm SF} = D_j SFR$, where $D_j$
marks the DTG ratio for element $j$.
\item SN and AGN can heat up and expel gas and dust from the ISM into
  the halo or even further out. We assume that the DTG ratio of
  the heated ISM and outflows equals the average DTG ratio of the ISM. The rate
  $\dot{M}_{j,\rm{dust}}^{\rm out}$ at which dust is removed from the galaxy is therefore directly proportional to
  the total ISM mass heated up or blown out by AGN and SNe through $D_j$. Similar
  to the metals, dust can also be ejected from and reaccreted into
  the halo.
\item The rate at which dust accretes onto the galaxy is proportional
  to the cooling rate of the gas through $D_{\rm j,hot}$, the dust
  abundance of element $j$ in the hot gas.
\item Whenever a central and a satellite galaxy merge, 
  the dust undergoes exactly the same processes as the cold gas,
  scaled by the DTG ratio $D_j$ for element $j$. A detailed
  description of the processes acting on the cold gas during mergers is given in S08.
\end{itemize}

\subsection{Dust in the hot halo}
\label{sec:hot_halo}
Once dust is ejected into the hot halo of a galaxy, it can be
destroyed by thermal sputtering and grain-grain collisions \citep{Draine1979hot}. We follow
the work by \citet{Tsai1995}, \citet{Hirashita2015}, and \citet{McKinnon2016} to include the
effects of thermal sputtering. The sputtering rate for a grain of
radius $a$ in gas with a density $\rho$ and temperature $T$ is
\begin{equation}
\frac{\rm{d}a}{\rm{d}t} = -(3.2\times
10^{-18}\rm{cm}^4\rm{s}^{-1})\bigg(\frac{\rho}{m_p}\bigg)\bigg[\bigg(\frac{T_0}{T}\bigg)^\omega + 1\bigg]^{-1},
\end{equation}
where $m_p$ is the proton mass, $T_0=2\times 10^6\rm{K}$ is the
temperature above which the sputtering rate flattens,
and $\omega=2.5$ controls the low-temperature scaling of the
sputtering rate. The associated sputtering time-scale for the grain is
\citep{Tsai1995}
\begin{equation}
\label{eq:sputter_halo}
\tau_{\rm sp} = 0.17\rm{Gyr}\bigg(\frac{a_{-1}}{\rho_{-27}}\bigg) \bigg[\bigg(\frac{T_0}{T}\bigg)^\omega + 1\bigg],
\end{equation}
where $a_{-1}$ is the grain size in units of 0.1$\mu$m and
$\rho_{-27}$ is the gas density in units of $10^{-27}$g cm$^{-3}$. For the
  temperature $T$ we take the virial temperature of the
  halo. Following \cite{McKinnon2016} we now estimate the
  destruction rate of dust species $j$ in the hot halo due to thermal
  sputtering as 
\begin{equation}
\dot{M}_{j,\rm{dust}}^{\rm sputtering} = -\frac{M_{j,\rm{dust}}}{\tau_{\rm{sp}}/3}.
\end{equation}
The dust that is destroyed by thermal sputtering is added to the
metals in the hot halo.

 As discussed earlier, the reservoir of dust in the hot
  halo is distinct from the dust in the ejected reservoir. Because of the
  poorly defined nature of ejected reservoir (a combination of the CGM
  and IGM), it is unclear what density and/or temperature distribution
  to assume. For simplicity, we assume here that the ejected reservoir has
  the same properties in terms of density and temperature as the hot
  halo, resulting in the same timescales for sputtering as presented
  in Equation \ref{eq:sputter_halo}. In reality the densities and
  temperature of the IGM might be lower and higher, respectively,
  allowing for even more efficient sputtering. The calculated
  sputtering rates for the ejected reservoir should therefore be regarded
  as lower limits.

We do not self-consistently include dust cooling
  channels in our models in the present work
  \citep{Ostriker1973,Cantalupo2010,GnedinHollon2012}. Instead,
  we treat the dust as `normal metals' when calculating the
  cooling rate of the hot gas (i.e., the cooling rates are based on
  the temperature and the sum of the metals and dust in the hot
  gas). We will include a self-consistent treatment of the dust
  cooling physics in a future work.

\begin{table*}
 \caption{\label{tab:model_variants} Summary of the model
   variants. Unless listed in the changed parameters column, all parameters are as
   listed in Table \ref{tab:free_parameters}. The `dwek98' and
   `dwek-evol' model variants are briefly discussed in the main body
   of this work and further presented in the appendix.}
 \begin{tabular}{ccc}
\hline
 Name & Changed Parameters & Growth model\\
\hline
fiducial & ... & Zhukovska\\
&\\
no-acc &  no dust growth, $\delta^{\rm AGB}_{\rm carbon} = 1.0,\,\delta^{\rm
            AGB}_{\rm other} = 1.0,\,\delta^{\rm SN}_{\rm carbon} =
            1.0,\,\delta^{\rm  SN}_{\rm other} = 1.0$& Zhukovska\\
&&\\fix-tau & $\tau_{\rm acc} = $100 Myr & Zhukovska\\
&&\\high-cond & $\delta^{\rm AGB}_{\rm carbon} = 1.0,\,\delta^{\rm
            AGB}_{\rm other} = 0.8,\,\delta^{\rm SN}_{\rm carbon} =
            1.0,\,\delta^{\rm  SN}_{\rm other} = 0.8 $& Zhukovska\\
\hline
\hline
dwek98 & $\delta^{\rm AGB}_{\rm carbon} = 1.0,\,\delta^{\rm
            AGB}_{\rm other} = 0.8,\,\delta^{\rm SN}_{\rm carbon} =
            1.0,\,\delta^{\rm  SN}_{\rm other} = 0.8, \tau_{\rm acc} =
         $150 Myr & Dwek\\
&&\\dwek-evol & $\tau_{\rm acc,0} =
         $7 Myr& Dwek\\
\hline
\hline
 \end{tabular}
 \end{table*}

\subsection{Physical parameters}
\label{sec:free_parameters}
We chose the values of the main physical parameters in this work
either based on theoretical work or
by tuning our model to observations in the local
Universe. The parameters and their respective values are all listed in
Table \ref{tab:free_parameters}.

In our fiducial model, we take a fixed dust condensation efficiency for AGB
stars of $\delta_{j,\rm{dust}}^{\rm AGB} = 0.2$. 
Theoretical models have begun to explore the dust condensation
efficiency as a function of a star's mass and metallicity, and found
variations around $\delta_{j,\rm{ dust}}^{\rm AGB} = 0.2$ \citep[from
  $\sim 0.05$ to 0.6 or even higher,
  e.g.,][]{Ferrarotti2006,Zhukovska2008,
  Valiante2009,GallReview2011,Piovan2011,Ventura2012,Nanni2013,Gioannini2017}. While
not included in our model, we acknowledge that varying condensation
efficiencies depending on the stellar type may be more realistic,
especially in very low-metallicity regimes. We will use
  metallicity dependent condensation efficiencies in a forthcoming
  work.  We note that the adopted dust-condensation efficiency for
AGB stars is significantly lower than the numbers used in
\citet{Dwek1998} and some other recent works that assume
$\delta_{j,\rm{dust}}^{\rm AGB} = 1.0$
\citep[e.g.,][]{Bekki2013,Bekki2015,McKinnon2015,McKinnon2016}.

The dust-condensation efficiency that we adopt for SNe corresponds to the
condensed dust that survives the passage of the reverse SN
shock. Observational work on the condensation of dust in SN ejecta
typically probes condensation before the passage of the reverse
shock. A direct comparison between the different condensation
efficiencies should therefore be treated with caution. Theoretical
work by \citet{Bianchi2007} suggests dust condensation efficiencies of
$\sim 40-100$ percent before the reverse SN shock. Only $\sim 2-20$
percent of the initial dust mass survives the reverse shock,
corresponding to a condensation efficiency $\delta_{\rm dust,j}^{\rm
  SN}$ of $\sim 1 - 20$\%.  The dust content of low-mass
and low-metallicity galaxies is fully determined by the condensation
efficiency of SNe (and the destruction rate of the dust). These objects are too young for AGB stars to
contribute significantly to the dust mass and the growth of dust in
their ISM is not yet efficient \citep{Zhukovska2014}. We choose a SN
dust condensation efficiency of $\delta_{\rm
  dust,j}^{\rm SN} =0.15$, which yields good agreement with the dust
mass in low mass galaxies at $z=0$ (Figure
\ref{fig:mstar_mdust}). The
resulting dust masses per SN event are in the same range of dust yields
predicted for SNII events as presented in for example \citet{Bianchi2007} and
\citet{Piovan2011}. For simplicity, we have assumed that SNIa have the same condensation
efficiency as SN type II (though see for example \citet{Nozawa2011},
\citet{Dwek2016}, and \citet{Gioannini2017} for arguments againts dust
production in SNIa). We will adopt
yield tables based on supernova type and mass in a future work.

We calibrate the normalisation of the time-scale for dust growth
$\tau_{\rm acc,0}$ (Eqn.~\ref{eq:tau_acc}) using the observational
constraints on dust mass in massive galaxies 
\citep{Ciesla2014,RemyRuyer2014}.
We find a time scale of $\tau_{\rm acc,0} = 15$ Myr, consistent with
values adopted in earlier simulations using a similar approach for
dust accretion in the ISM
\citep[e.g.,][]{Hirashita2000,Asano2013,deBennassuti2014,Feldmann2015,Schneider2016}.

The adopted values for the physical parameters that describe the
destruction of dust by SN blast waves are based on work by
\citet{Slavin2015}. The authors estimate the gas mass cleared of dust
by a SN event ($M_{\rm cleared}$) based on new calculations of grain
destruction in evolving, radiative SN remnants. A distinction has to
be made between carbonaceous grains (in our model made up by carbon) and
silicates (in our model made up by the other refractory elements tracked in our
model). \citet{Slavin2015} find $M_{\rm cleared,carbonaceous} =
600\,\rm{M}_\odot$ and $M_{\rm cleared,silicate} = 980
\,\rm{M}_\odot$ for carbonaceous and silicate grains,
respectively. These numbers are about a factor of 1.5 lower than
estimates for the LMC \citep{Zhukovska2008}. The parameter that
accounts for the correlated nature of supernova events and
supernovae occurring outside of the plane of galaxies is set to
$f_{\rm SN}=0.36$. Observations have found similar estimates for
$f_{\rm SN}$ in the Milky Way and LMC
\citep{McKee1989,Zhukovska2013AGB,vanLoon2015}.

\subsection{Model variants}
We consider a number of different variants of our model. These variants are
chosen to illustrate different scenarios for dust formation, and to
provide a comparison with other models and simulations in the
literature. We summarise the model variants in Table
\ref{tab:model_variants}.

The first variant is our fiducial model. This variant adopts the
free parameters as discussed in the previous section and listed in
Tabel \ref{tab:free_parameters}. 

\begin{figure*}
\includegraphics[width = 0.95\hsize]{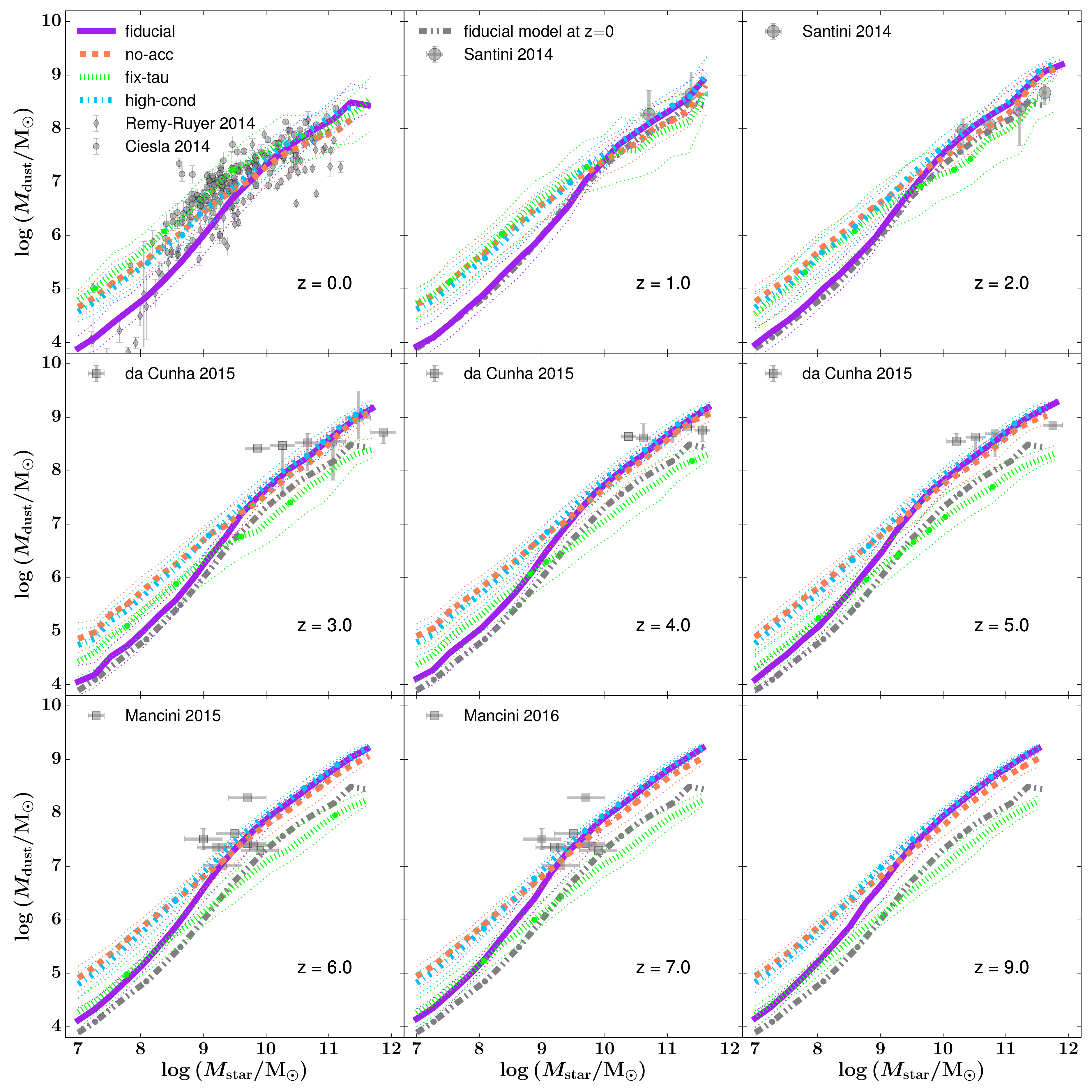}
\caption{\label{fig:mstar_mdust} The dust mass of galaxies as a function of their
  stellar mass from redshift $z=9$ to $z=0$, for our four model
  variants. Thick lines mark the 50th percentiles, whereas the lower
  and upper thin dotted lines mark the 16th and 84th percentiles. The $z=0$
  prediction of our fiducial model is shown as a dashed double dotted grey line in
  the higher redshift bins for comparison. Model predictions
  are compared to observations from \citet{Ciesla2014}  \citet{Remy-ruyer2014} at $z=0$,
  \citet{Santini2014} at $z=1$ and $z=2$, \citet{daCunha2015} at
  $z=3$, 4, and 5, and a compilation of data in
  \citet{Mancini2015} at $z=6$ and $7$, taken from \citet{Kanekar2013}, \citet{Ouchi2013}, \citet{Ota2014},
    \citet{Maiolino2015}, \citet{Schaerer2015}, and
    \citet{Watson2015}.}
\end{figure*}

The second model variant, `no-acc', is motivated by recent work by
\citet{Ferrara2016}. These authors suggested that the contribution of
dust growth on grains to the dust mass of galaxies is negligible. They
argue that accretion does take place in dense environments, but the
accreted materials are locked up in icy water mantles, which
photo-desorb quickly after the grains return to the diffuse ISM. To
mimic this process, we completely turn off the growth of dust through
accretion onto grains. In order to still reproduce the $z=0$ relation
between stellar mass and dust mass, we increase the efficiency of dust
condensation in stellar ejecta. We therefore test the extreme case in which the
condensation efficiency is 100\%, for AGB ejecta and SN ejecta in
order to reproduce the dust content of local galaxies (Figure \ref{fig:mstar_mdust}).

The third variant, `fix-tau', assumes a fixed timescale for dust
accretion in the ISM. Cosmological simulations of galaxy formation
that include dust chemistry have so far often assumed a fixed
timescale for the accretion of dust, independent of gas density and/or
gas phase metallicity \citep{Bekki2013,McKinnon2015}. We choose a
value of $\tau_{\rm acc} = 100$ Myr, in agreement with the accretion
time scales adopted in \citet{Bekki2013} and \citet{McKinnon2015}. For
clarity, we plot the accretion timescale $\tau_{\rm{acc}}$ as a
function of gas surface density and metallicity in Figure
\ref{fig:accretion_time}, and mark the fixed timescale of 100 Myr as a
dashed horizontal line. This figure immediately shows that the
metallicity and density dependent accretion timescale can differ
significantly from the fixed timescale for the range of densities and
metallicities expected in different galaxies and sub-galactic
environments.

The fourth variant, `high-cond', assumes a much higher condensation
efficiency for dust in stellar ejecta, as adopted in the fiducial
models of \citet{Bekki2013}, \citet{McKinnon2015}, and
\citet{McKinnon2016}. We take $\delta_{j}^{\rm AGB} = 1.0$ for carbon
and $\delta_{j}^{\rm AGB} = 0.8$ for the other elements. Similarly, we
assume $\delta_{j}^{\rm SN} = 1.0$ for carbon and $\delta_{j}^{\rm SN}
= 0.8$ for the other elements. The condensation efficiencies are also
close to the values adopted in \citet{Dwek1998}. In this variant the
timescale for dust accretion is a function of gas density and
gas-phase metallicity, as in our fiducial model.

We also run two model variants that adopt the recipe for
  dust growth in the ISM as adopted in for instance \citet{Dwek1998},
  \citet{Calura2008}, and \citet{McKinnon2015}. For the first
  model variant, `dwek98' we adopt exactly the same parameters as those presented
  in \citet[also used in for example \citet{McKinnon2015}]{Dwek1998}. This
  model variant adopts high condensation efficiencies, identical to
  the `high-cond' model. Furthermore it assumes a fixed timescale for
  dust-accretion of 150 Myr. In the second model variant, 'dwek-evol', we adopt the
  same condensation and SN destruction efficiencies as in our
  fiducial model. We furthermore let $\tau_{\rm acc}$ evolve as a function
  of metallicity and density, similar to our fiducial model. We adopt $\tau_{\rm acc,0} = 7$ Myr,
  which yields the best agreement with the observational constraints at
  $z=0$ for the dust mass of galaxies as a function of their stellar
  mass. The results of the latter two model variants are briefly
  discussed in the main body of this work and further presented in
  Appendix \ref{sec:appendix_dwek}. 

\section{Measuring dust masses}
\label{sec:observing_dust}

In this work we compare a wide range of observational estimates of the
dust content of galaxies. These estimates are obtained in several
different ways. Here we summarize the main existing approaches for
obtaining observational estimates of the dust mass in galaxies.

The infrared emission of galaxies is widely used to estimate their
dust content.  The modelling of their IR spectral energy distribution
(SED) has been especially improved within the past 10-20 years with
the arrival of far-IR (Spitzer, Herschel) and sub-mm (Herschel,
Submillimetre Common-User Bolometer Array (SCUBA), Balloon-borne Large
Aperture Submillimeter Telescope (BLAST), and ALMA ground
instrumentation and space telescopes, adding much better constraints
on the cold dust regime.

Assuming that galaxies behave as optically thin single or double
temperature sources, the SEDs of local and high redshift galaxies have
been (and still are) widely modeled using 1 or 2 component modified
blackbody (MBB; with $I_{\nu} = A B_{\nu}(T) \nu^\beta$) fitting
techniques. This is the case for half of the observationally derived
dust masses we will be quoting in this paper \citep{Dunne2003,
  Eales2009, Clark2015} as well as papers using the \texttt{MAGPHYS}
model \citep{daCunha2008,daCunha2015, Clemens2013} in which the dust
components in thermal equilibrium are considered as two MBBs with
different dust power-law emissivity indices ($\beta_{\rm warm}$ and
$\beta_{\rm cold}$).

Dust in galaxies is, in reality, heated by a distribution of starlight
intensities and strong local variations of the interstellar radiation
field are expected, especially affecting the emission in the mid-IR
regime. More complex dust models have been developed to account for
the distribution of radiation field intensities as well as to propose
a more physical treatment of the dust composition, using a mixture of
amorphous silicate and carbonaceous grains. Such models include, for
example, the \citet{Draine2007}, \citet{Galliano2011}, \texttt{GRASIL}
\citep{Silva1998}, \texttt{CIGALE} \citep{Burgarella2005}, and
\texttt{THEMIS} \citep{Jones2016} models. In this paper, we will also
make use of dust mass estimates derived using these more complex
models, such as in \citet{Sandstrom2013}, \citet{RemyRuyer2014},
\citet{Ciesla2014} and \citet{Santini2014}.

\citet{Dale2012} showed that dust masses derived from single
temperature MBB fits on far-IR fluxes are about a factor of two lower
than those derived using a more complex formalism on the full IR
SED. These dust mass discrepancies could be larger for galaxies with
colder dust. One has to keep this in mind when comparing dust masses
derived using different methods in the same diagram.

Significant uncertainties still remain on the dust opacity
itself. Observations with Herschel and Planck have shown that the dust
emissivity might change on local scales, resulting in a more difficult
interpretation/modelling of the submm slope of the galaxy global
SED. \citet{Galliano2011} showed for instance that replacing graphite
grains (standardly used to model carbon dust, with $\beta\sim2$) by
more amorphous carbon grains \citep[for instance][$\beta \sim
  1.7$]{Zubko1996} would lead to a decrease of the dust mass estimates
by a factor of 2.5--3 because amorphous carbon grains absorb more
light, and thus require less mass to reproduce the same luminosity.
Recently, Planck observations have shown that in the diffuse ISM,
estimates for the attenuation in the V band, $A_{\rm
    V}$, from the \citet{Draine2007} model are a factor of three
larger than values of $A_{\rm V}$ derived from optical estimates from
quasars observed in the Sloan Digital Sky Survey
\citep{Planck2014XVII}. These new results indicate that some of the
physical assumptions regarding the dust opacity that are incorporated
in current SED models should probably be revised.

Most of the dust mass estimates quoted in this paper were derived
using MBB with $\beta_{\rm cold}=2$ or more complex models with
graphite, except results taken from \citet[MBB, with $\beta$
  free]{Vlahakis2005}, \citet[MBB, with $\beta =1.5$]{Dunne2011} and
\citet[ MBB, $\beta=1.5$]{Mancini2015}. Again, one has to keep in mind
that this choice of the emissivity index directly impacts the derived dust
mass.

For high-redshift galaxies, the contribution of a cold dust component
(T $<$ 20K) is still not well constrained but submm/mm observations
using ALMA now enable us to extend the SED coverage toward the cold
regime, lowering the uncertainties on the dust masses of these objects.
In our study, we will use ALMA-based dust estimates from
\citet{daCunha2015} and \citet{Mancini2015}.

Finally, the DTM ratio can be measured indirectly via optical/UV
absorption-line spectroscopy, for example using Fe or Zn
lines. Abundance ratios of elements that strongly deplete onto dust
grains and elements that are weakly depleted provide an estimate for
the DTM ratio. This paper will use the DTM estimates obtained by
\citet{DeCia2013,DeCia2016} and \citet{Wiseman2016}.

\begin{figure*}
\includegraphics[width = 0.95\hsize]{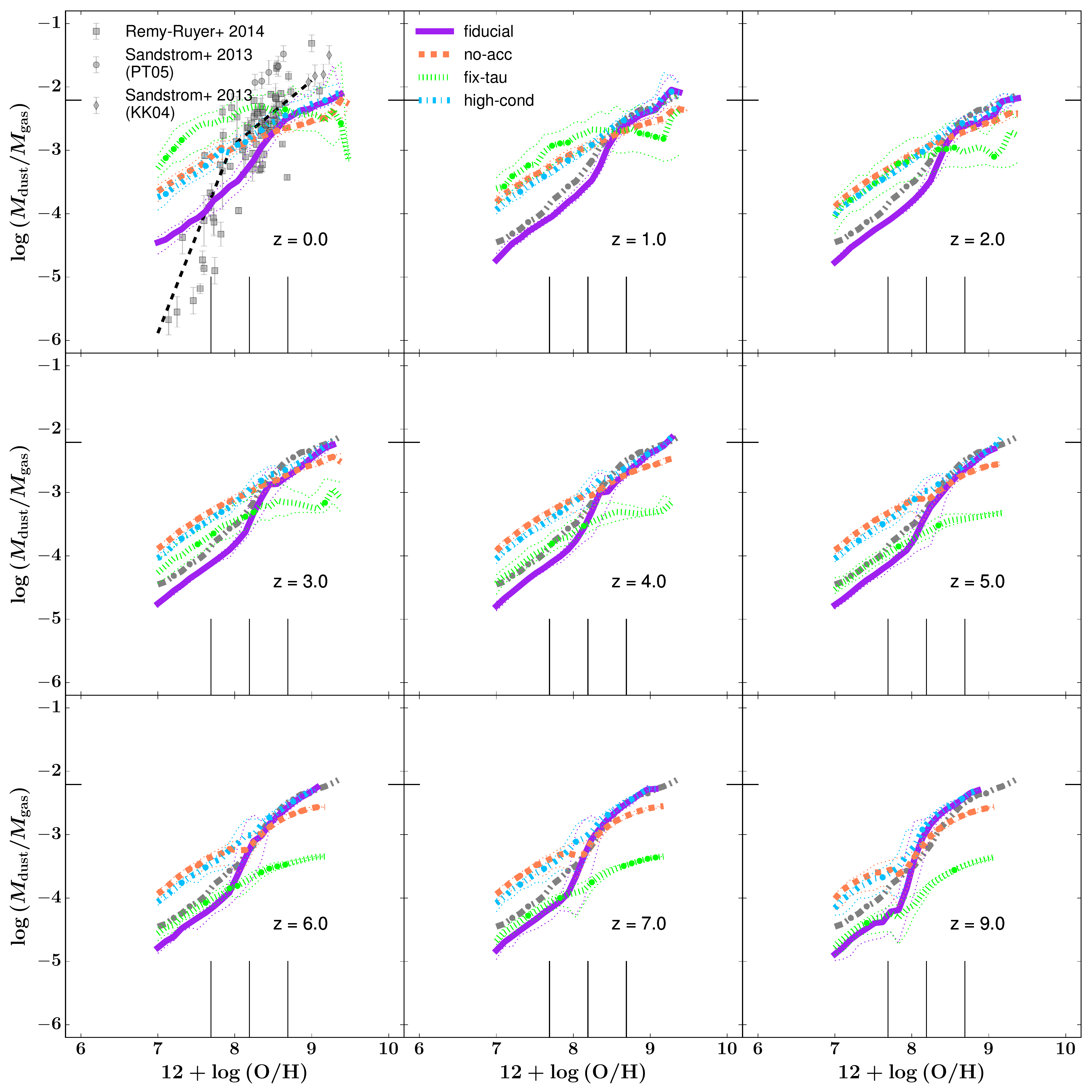}
\caption{The DTG ratio of galaxies as a function of their
  metallicity from redshift $z=9$ to $z=0$, for our four model
  variants. Thick lines mark the 50th percentiles, whereas the narrow
  dotted lines mark the 16th and 84th percentiles (note that the scatter
  predicted by the different model variants for this plot is very small). The $z=0$
  prediction of our fiducial model is shown as a dashed double dotted grey line in
  the higher redshift bins for comparison. Model predictions at $z=0$
  are compared to the observations by \citet{Remy-ruyer2014} and
  \citep{Sandstrom2013}, as well as the best-fit to the observations
  (black dashed line) 
  presented in \citet{Remy-ruyer2014}. The 12
  + log(O/H) values corresponding to metallicities of 0.1, 0.5, and
  1.0 Z$_\odot$ are represented by small black vertical lines. The
  solar DTG ratio \citep[0.006;][]{Zubko2004} is represented
  for comparison as small horizontal black lines in every panel.\label{fig:dust_to_gas}}
\end{figure*}

\section{Results}
\label{sec:results}
In this section we present our predictions for the evolution of the
dust content of galaxies over a redshift range from $z=0$ to
$z=9$. Unless stated otherwise, we restrict our analysis to central star
 forming galaxies, selected using the criterion $\rm{sSFR} >
 1/(3t_H(z))$, where $\rm{sSFR}$ is the galaxy specific star-formation
 rate and $t_H(z)$ the Hubble time at the galaxy's redshift. This
 approach selects galaxies in a similar manner to commonly used
 observational methods for selecting star-forming galaxies, such as
 color-color cuts \citep[e.g.,][]{Lang2014}. In most figures we present the 14th,
 50th, and 86th percentile of the different model variants. The 50th
 percentile corresponds to the median, the 14th percentile
 corresponds to the line below which 14 per cent of the galaxies are
 located, whereas the 86th percentile corresponds to the line below
 which 86 per cent of the galaxies are located. Whenever we discuss
 the scatter of a model, we refer to the area between the 14th and
 86th percentile.

\subsection{Dust masses in galaxies}
In Figure \ref{fig:mstar_mdust} we present the dust masses of galaxies
as a function of their stellar mass from $z=0$ to $z=9$. We find good
agreement between the predictions of our fiducial model and the observed dust masses
at $z=0$ over the entire mass range probed. Our predictions are $\sim 0.5$ dex lower than the
observations by \citet{Ciesla2014} at $M_* < 10^9\,\rm{M}_\odot$, but
in good agreement with the \citet{Remy-ruyer2014} observations in this
mass range. We find that the
increasing slope between galaxy dust mass and stellar mass flattens a bit at
stellar masses $\sim10^{9.5}\,\rm{M}_\odot$, as also seen in the
observations by \citet{Remy-ruyer2014}.

We immediately notice differences at low stellar masses ($M_* <
10^9\,\rm{M}_\odot$) between our fiducial model and the other model
variants at $z=0$. All model variants predict dust masses
approximately 0.5 dex larger than those predicted by our fiducial
model for galaxies in this stellar mass range. This is driven by
higher efficiencies for the condensation of dust in stellar ejecta in
the `no-acc' and `high-cond' model variant and shorter dust-growth
timescales in low mass galaxies with low metallicities for the
`fix-tau' model variant. All model variants are in good agreement with
the observed dust masses for galaxies with $M_* >
10^{8.5}\,\rm{M}_\odot$.

We find that the relationship between galaxy dust
mass and stellar mass as predicted by our fiducial model is roughly
constant  from $z=2$ to $z=0$. It decreases a bit between redshifts
$z=3$ and $z=2$. The decrease is weak for galaxies with
$M_*<10^9\,\rm{M}_\odot$ ($\sim 0.1$ dex), and much stronger for more
massive galaxies ($\sim$0.5 dex;  we will show later that this is the
regime where the growth of dust in the ISM dominates).  At higher
redshifts, the relation between galaxy dust mass and stellar mass
remains constant with time. Our fiducial model successfully reproduces the dust masses
observed in galaxies from redshift $z=0$ to $z=7$. 

The `no-acc' and `high-cond' model variants predict a similar
evolution in galaxy dust masses as our fiducial model. In
  the case of the `high-cond' model this is driven by accretion of
  metals onto dust grains, which dominates over the dust from stellar
  ejecta, similar to our fiducial model (see Figure
  \ref{fig:formation_channel_high_cond}). In case of the `no-acc'
  model variant this is driven by a fairly constant ratio between dust
  destruction by SNe and the production of dust in stellar ejecta
  (Figure \ref{fig:formation_channel_no_acc}). The `no-acc' model
predicts dust masses slightly lower than those found by \citet{Santini2014},
but is in good agreement with the observations by \citet{daCunha2015}
and \citet{Mancini2015}. The `fix-tau' model on the other hand
predicts a reverse trend in the evolution between galaxy dust mass and
stellar mass. The dust mass of galaxies is constant from $z=9$ to
$z=4$, and increases by $\sim0.5$ dex from $z=4$ to $z=0$. This model
variant fails to reproduce the observational constraints in the
highest redshift bin.

We present and discuss our predictions for the `dwek98'
  and `dwek-evol' model variants in some detail in Appendix
  \ref{sec:appendix_dwek}. In summary, we find that the `dwek98' model
  variant predicts dust masses similar to the `high-cond' variant,
  driven by the high condensation efficiencies assumed for stellar
  ejecta. The `dwek-evol' model variant predicts dust masses in
  reasonable agreement with the constraints at $z=0$, but predicts
  dust masses up to 0.3 dex larger than the observational constraints
  in the redshift range $1\leq z \leq 5$. We disfavour the `dwek-evol'
  model variant based on this strong disagreement at $z>0$.

\begin{figure*}
\includegraphics[width = 0.95\hsize]{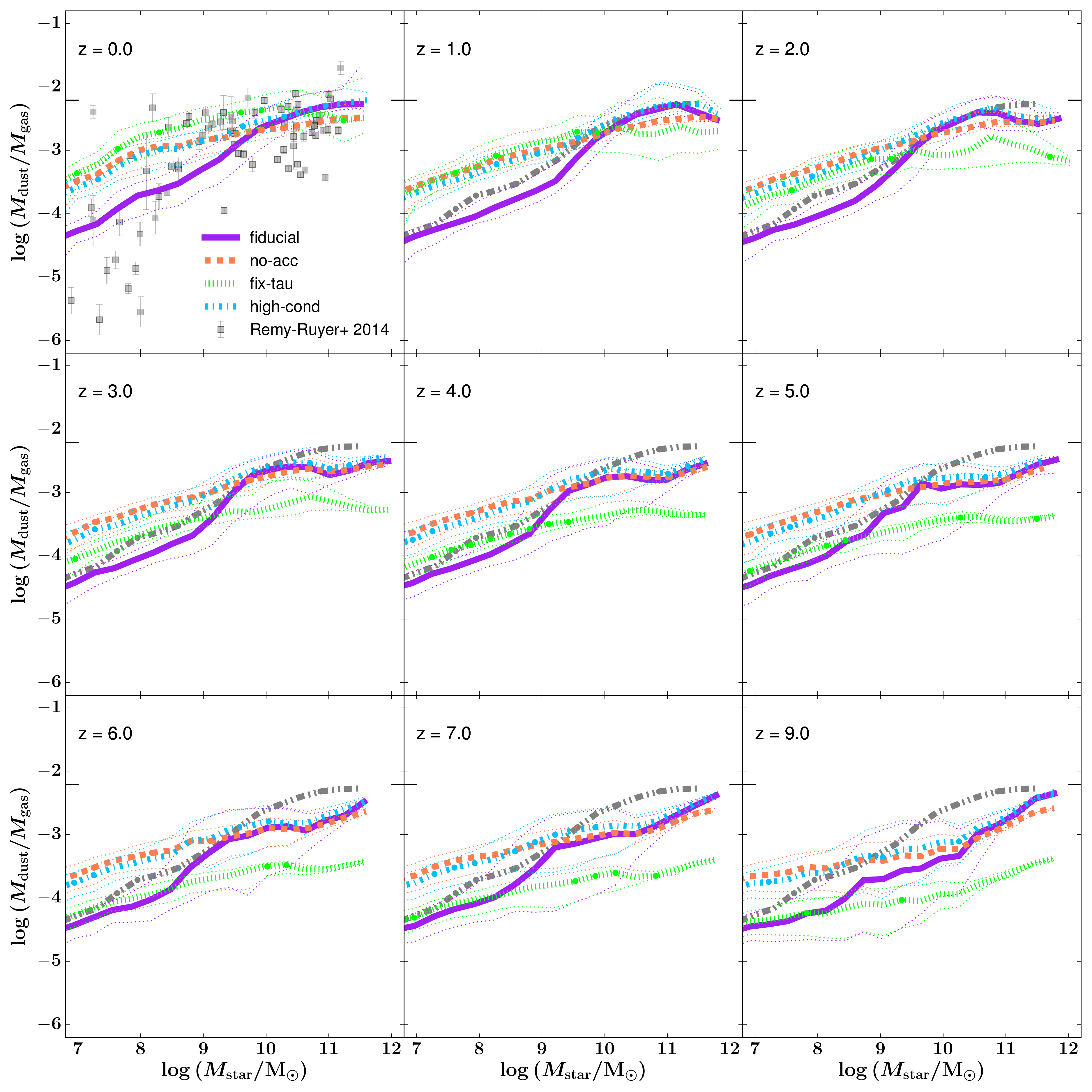}
\caption{The DTG ratio of galaxies as a function of their
 stellar mass from redshift $z=9$ to $z=0$, for our three model
  variants. Thick lines mark the 50th percentiles, whereas the narrow
  dotted lines mark the 16th and 84th percentiles. The $z=0$
  prediction by our fiducial model is shown as a dashed double dotted grey line in
  the higher-redshift bins for comparison. Model predictions at $z=0$
  are compared to the observations by \citet{Remy-ruyer2014}. The
  solar DTG ratio \citep[0.006;][]{Zubko2004} is represented
  for comparison as small horizontal black lines in every panel.\label{fig:mstar_dust_to_gas}}
\end{figure*}

\begin{figure*}
\includegraphics[width = 0.95\hsize]{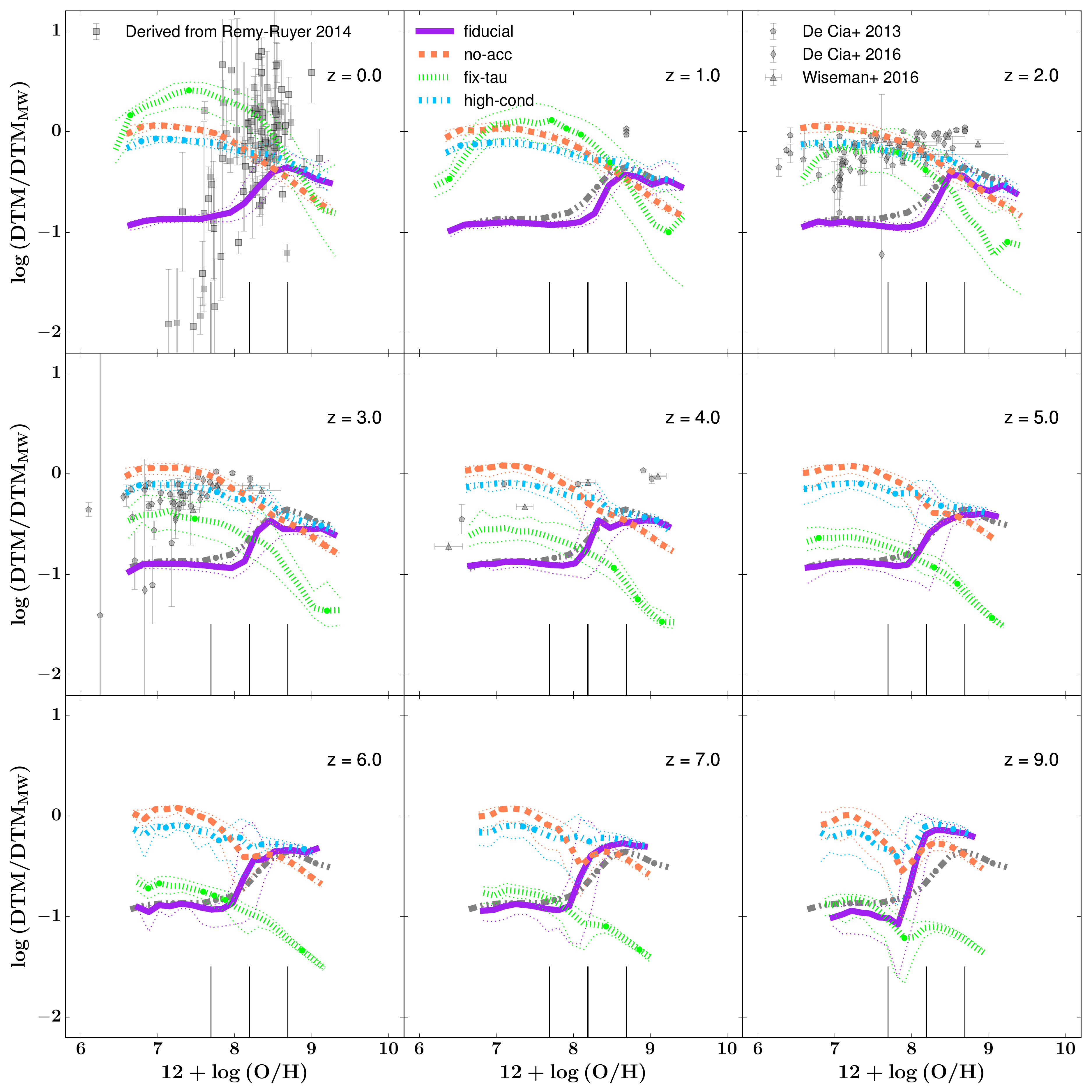}
\caption{The DTM ratio of galaxies as a function of their
  metallicity from redshift $z=9$ to $z=0$, for our four model
  variants. Thick lines mark the 50th percentiles, whereas the narrow
  dotted lines mark the 16th and 84th percentiles (note that the scatter
  predicted by the different model variants for this plot is very small). The $z=0$
  prediction by our fiducial model is shown as a dashed double dotted grey line in
  the higher-redshift bins for comparison. Model predictions at $z=0$
  are compared to the DTM ratios derived from \citet{Remy-ruyer2014}. The 12
  + log(O/H) values corresponding to metallicities of 0.1, 0.5, and
  1.0 Z$_\odot$ are represented by small black vertical lines. \label{fig:dust_to_metal}}
\end{figure*}

\begin{figure*}
\includegraphics[width = 0.95\hsize]{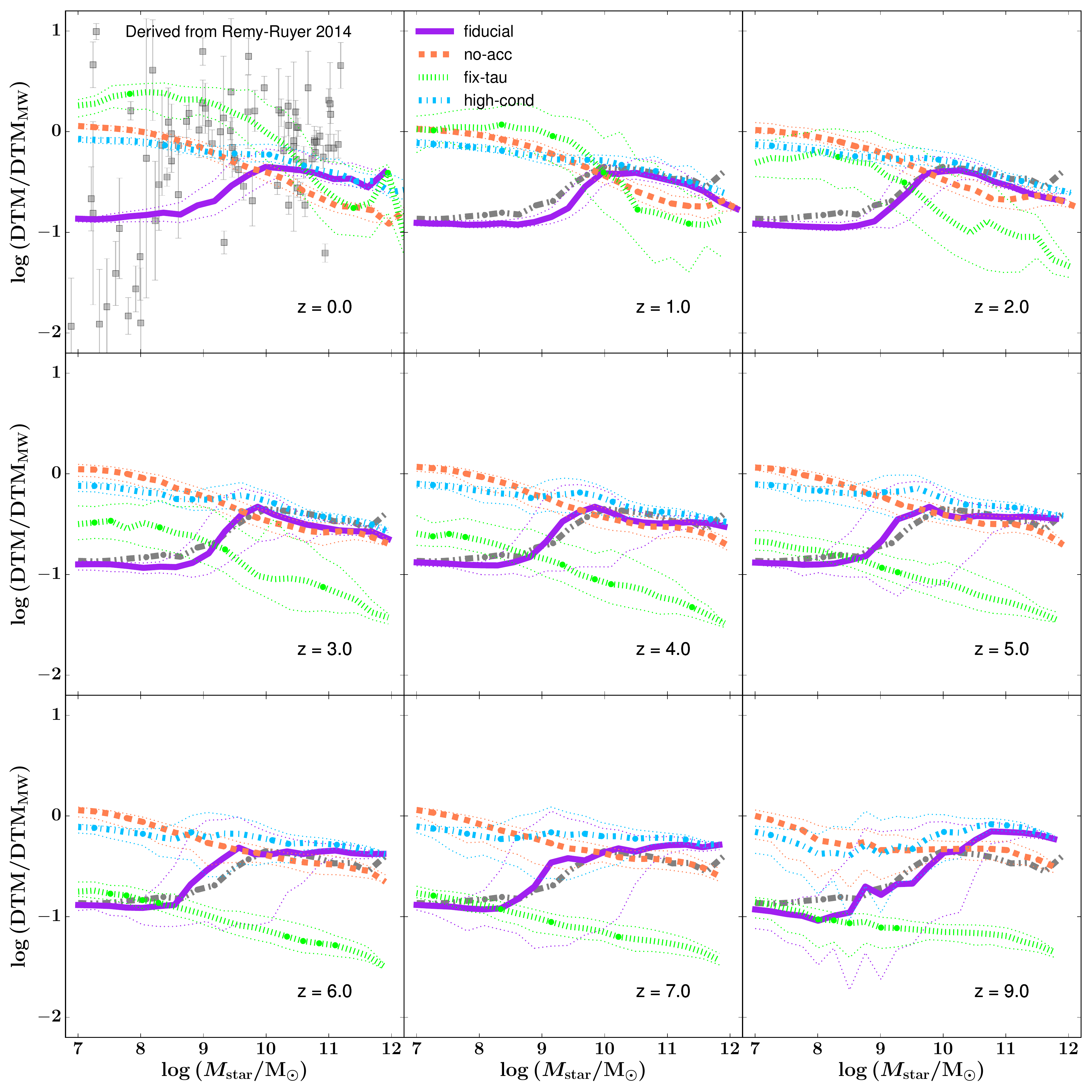}
\caption{The DTM ratio of galaxies as a function of their
 stellar mass from redshift $z=9$ to $z=0$, for our four model
  variants. Thick lines mark the 50th percentiles, whereas the narrow
  dotted lines mark the 16th and 84th percentiles. The $z=0$
  prediction by our fiducial model is shown as a dashed double dotted grey line in
  the higher-redshift bins for comparison. Model predictions at $z=0$
  are compared to the DTM ratios derived from \citet{Remy-ruyer2014}.\label{fig:mstar_dust_to_metal}}
\end{figure*}

\subsection{Dust-to-gas and dust-to-metal ratios}
\subsubsection{Dust-to-gas ratio}
We present the DTG ratio ($M_{\rm dust}/(M_{\rm HI} + M_{\rm H2})$) of
galaxies in Figure \ref{fig:dust_to_gas}. We find that the DTG ratio
predicted by our fiducial model rapidly increases with metallicity up
to a gas-phase metallicity of $\sim 0.7$ $Z_\odot$. Above this
metallicity the DTG ratio still increases, but the slope of this trend
is less steep. The predicted DTG ratios are 0.2 dex below the mean
trend in the observations and have a shallower slope as a
  function of metallicity. Furthermore, the data suggests
  that the turnover between the two power-law relations occurs at a
  metallicity of $\sim 0.7$ $Z_\odot$. Only at the lowest
metallicities ($< 0.1$ Z$_\odot$) do we predict DTG ratios slightly
higher than those suggested by observations. The use of
  condensation efficiencies that are a function of stellar metallicity
  could possibly alleviate this tension
  \citep{Ferrarotti2006,Gioannini2017}.

We find less good agreement between model predictions and observations
for the other model variants. Especially at metallicities lower than
0.1 Z$_\odot$, the other model variants predict DTG ratios that are
much too high. In the case of the `no-acc' and `high-cond' models this
is easily explained by the high condensation efficiencies in
SNe. In the case of the `fix-tau' model variant, this is
  explained by the short time-scale for dust growth in low metallicity
  environments compared to our fiducial model (for \h2 surface
  densities less than 300 M$_\odot\,\rm{pc}^{-2}$, see Figure
  \ref{fig:accretion_time}). Therefore, the formation of dust via
accretion plays a more important role and increases the DTG ratios at
low metallicities rapidly. At the highest metallicities, on the other
hand, the 'fixed-tau' model variant predicts DTG ratios that are lower
than our fiducial model. This is because the accretion times at these
high metallicities are never short enough to deplete
enough metals onto dust grains (see Figure
\ref{fig:accretion_time}). Similarly, the `no-acc' model also predicts
DTG ratios that are too low, due to the lack of growth of dust in the
ISM in general.

Our fiducial model predicts weak evolution in the relation between DTG
ratio and gas-phase metallicity. At metallicities less than 0.5
Z$_\odot$ the dust to gas ratio increases with $\sim$0.3 dex from
$z=9$ to $z=0$. At higher metallicities the relation between DTG ratio
and metallicity at $z>5$ is similar to the $z=0$ relation. The
relation then decreases by approximately 0.2 dex to $z=4$ and slowly
increases again with cosmic time to its $z=0$ value. We will discuss
the origin of this strange behaviour in Section
\ref{sec:disc_dust_to_gas}. \citet{Santini2010} discusses
  a population of sub-mm galaxies at a median redshift of $z=2$ with
  dust-to-gas ratios larger than 0.01 at metallicities lower than 0.5
  Z$_\odot$. Such galaxies are in tension with our model predictions.

We find basically no evolution in the relationship between DTG ratio and
gas-phase metallicity for the `no-acc' model variant. The predictions
by the `high-cond' model variant are very similar to our fiducial
model. The `fix-tau' model variant on the other hand predicts a strong
evolution in the DTG ratio of galaxies. This variant predicts an
increase in the DTG ratio of galaxies from $z=9$ to $z=0$ of almost an
order of magnitude, independent of gas-phase metallicity. This
increase is especially pronounced at redshifts $z<3$.

We show the DTG ratio of galaxies as a function of their stellar
masses in Figure \ref{fig:mstar_dust_to_gas}. The DTG ratio increases
with stellar mass. Our fiducial model reproduces the DTG
  ratios observed in local galaxies more massive than
  $10^{9}\,\rm{M}_\odot$. We find a rapid increase in the DTG ratio
with stellar mass in the mass range $10^8<M_*<10^{9.5}\,\rm{M}_\odot$
and a shallower increase at lower and higher stellar masses. The other
model variants also predict an increase in DTG ratio, although one
that is not as strong. They furthermore predict DTG ratios almost an
order of magnitude higher than our fiducial model in the mass range
$M_* < 10^{9.5}\,\rm{M}_\odot$. The agreement with the observations
seems to be worse in this mass range.

Our fiducial model predicts very weak evolution in the relation
between DTG ratio and stellar mass from $z=0$ to $z=2$. At higher
redshifts, the relation between DTG ratio and stellar mass gradually
increases by almost an order of magnitude from $z=9$ to $z=2$. The
scatter in the relation decreases significantly in this redshift
range. The `high-cond' model variant predicts an evolution that is
very similar to that seen in our fiducial model. The `no-acc' model variant
predicts hardly any evolution in the DTG ratio of galaxies as a
function of stellar mass. The `fix-tau' model variant predicts an
order of magnitude decrease of the DTG between $z=9$ and $z=0$.

We present and discuss the DTG ratios of galaxies as
  predicted by the `dwek98' and `dwek-evol' model variants in detail
  in Appendix \ref{sec:appendix_dwek}. In summary, we find that the
  `dwek98' model variant predicts DTG ratios in metal-poor galaxies
  that are too high, similar to the `high-cond' model variant. Based
  on the poor agreement with observational constraints and the
  unrealistic condensation efficiencies we disfavour this model
  variant. The `dwek-evol' model variant predicts DTG ratios similar
  to those yielded by our fiducial model in galaxies with gas-phase
  metallicities less than half of the solar value. It predicts higher
  DTG ratios in more metal-rich galaxies, in good agreement with the
  observational constraints at $z=0$. The relation between DTG and
  gas-phase metallicity curves backwards at redshifts $z\geq4$. This
  is driven by galaxies with very high DTM ratios, drastically
  increasing the DTG and decreasing the gas-phase metallicity.

\subsubsection{Dust-to-metal ratio}
We present the DTM ratio of galaxies \citep[normalized to the Galactic
  value of 0.44, based on ][]{Remy-ruyer2014}) as a function of their
gas-phase metallicity in Figure \ref{fig:dust_to_metal} and compare
our predictions to DTM ratios derived from
\citet{Remy-ruyer2014}.\footnote{To derive DTM ratios from the
  \citet{Remy-ruyer2014} work, we first converted the listed values
  for 12 + log(O/H) into gas phase metallicities Z assuming that 12 +
  log(O/H) $=$ 8.69 corresponds to Z$_\odot$. We derived the total
  mass of metals by multiplying the gas-phase metallicity Z with the
  total cold gas mass (\hi and \h2) including Helium (a factor of
  1.36). The total metal mass equals $M_{\rm metal} = 1.36\, M_{\rm HI
    + H2} \times \bigg(10^{\log{(\rm{O/H})}} / 10^{8.69 -
    12}\bigg)$. The DTM ratio was derived by dividing the dust masses
  presented in \citet{Remy-ruyer2014} by the derived metal mass.}  The
DTM ratio predicted by our fiducial model remains relatively constant
at $\sim 0.07$ up to metallicities of $\sim$ 0.5 Z$_\odot$. It then
increases towards a value of 0.4 at a gas-phase metallicity of 0.7
$Z_\odot$. Above this metallicity the DTM ratio decreases
slightly. Our fiducial model predicts DTM ratios in low-metallicity
galaxies that are too high compared to the observations. Above
metallicities of 0.5 $Z_\odot$ our model predictions for the DTM ratio
appear to be a bit too low \citep[though systematic uncertainties
on th emetallicity estimates in the ISM are quite
large,][]{Kewley2008}. 

The plateau in DTM ratios that we predict at low metallicities is set by
the constant condensation efficiencies of dust in stellar ejecta. The
increase in DTM ratio is then driven by the accretion of metals onto
dust grains. The decrease in the DTM ratio at the highest
metallicities is driven by the increasing importance of destruction of
dust with respect to the growth of dust in the ISM. These
  distinct trends are
not reflected in the \citet{Remy-ruyer2014} observations. We will discuss
the contribution of different formation channels to the dust content
of galaxies in more detail in Section \ref{sec:formation_mechanism}.

We find very different trends in the DTM ratios for the other model
variants. The `no-acc' model variant predicts DTM ratios at low
metallicities that are much too high compared to observations. The DTM
ratio decreases gradually towards higher metallicities. The
`high-cond' model variant predicts a roughly constant DTM ratio as a
function of gas-phase metallicity. The 'fixed-tau' model variant
predicts that the DTM ratio gradually decreases with increasing
gas-phase metallicities above metallicities of $\sim$ 0.1
Z$_\odot$. These redshift zero trends seem to be in contradiction with
observations in our local Universe.

Similar to the DTG ratio, we find weak evolution in the relation
between DTM ratio and gas-phase metallicity at $z<5$. At metallicities
larger than 0.1 Z$_\odot$, the DTM ratio increases gradually towards
the $z=0$ relation. At higher redshift and metallicities larger than
$\sim$0.5 Z$_\odot$, the DTM ratio is slightly higher than at $z=0$.
We will discuss the DTG ratio further in Section
\ref{sec:disc_dust_to_gas}.

The predictions of our fiducial model for the DTM ratios of galaxies
at $z=1$--$4$ are in poor agreement with the observational constraints
from Damped Lyman-alpha and GRB absorbers by \citet{DeCia2013},
\citet{DeCia2016}, and \citet{Wiseman2016}. Our fiducial model
predicts DTM ratios systematically lower than those found in the absorbers,
especially at low metallicities. Although the disagreement is
discouraging, it is important to remember that we did not try to
select for absorbers in any way and that the exact nature of Damped
Lyman-alpha absorbers and their host galaxy properties remain
unclear. We will perform a more fair comparison between absorbers and
our model results in a future work, employing selection techniques to
mimic the observational selection of DLAS as in
\citet{Berry2014}. Moreover, the method for determining both dust mass
and metallicity in absorbers is quite different from that used for
galaxies that are selected via their stellar or dust emission. It is
unknown whether these measurements can be compared on a consistent
scale. Certainly, the strong change from a very flat DTM ratio with
metallicity seen in DLAS at high redshift and the strong dependence on
metallicity seen in nearby galaxies is intriguing, if true. DLAS are
thought to arise from the outskirts of gas disks in galaxies and
perhaps even from the circumgalactic medium \citet[see the discussion
  in][]{Berry2014}. This discrepancy may reflect a difference in dust
growth or destruction timescales in different environments rather than
an evolutionary effect.

The `no-acc' and `high-cond' model variants show very weak evolution
in their predicted DTM ratios. The DTM ratios predicted by the
`fixed-tau' model variant increase by an order of magnitude from $z=9$
to $z=2$ for galaxies with gas-phase metallicities of 12+log(O/H)
$\sim$8. An interesting difference compared to our fiducial model is
that the other model variants seem to agree much better with the DTM
ratios found in absorbers, especially at low metallicities. This is
driven by the high-condensation efficiencies for the `no-acc' and
`high-cond' model, and the lower accretion timescale at low
metallicities for the `fix-tau' model than in our fiducial model
(Figure \ref{fig:accretion_time}). The trend with metallicity on the
other hand is the opposite from what the observations suggest. The
absorbers show a shallow increase in DTM ratio with increasing
metallicity, whereas the model variants all show a decreasing trend
with metallicity.

We plot the DTM ratio of galaxies as a function of their stellar mass
in Figure \ref{fig:mstar_dust_to_metal}. We find that the DTM ratio at
$z=0$ predicted by our fiducial model is constant up to stellar masses
of $10^{8.5}\,\rm{M}_\odot$ at 0.07, then increases till 0.3 at $M_* =
10^{10}\,\rm{M}_\odot$, and slowly decreases again at higher stellar
masses. Just as in the previous plot, these phases represent the
regimes where only condensation in stellar ejecta is relevant (at low
masses), dust growth in the ISM starts to become important (at
intermediate masses), and the destruction becomes more efficient (at
the highest masses). Our model predictions are in the same range as
the observations, though the shape of the trend appears to be very
different and on average our model predicts DTM ratios that are a bit
too low for galaxies with stellar masses larger than
$10^{8.5}\,\rm{M}_\odot$. Furthermore, the scatter in the
  data is not reproduced by our models.

The other model variants show very different trends, where the highest
DTM ratios are found at the lowest stellar masses, and the
DTM ratio gradually decreases towards lower values at higher
stellar masses. The latter is due to more efficient destruction.
Although the agreement with the data seems better at
stellar masses larger than $10^{8.5}\,\rm{M}_\odot$, at lower stellar
masses the predicted DTM ratios are an order of magnitude
(or more for the `fix-tau' model variant) too high. The high
DTM ratios at lower stellar masses are driven by the high
condensation efficiencies for the `no-acc' and `high-cond' model and
the shorter accretion times for dust growth in low-metallicity
environments for the `fix-tau' model variant.

There is only very weak evolution in the relation between stellar mass
and DTM ratio predicted by our fiducial model up to $z=6$. We find
that at redshifts $z>3$ the DTM ratio tends to be approximately 0.1
dex higher in the mass range $10^9<M_*/\rm{M}_\odot < 10^{10}$ than at
$z=0$, the regime where the growth of dust in the ISM becomes
important. At higher redshifts we predict DTG ratios in the most
massive galaxies that are up to 0.2 dex higher than at $z=0$. Although
the mean trend is very similar, the scatter in the relation increases
significantly when going to larger lookback times. The `high-cond' and
`no-acc' model variants behave in the same way. The `fix-tau' model
variant on the other hand predicts a strong increase in the DTM ratio
of galaxies with cosmic time of an order of magnitude (or even more at
the lowest stellar masses) from $z=9$ to $z=0$. We will discuss this
further in Section \ref{sec:discussion}.

\begin{figure*}
\includegraphics[width = 0.95\hsize]{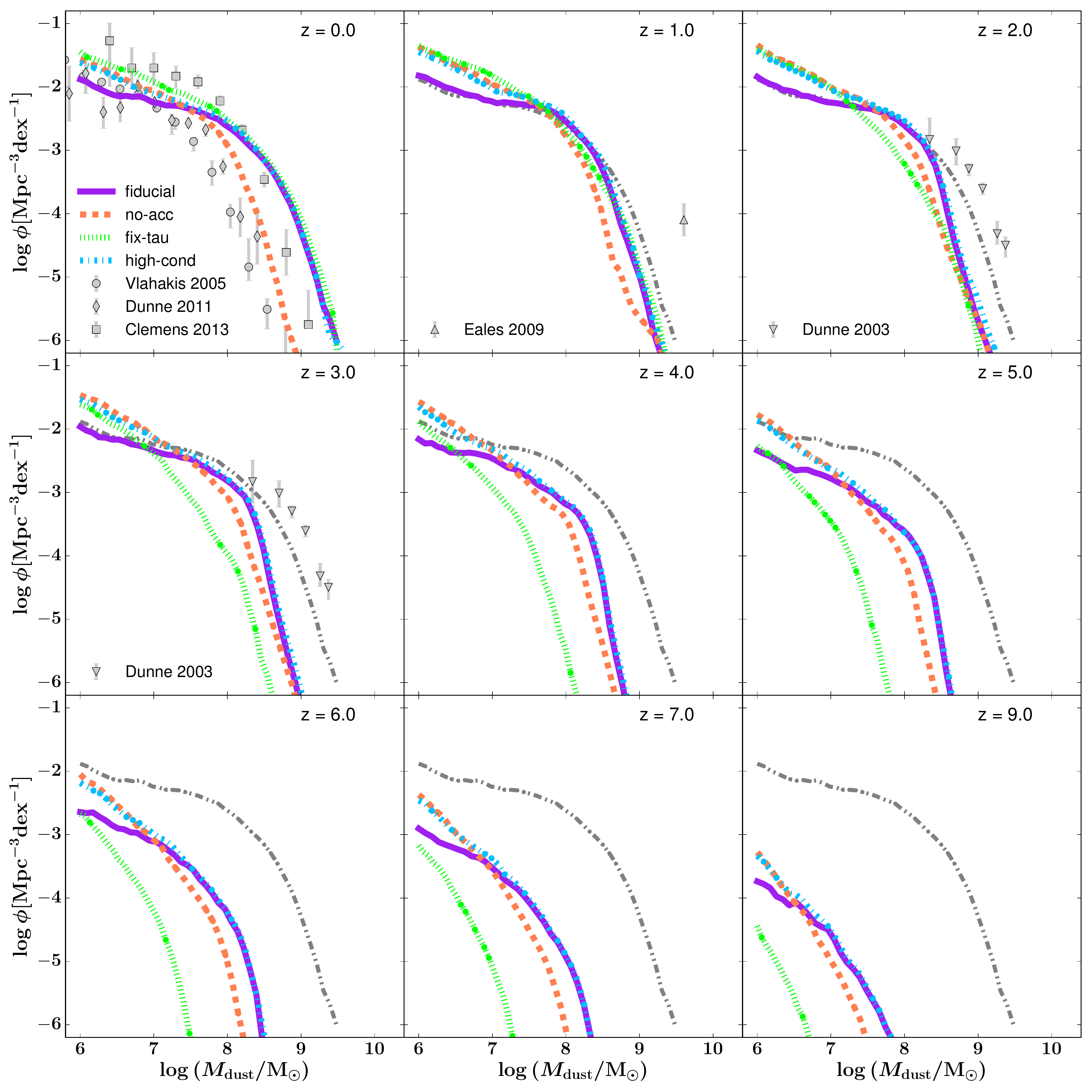}
\caption{The redshift evolution of the dust mass function for our four
  model variants. Predictions are compared to dust mass functions from
  the literature
  \citep{Dunne2003,Vlahakis2005,Eales2009,Dunne2011,Clemens2013}. The
  $z=0$ prediction of our fiducial model is shown as a dashed double
  dotted grey line in the higher-redshift bins for
  comparison.\label{fig:mass_function}}
\end{figure*}

\subsection{Dust mass functions}
Figure \ref{fig:mass_function} shows our predictions for the dust mass
function of galaxies. No selection criteria were applied for this
Figure, so all galaxies are included here. We compare our predictions
to the observed dust mass function in the local Universe and up to
$z\sim3$. We warn the reader that different groups have used different
approaches to infer the dust mass of a galaxy based on its IR and
sub-mm fluxes (see Section \ref{sec:observing_dust}). This can lead to
systematic uncertainties in the observed dust masses up to a factor of
three.

We find that our predicted dust mass functions closely follow a
Schechter \citep{Schechter1976} function with a characteristic dust
mass of $\sim10^{8.3}\,\rm{M}_\odot$ at $z=0$.  This characteristic
dust mass is similar for the `high-cond' and `fix-tau' model
variants. The dust mass function predicted by our fiducial model is in
good agreement with the observed dust mass functions at dust masses
less than $10^{8.3}\,\rm{M}_\odot$. Our fiducial model predicts number
densities for galaxies with larger dust masses that are too high.

The `no-acc' model variant predicts a dust mass function with slightly
higher number densities than our fiducial model at dust masses lower
than $10^8\,\rm{M}_\odot$. At higher dust masses the number densities
predicted by the `no-acc' model variants are lower than those of our
fiducial model and right between the \citet{Vlahakis2005} and
\citet{Dunne2011} observations.  The predictions of the fiducial and
the `high-cond' model variants are very similar. Only at dust masses
smaller than $10^7\,\rm{M}_\odot$ does the `high-cond' variant predict
a number density $\sim0.3$ dex higher than our fiducial model. The
'fix-tau' model variant predicts number densities approximately 0.5
dex larger for galaxies with dust masses smaller than $10^{7.5}$ and
larger than $10^9\,\rm{M}_\odot$. The elevated number densities at low
dust masses with respect to our fiducial model are driven by the high
condensation efficiencies in the `high-cond' and `no-acc' model
variants, and the short timescales for dust growth in the ISM in
low-metallicity environments in the `fix-tau' model variant.

Our fiducial model predicts a rapid increase in the dust mass function
from $z=9$ to $z=3$, independent of galaxy dust mass. The dust mass
function is remarkably constant from $z=2$ to $z=0$ for galaxies with
dust masses less than $10^{8.3}\,\rm{M}_\odot$, whereas the number
densities keep increasing from $z=2$ to $z=0$ for galaxies with larger
dust masses. Our fiducial model predicts too few galaxies with dust
masses larger than $\sim10^9\,\rm{M}_\odot$ at redshifts $z>2$.  We
note that the observational constraints by \citet{Eales2009} and
\citet{Dunne2003} are based on surveys of sub-mm sources with large
beam sizes (larger than 14 arcsec). High spatial resolution
observations with ALMA have suggested that the brightest sub-mm
sources in such surveys consist of multiple lower luminosity objects,
blended within one large beam \citep{Karim2013,Hayward2013SHAM}. 

The evolution of the dust mass function predicted by the `high-cond'
model variant is very similar to our fiducial model. The same is true
for the `no-acc' variant, although with lower number densities at the
highest dust masses. The `fix-tau' model variant also predicts an
increase in the number density with time, but the rate of increase is
much slower than for our fiducial model and continues up to $z=1$. The
slow growth of dust masses is in accordance with the evolution in the
relation between dust mass and stellar mass (Figure
\ref{fig:mstar_mdust}).

\begin{figure*}
\includegraphics[width = 0.95\hsize]{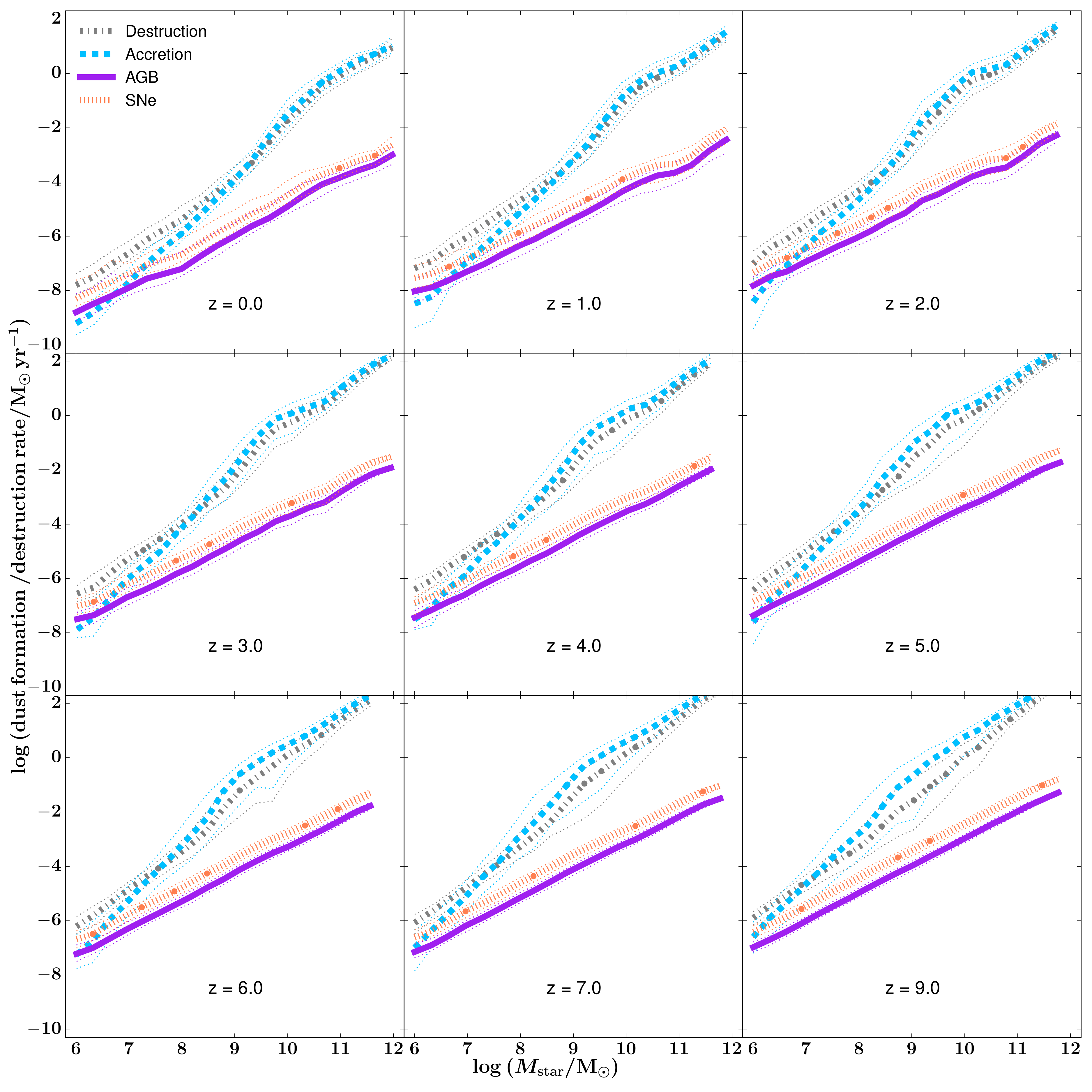}
\caption{The formation and destruction rate of dust as a function of
  stellar mass from redshift $z=9$ to $z=0$, for our fiducial
  model. Dust formation rates are separated into formation due to AGB
  stars, SNe, and growth of dust in the ISM. Thick lines mark the 50th
  percentiles, whereas the thin dotted lines mark the 16th and 84th
  percentiles.\label{fig:formation}}
\end{figure*}

\subsection{Dust formation and destruction rates}
\label{sec:formation_mechanism}
We present the dust formation and destruction rate (only in the cold
ISM, not in the hot halo) of galaxies as a
function of their stellar mass in Figure \ref{fig:formation}. We
only focus on our fiducial model, as this model most successfully
reproduces the trends between stellar mass and dust mass at low and
high redshifts, together with the DTG ratio in local galaxies. We present the formation
rate through the individual channels for the other model variants in
Appendix \ref{sec:appendix}. We find
that the formation rate through the different dust formation channels
(SNe, AGB stars, and dust growth in the ISM) all increase as a
function of stellar mass. The formation rate by stellar ejecta
can be described by one linear relation as a function of stellar
mass. The relation between formation rate through accretion and stellar mass is made up of
multiple components. It is steepest in the stellar mass
range $10^{8-10}\,\rm{M}_\odot$ and flattens at higher stellar
masses. The dust destruction rate is systematically $\sim 0.2$ dex lower than the formation rate
through dust accretion for galaxies with stellar masses larger than
$10^{8}\,\rm{M}_\odot$. In galaxies with lower stellar masses, the
destruction rate of dust roughly equals the formation rate through
accretion.

The normalisation in the relation between dust formation rate
by stellar ejecta and stellar mass slowly decreases from $z=9$ to
$z=0$ by approximately 2 dex and 1 dex for SNe and AGB stars,
respectively. We also find a 2 dex decrease in
this redshift range for the formation rate by accretion and the
destruction rate of dust. The characteristic stellar mass at which the
relation between formation rate through accretion and stellar mass
flattens also evolves with time, from $\sim10^{9}\,\rm{M}_\odot$ at
$z=9$ to $\sim 10^{10}\,\rm{M}_\odot$ at $z=0$. 

An interesting prediction of our fiducial model is that the formation
rate of dust through accretion is almost always higher than the
formation rate by SNe and by AGB stellar ejecta. Only at stellar
masses of $\sim 10^7\,\rm{M}_\odot$ is the formation rate through SNe
as high as the formation rate through growth in the ISM, although not
at all redshifts.  We will discuss this further in Section
\ref{sec:discussion}.

\begin{figure*}
\includegraphics[width = 0.95\hsize]{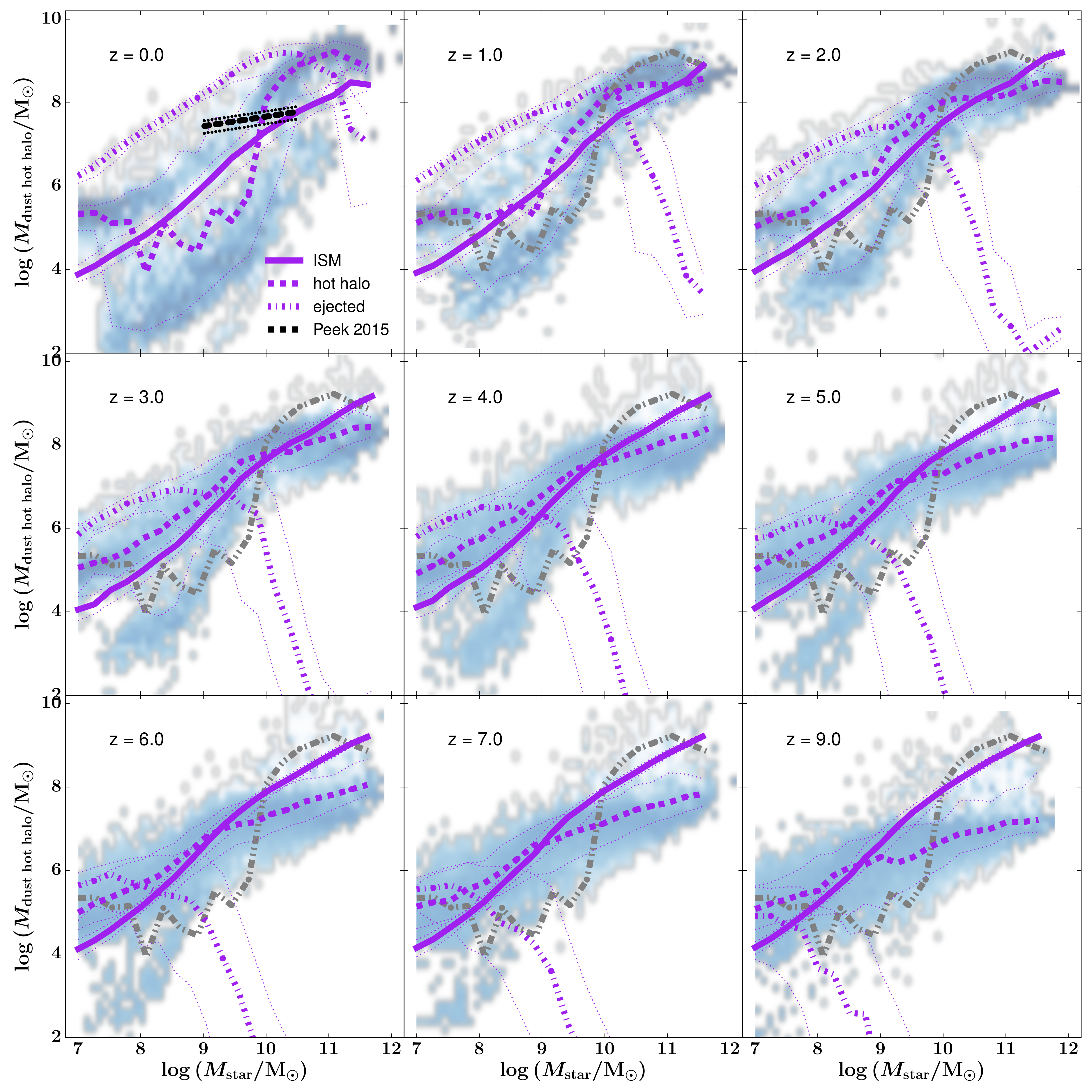}
\caption{\label{fig:mstar_mdustCGM} The mass of dust in the ISM (solid
  line), hot halo (dashed line), and the ejected reservoir (dotted
  line) as a function of host galaxy stellar mass from redshift $z=9$
  to $z=0$, as predicted by our fiducial model. Thick lines mark the
  50th percentiles, whereas the narrow dotted lines mark the 16th and
  84th percentiles. The color scale shows the logarithm of the
  conditional probability distribution P($M_*|M_{\rm dust}\,_{\rm
    hot}\,_{\rm halo}$), which represents the likelihood to find a
  dust mass $M_{\rm dust}\,_{\rm hot}\,_{\rm halo}$in the hot halo as
  a function of central galaxy stellar mass. The $z=0$ prediction of
  our fiducial model for the dust mass in the hot halo is shown as a
  dashed double dotted grey line in the higher-redshift bins for
  comparison. We compare our predictions to the CGM constraints at
  $z=0$ by \citet{Peek2015}.}
\end{figure*}

\begin{figure*}
\includegraphics[width = 0.95\hsize]{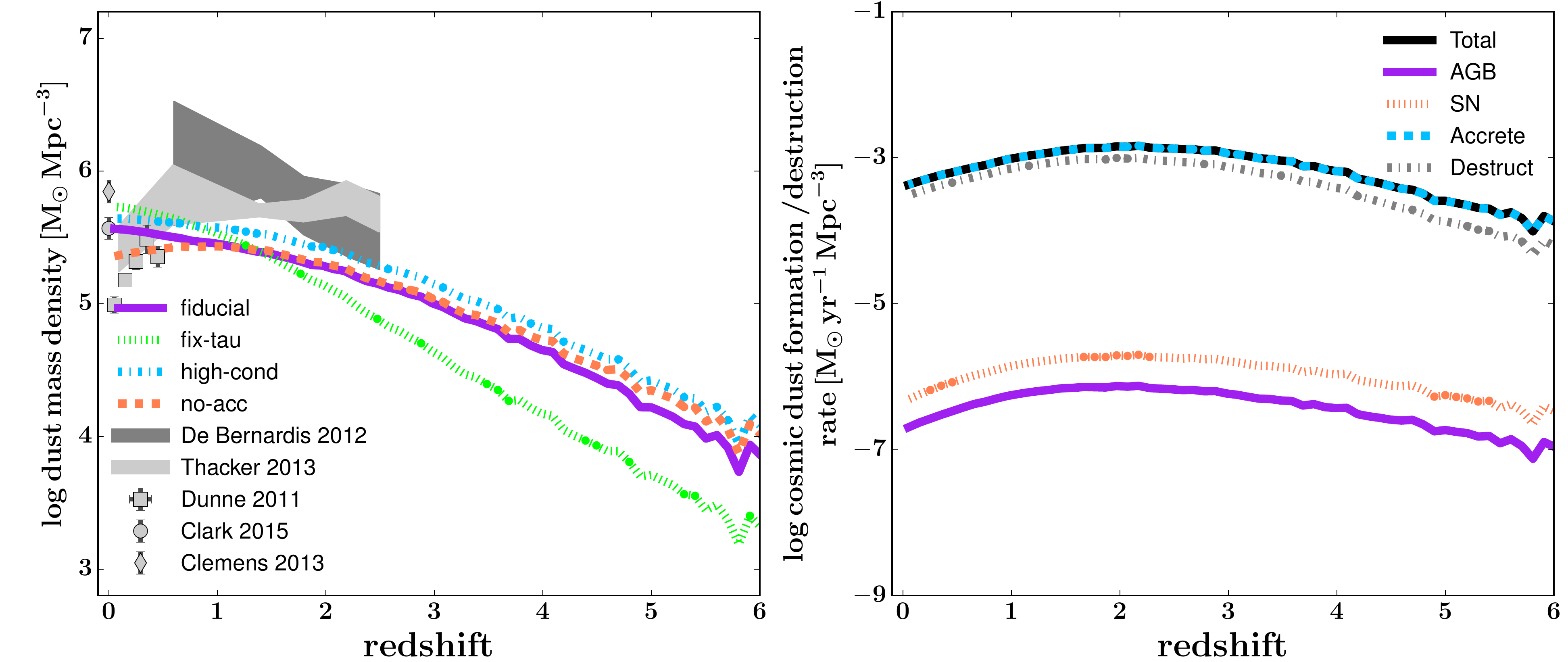}
\caption{Left: The cosmic density of dust in the ISM of galaxies as a function of redshift
  for the four model variants. Model predictions are compared to
  constraints from the literature \citep{Dunne2011,deBernardis2012,Clemens2013,Thacker2013,Clark2015}. Right:
  The cosmic formation rate (by AGB stars, SNe, and metal accretion)
  and destruction rate of dust in the ISM as a function
  of redshift.\label{fig:cosmic_evol}}
\end{figure*}

\subsection{Dust in the hot halo and ejected reservoir}
Besides dust in the ISM of galaxies, our model also tracks the dust in
the hot halo and the ejected reservoir. If we make the simple
assumption that dust in the hot halo corresponds to the dust in the
CGM, we can compare our model predictions to recent observations
\citep{Menard2010,Peek2015}.

We present the relation predicted by our fiducial model between host
galaxy stellar mass and dust in the hot halo in Figure
\ref{fig:mstar_mdustCGM} (the predictions of the other model variants
are very similar). For this figure we required galaxies to be the
central galaxy in their halo. The color scale shows the logarithm of
the conditional probability distribution P($M_*|M_{\rm dust}\,_{\rm
  hot}\,_{\rm halo}$), which represents the likelihood to find a given
dust mass in the hot halo as a function of central galaxy stellar
mass. We also plot the dust mass in the ISM of the central galaxies,
as well as the mass of dust in the ejected reservoir.

We first of all find that at $z\leq5$ there seem to be two distinct
`branches' on the plot of stellar mass and hot halo dust mass, one
with high dust masses and the other with low dust masses. These two
branches reflect two different regimes for cooling and accretion in
our model. When the predicted ``cooling radius'' is larger than the
virial radius, we assume that gas (and dust) accrete directly onto the
central galaxy on a dynamical timescale, so little gas and dust
collect in the ``hot'' reservoir. When the predicted cooling radius is
smaller than the halo virial radius, gas and dust cool onto the
central galaxy on a cooling timescale, so more material builds up in
the hot halo reservoir \citep[see][for more
  details]{Somerville2008}. These two regimes correspond to the ``cold
mode'' and ``hot mode'' accretion also seen in numerical hydrodynamic
simulations \citep{birnboim_dekel:03,dekel_birnboim:06,keres:05}. 
One should keep in mind that the plotted means in Figure
\ref{fig:mstar_mdustCGM} represent the mean of these cooling
modes. Comparisons of the relative importance of these different
cooling modes in SAMs and
hydrodynamic models showed that cooling rates are similar to each
other within 20 per cent \citep{Hirschmann2012,Monaco2014}. The details of where and when accretion modes set in can be
sensitive to the details of the numerical scheme used \citep{Nelson2013}.

The mass of dust in the hot halo of galaxies at $z=0$ predicted by our
fiducial model increases with host galaxy stellar masses and flattens
above stellar masses of $\sim10^{10.5}\,\rm{M}_\odot$. The flattening
is driven by an increased efficiency for the sputtering of dust in
hot gas as a function of host halo mass (virial temperature). The amount of
dust in the hot halo is comparable to or even larger than the amount of
dust in the ISM for galaxies with stellar masses of
$\sim10^{10}\,\rm{M}_\odot$ and larger. Our predictions are in the
same mass range as the observational constraints derived by
\citet{Peek2015}. However, we find a distinct difference between our
model predictions and the Peek et al. results for the slope between
central galaxy stellar mass and CGM dust mass. The Peek et al. slope
is significantly shallower than our model predicts.

Our model predicts that the mass of dust in the hot halo gradually
increases with cosmic time. Especially in the most massive central
galaxies ($M_* > 10^{10}\,\rm{M}_\odot$), we see an increase in hot
halo dust mass from $z=9$ to $z=0$ of two orders of magnitude. In this
mass range, the mass of dust in the hot halo starts to dominate over
the mass of dust in the cold gas at redshifts $z<2$. We find a very
similar evolution for the dust in the ejected reservoir.

The mass of dust in the ejected reservoir 
gradually increases with central galaxy stellar mass up to some
characteristic mass, after which it rapidly decreases again. This
characteristic mass increases from $\sim 10^7\,\rm{M}_\odot$ at $z=9$
to $\sim10^{10}\,\rm{M}_\odot$ at $z=0$. This characteristic mass
marks the point where most of the gas/dust heated up by SNe is trapped within the potential
well of the dark-matter halo and does not contribute significantly to
the ejected reservoir anymore. At stellar masses below the
characteristic turn-over mass, the mass of dust in the ejected
reservoir is larger than the mass of dust in the cold gass component
and hot halo. This is independent of redshift. As discussed in Section
\ref{sec:hot_halo}, the amount of dust destruction through sputtering
in the ejected reservoir calculated in our model is likely a lower
limit. In reality the mass of dust in the ejected reservoir may
therefore be smaller.
Our results should thus be treated with caution, but demonstrate that
very large amounts of dust may be removed from galaxies by winds in
these models (although it is an open question whether dust can survive
in these winds). We have performed some initial tests in which we
assume that all the dust in the ejected reservoir is destroyed, and
find that this doesn't affect the mass of dust in the ISM.

\subsection{The cosmic density and formation history of dust}
\subsubsection{Evolution of the dust density of our Universe}
We show the cosmic density of dust in the ISM of galaxies in the left
panel of Figure \ref{fig:cosmic_evol}. In this figure, all modeled
galaxies are included, with no selection criteria applied. The cosmic
density of dust in the ISM in our fiducial model gradually increases
with time up to $z=0$. At $z=0$ our model predictions are in decent
agreement with the findings by \citet{Clark2015}, but overpredict the
constraints by \citet{Dunne2011} by a factor of four. We again note
that these different studies adopted different methods to estimate the
dust mass of galaxies, which accounts for some of the discrepancy
between them (see Section \ref{sec:observing_dust}).  Our model
predicts too much dust at $z>0$ compared to the constraints by
\citet{Dunne2011}. It is in good agreement with the constraints by
\citet{Thacker2013} at $z<0.5$, but predicts too little dust compared
to the observations by \citet{deBernardis2012} and \citet{Thacker2013}
at higher redshifts.

\begin{figure}
\includegraphics[width = 0.95\hsize]{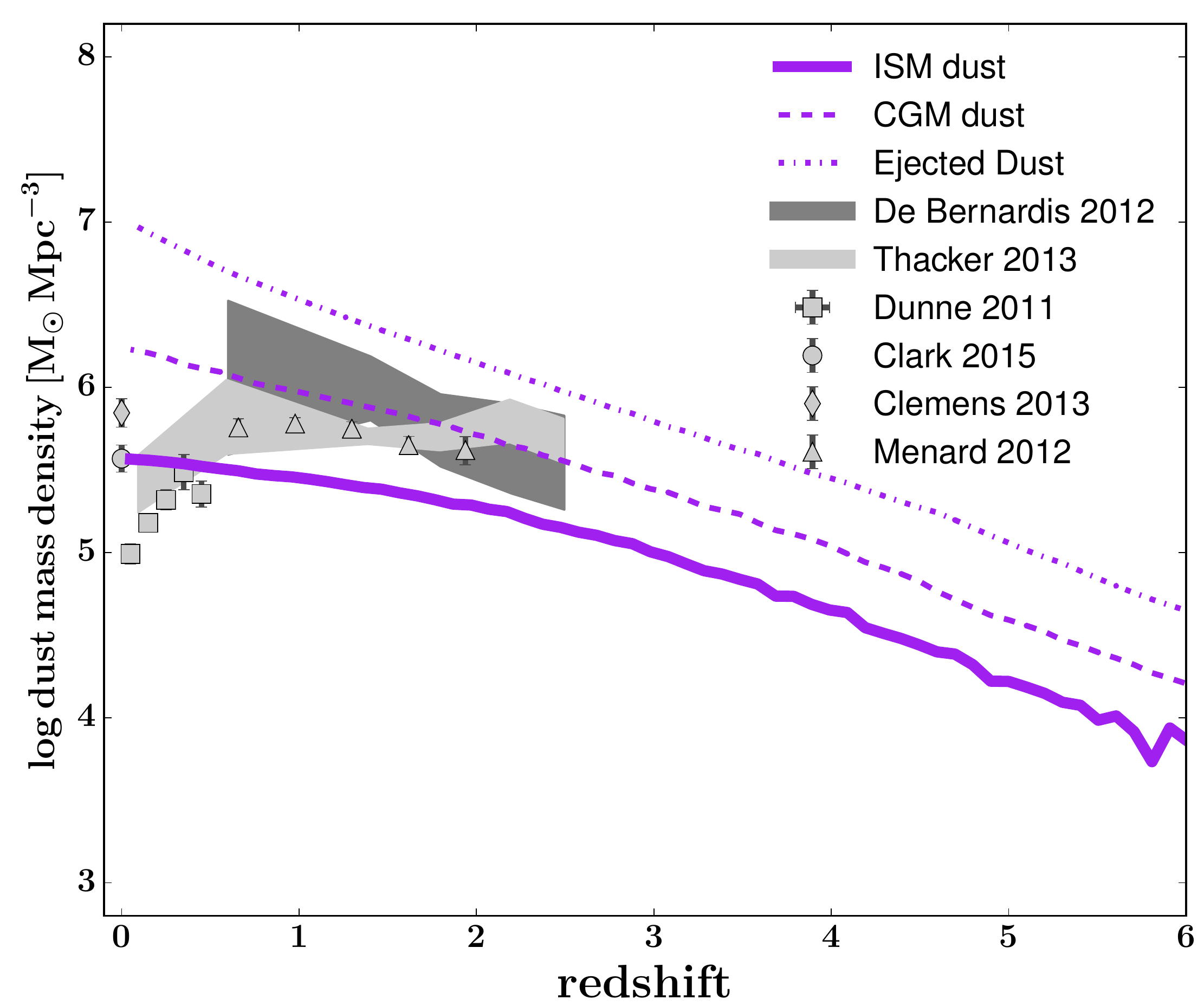}
\caption{The cosmic density of dust in the ISM, the CGM, and the
  ejected reservoir, all as a function of redshift
  for the three model variants. Model predictions are compared to
  constraints from the literature \citep{Dunne2011,deBernardis2012,Menard2012,Clemens2013,Thacker2013,Clark2015}.\label{fig:cosmic_evol_ejected}}
\end{figure}
\begin{figure*}
\includegraphics[width = 0.95\hsize]{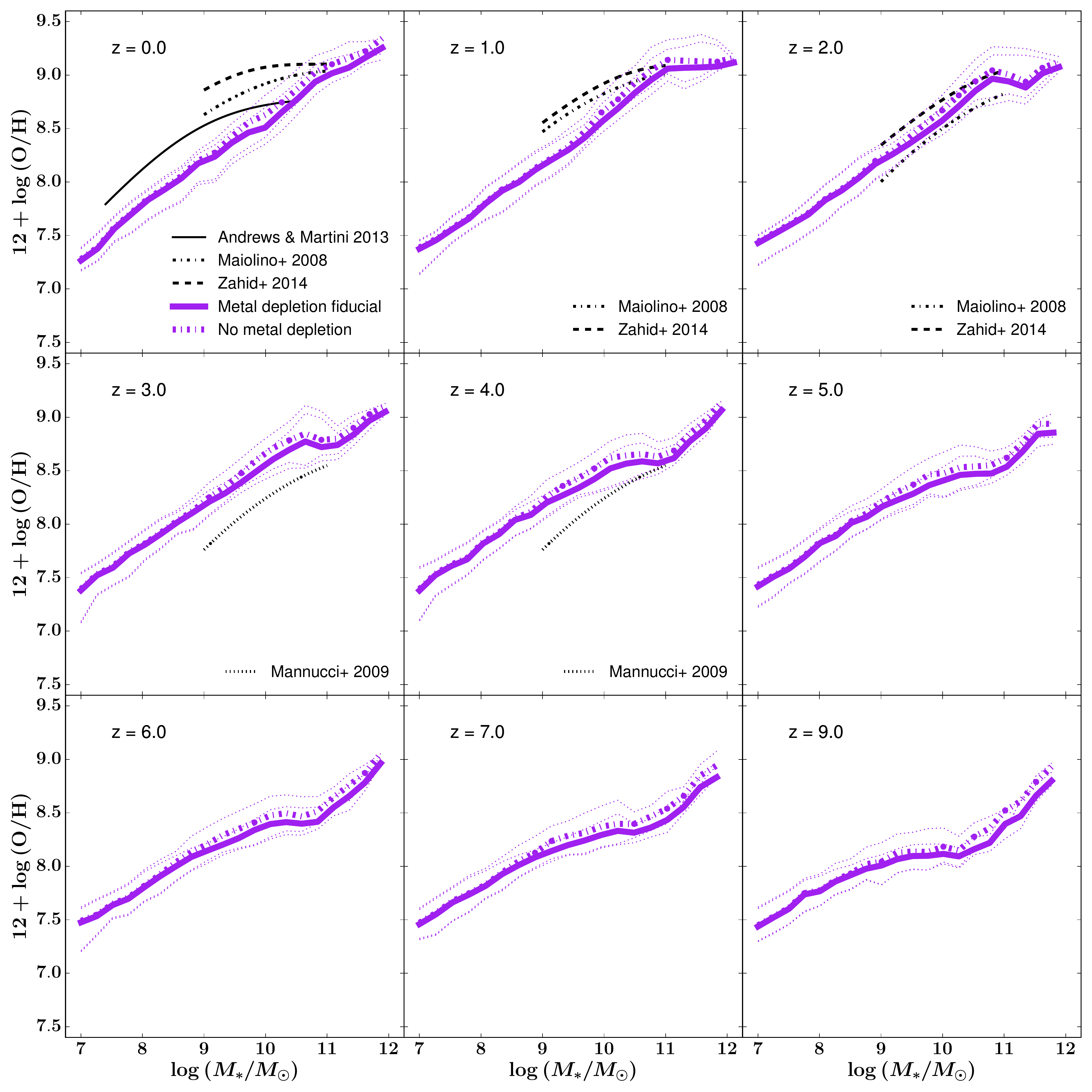}
\caption{The gas-phase metallicity of galaxies as a function of their
  stellar mass from redshift $z=9$ to $z=0$. Plotted is our fiducial
  model where metals are depleted onto dust grains and a model
  variant in which dust depletion is not included.  Thick lines mark
  the 50th percentiles, whereas the narrow dotted lines mark the 16th
  and 84th percentiles. Model predictions are compared to observations
  from \citet[dahsed-dotted line]{Maiolino2008},\citet[dotted
    line]{Mannucci2009}, \citet[solid line]{Andrews2013}, and
  \citet[dashed line]{Zahid2014}.\label{fig:metal_depletion}}
\end{figure*}
The cosmic dust density predicted by the `no-acc' model variant
deviates from from our fiducial model at redshifts $z\leq1$, where it predicts lower
dust mass densities. These predictions are still too high compared to
the \citet{Dunne2011} results at $z<0.5$, but in better agreement at
redshifts $0.5<z<1$. The predictions are too low compared to the densities found by
\citet{deBernardis2012} and \citet{Thacker2013}. The predictions of
the `high-cond' and fiducial model variants are almost identical. The
`high-cond' model predicts slightly higher cosmic densities for dust,
driven by the high condensation efficiencies of dust in the stellar
ejecta. The `fix-tau' model variant predicts lower densities for dust
at $z>1.5$, up to approximately 0.8 dex less dust at $z=6$ than
our fiducial model. As the other model variants, the `fix-tau' model predicts too low dust
densities compared to the \citet{deBernardis2012} and
\citet{Thacker2013} results at $z>1$.

We present the integrated density of dust in hot halos and ejected
reservoirs as predicted by our fiducial model in Figure
\ref{fig:cosmic_evol_ejected}. We find that the cosmic density of dust
in hot halos and ejected reservoirs increases gradually with cosmic
time. The mass of dust in hot halos is larger than the mass of dust in
cold gas, independent of redshift. We find that the density of dust in
ejected reservoirs is 1--1.5 dex higher than in cold gas, independent
of redshift. We will further discuss this large reservoir in Section
\ref{sec:disc_CGM_dust} and Section
\ref{sec:disc_ejected_reservoir}. We compare our predictions to the
observations of dust in the IGM by \citet{Menard2012}.The cosmic
density of dust that we predict for the CGM is 0.1--0.2 dex higher
than the constraints by \citet{Menard2012} for the cosmic density of
dust. The predicted cosmic density of dust in the ejected reservoirs
is significantly larger than the \citet{Menard2012}
constraints. Our predictions for the density of dust in
  the hot halo are similar to the model predictions by of \citet{Grieco2014} for
  the cosmic density of dust.

\subsubsection{Formation history of dust}
We present our predictions for the global dust formation and
destruction rates in the Universe in the right hand panel of Figure
\ref{fig:cosmic_evol}. Here we show only the predictions of our
fiducial model.

The total dust formation rate increases from redshift $z=6$ to $z=2$
and decreases again at later times. At its maximum, the formation rate
is almost four times higher than at $z=0$. The evolution of this trend
is similar to the cosmic SFR density \citep{Madau2014}, though the
difference between its maximum and the $z=0$ value is not as large. The
accretion of metals onto dust grains in the ISM is the dominant mode
of dust formation/growth, independent of redshift. The contribution by
SNe and AGB stars to the dust formation rate density of our Universe
is more than two orders of magnitude less. We emphasise that the dust
formation rate density will be dominated by galaxies with dust masses
close to the knee of the dust mass function at the respective
redshifts. As seen in the previous section, the relative importance of
the different dust formation channels may be different in individual
galaxies.

The cosmic rate of dust destruction follows a trend very similar to
the formation rate; it increases up to $z=2$ and decreases again at
later times. These results are similar to the findings by \citet{Grieco2014}.

\subsection{Depletion of gas-phase metals}
The inclusion of dust in our galaxy formation model
impacts the gas-phase metallicity of galaxies. A fraction of the
metals in the ISM will be depleted onto dust grains, effectively
lowering the gas-phase metallicity of galaxies.

We show the effects of the depletion of metals onto dust in Figure
\ref{fig:metal_depletion}. We present the gas-phase metallicity of
galaxies using our fiducial model and the gas-phase metallicity of
galaxies when not including dust depletion.  We find that the
depletion of metals typically only becomes relevant in galaxies with
stellar masses larger than $10^9\,\rm{M}_\odot$. The depletion
of metals lowers the gas-phase metallicity of galaxies by
approximately 0.1 dex.

Now that metal depletion is properly accounted for, we can assess how
well our model reproduces the observed relationship between stellar
mass and gas-phase metallicity
\citep{Maiolino2008,Mannucci2009,Andrews2013,Zahid2014}. It is
important to note that gas-phase abundance measures are sensitive to
calibration \citep{Kewley2008}, but usually relative abundances are
consistent among various indices. Hence the slope of the observed
gas-phase metallicity relation is more robustly known than the
amplitude, though the amplitude should still be accurate to within a
factor of 2--3. We find that at $z\leq2$ our model predicts
metallicities systematically below the observed relation in our local
Universe, up to as much as 0.4 dex in low-mass galaxies at $z=1$. 

The slope of the mass-metallicity relation is in decent agreement with
the observations at $z>0$ for galaxies with stellar masses less than
$10^{10}\,\rm{M}_\odot$. We find some flattening of the
mass-metallicity relation in galaxies with stellar masses larger than
$10^{10.5}\,\rm{M}_\odot$ at $z=1$ and $z=2$. At $z=0$ on the other
hand the slope is correct in low mass galaxies, but we don't predict
the flattening observed in galaxies with stellar masses larger than
$10^{10}\,\rm{M}_\odot$.

\section{Discussion}
\label{sec:discussion}
In this paper we have included tracking of the formation and
destruction of dust in the Santa Cruz semi-analytic model of galaxy
formation. We discuss the conclusions that can be drawn from our model
predictions for dust evolution in galaxies in this section.

\subsection{Insights from comparing model variants}
In our fiducial model, dust is produced in the ejecta of AGB stars and
supernovae, and also through accretion onto grains in the dense
ISM. The timescale for dust growth by accretion is a function of the
density, temperature, and metallicity of the molecular clouds in which
this process is efficient.
In addition to our fiducial model we have also explored other
model variants: one in which no growth of dust through accretion onto
grains in the ISM occurs, and condensation efficiencies in stellar
ejecta are large; one in which the only change is that the dust
condensation efficiencies in stellar ejecta are high; and finally one
in which a fixed time scale for the accretion of metals onto dust
grains is assumed. In Appendix \ref{sec:appendix_dwek} we explore in
  more detail two model variants based on the \citet{Dwek1998}
  approach for dust growth in the ISM, also adopted by e.g.
  \citet{Calura2008} and \citet{McKinnon2015}.

Recent work by \citet{Ferrara2016} has suggested that the contribution
from dust growth by accretion onto grains in dense environments to the
dust mass of galaxies is negligible. Ferrara et al. argue that in
dense environments accreted materials form icy water mantles. These
mantels quickly photo-desorb once the grains return to the diffuse ISM
at the end of the cloud lifetime. This would have significant
implications for one of our conclusions, that dust growth in the ISM
is a key ingredient to reproduce both the observed dust masses in
galaxies and the shape of various scaling relations
\citep[e.g.,]{Dwek1998,Dwek2007,Zhukovska2008,Michalowski2015,Schneider2016}. Our
`no-acc' model variant is an attempt to explore the global
consequences of the \citet{Ferrara2016} picture. Assuming a
condensation efficiency of 100\% in the stellar ejecta, this model is
able to reproduce the dust masses observed in very high redshift
galaxies ($z\sim 4$--7). The model also reproduces dust masses in
massive galaxies at $z=0$.  However, the model predicts $z=0$ dust
masses in low-metallicity objects that are too high and DTG ratios in
metal-rich galaxies that are too low.

It is also possible that galactic
winds could preferentially eject dust relative to gas, or that dust
could be destroyed in these winds. However, these processes must lead
to an effective metallicity dependence in the net production of dust
in order to steepen the DTG and DTM ratio relation and bring this
model into agreement with observations. Moreover, clearly our
assumption of 100\% efficiency of dust production in stellar ejecta is
extreme. We carried out tests in which we adopted condensation
efficiencies of 50\% and found that our model is then unsuccessful in
reproducing the dust masses of local galaxies. Lower condensation
efficiencies would result in even larger discrepancies. As an example
of this we can take the `fix-tau' model variant that fails to
reproduce the dust-masses of high-redshift galaxies. If a fixed
timescale for dust growth of 100 Myr is not sufficient, than a model
variant with shorter accretion times or no accretion at all can
certainly not reproduce high-z observations (without increasing the
condensation efficiency in SNae and AGB stars).

There are significant uncertainties in other aspects of our modelling
that could plausibly bring this model into better agreement with
observations. The amount of dust around local supernova
remnants provides a constraint on the destruction efficiency of
dust and the net condensation efficiency after the reverse
shock. Recently, \citet{deLooze2017} combined \emph{Herschel} and
\emph{Spitzer} observations of Cassiopeia A and measured a dust mass of
0.3--0.5 $\rm{M}_\odot$ for silicate-type dust. \citet{Owen2015}
derived a dust mass in the range 0.3 -- 0.5
$\rm{M}_\odot$ for the Crab Nebula supernova
remnant \citep[see also][]{Gomez2012}. \citet{Temim2017} found a shell of dust around supernova
G54.1+0.3 of at least 0.3 $\rm{M}_\odot$. Using \emph{Herschel}
photometry, \citet{Matsuura2015} found a newly formed dust mass of 0.8
$\rm{M}_\odot$ for SN 1987A. \citet{deLooze2017} found a drop in dust
mass behind the reverse shock for Cassiopeia A, suggesting that
$\sim 70$ per cent of the mass of newly formed dust is destroyed by the reverse
shock. The authors conclude that if indeed approximately $\geq20 -30$
per cent of the initially condensed dust is capable of surviving the
reverse shock, this can explain the dust masses observed at high
redshifts. If dust destruction in galaxies is less efficient than
commonly thought, the observed condensation efficiencies could
support the `no-acc' model. For example, it is possible that the efficiency of dust
destruction depends on environment in a more complex manner than we
have represented in our model. Furthermore, there might be a
metallicity dependence to the efficiency of dust destruction
\citep{Yamasawa2011}.  \citet{Temim2015} found the
dust destruction rate in the Large Magellanic cloud to be lower than
in the Small Magellanic Cloud. Nevertheless, before a model without
growth of dust in the ISM can really be favoured, an improved understanding of the condensation
of dust in SN ejecta, the importance of the reversed shock, and a
better understanding of the destruction efficiency of SN in diffuse
media as a function of gas density and metallicity are necessary.

It is interesting that the `no-acc' model predicts very little
evolution in the DTG ratios of galaxies as a function of their stellar
mass, independent of redshift.  We do predict an evolution in these
trends for all the other model variants (that include metal
accretion). This suggests that the evolution seen in our fiducial
model is driven by the effective dependence of the dust growth
timescale on galaxy properties which evolve with cosmic time.  A more
complete census of the DTG ratio of galaxies as a function of cosmic
time and galaxy properties is therefore an important test of the
importance of dust growth in the ISM.

We find that differences between our fiducial model and the
`high-cond' model variants are only prevalent in galaxies with low
stellar masses and low metallicity. The same holds for the
  `dwek98' model variant. In massive and metal-rich galaxies, the
dust properties are dominated by the accretion of metals onto dust
grains (Figure \ref{fig:formation}). The `high-cond' model variant
overpredicts the dust abundance for galaxies with low metallicities
($< 0.2\, \rm{Z}_\odot$) and low stellar masses ($<
10^9\,\rm{M}_\odot$). Besides theoretical work on the condensation
efficiency of AGB stars and SNe \citep{Ferrarotti2006,Bianchi2007,
  Gioannini2017}, our predicted dust abundances also suggest that
constant high condensation efficiencies \citep[as in
  e.g.,][]{Bekki2013,McKinnon2016} are an unrealistic way of building
up large dust masses in galaxies. The buildup of dust is too efficient
in low-mass galaxies. This same conclusion applies to the `no-acc'
model variant.

Differences between our fiducial model variant and the `fix-tau' model
are prevalent across the metallicity and stellar mass range probed by
our models. The `fix-tau' model predicts dust abundances in low
metallicity galaxies at $z=0$ that are much
higher than those predicted by the fiducial model. Dust abundances are much lower
for galaxies with metallicities higher than 1 Z$_\odot$. The build up of dust in
galaxies when adopting the `fix-tau' model variant is much slower
(i.e., much lower dust masses as a function of stellar mass and
redshift) than for our fiducial model (see for example Figures
\ref{fig:mstar_mdust}, \ref{fig:mass_function}, and
\ref{fig:cosmic_evol}). As briefly mentioned before, the slow buildup
is driven by the long accretion time scales of 0.1 Gyr compared to
accretion time scales of 10 Myr or even less in dense environments for
our fiducial model (see Figure \ref{fig:accretion_time}). A variable
accretion time scale allows for much more efficient accretion of
metals onto dust grains in the early Universe, speeding up the
formation in dust. In low metallicity objects, on the other hand, the accretion time scale
is easily shorter than we would calculate with our density and
metallicity variable recipe (see Figure
\ref{fig:accretion_time}). This allows for a higher rate of dust
growth in the low-metallicity ISM in small objects, and an increase in
dust mass with respect to our fiducial model. This can also be seen
when looking at the formation rate of dust through growth in the ISM
in Figure \ref{fig:formation_channel_fix_tau}.

The poor agreement between the predictions by the `fix-tau' model and
observed dust masses in galaxies at $z=6$ and $z=7$ seems to rule out
fixed accretion times of 100 Myr. This is further supported by the
poor agreement between model predictions and observations for the DTG
ratio and DTM ratio of galaxies. \citet{McKinnon2016} adopted a fixed
accretion timescale and predicted dust reservoirs in galaxies at
$z=2.5$ that are much too small compared to observations. What this
tells us is that accretion times-scales should be much shorter than
100 Myr in the early Universe. We tested shorter fixed accretion times
of  $\sim$ 10 Myr to reproduce the $z=6$ and $z=7$ dust masses,
but this led to an even stronger disagreement between model
predictions and observations at $z=0$ at both low and high stellar
masses.

We thus conclude that a variable accretion time scale as adopted in
our fiducial model is necessary to reproduce the buildup of dust in
galaxies
\citep{Mattsson2011,Gall2011,Calura2014,Mancini2015,Schneider2016,Wang2017}. There
is a need for short depletion times in the early Universe, but these
short depletion times cannot be sustained, otherwise too much dust
grows in the low-redshift Universe. An accretion time-scale that is
solely a function of metallicity could not provide this
behaviour. Gas-phase metallicity increases with cosmic time, therefore
accretion time-scales would only become shorter with cosmic time, the
opposite from what is needed. With density as an additional parameter,
we can reverse this trend and reproduce the dust mass of low- and
high-redshift galaxies simultaneously \citep{Calura2014,Mancini2015}.

Both the fiducial and `dwek-evol' model variants invoke such a density dependent recipe for accretion of metals onto
  dust grains (the former adopts the \citet{Zhukovska2008} approach
  for the growth of dust in the ISM, whereas the latter adopts the
  \citep{Dwek1998} approach). Their first main difference is that the
  Zhukovska approach allows for more efficient accretion of metals in
  environments with low-molecular hydrogen fractions. The second main
  difference is that the Dwek approach allows for more efficient metal
  accretion in environments where the timescale for metal accretion is
  much longer than the time it takes to cycle all the ISM through
  molecular clouds. Of these two model variants, only our fiducial
  model successfully reproduces the dust content of galaxies from
  $z=0$ to $z=7$. Overall, we thus conclude that our fiducial model is
  the only variant that simultaneously reproduces the dust content of
  low-metallicity galaxies \emph{and} low- and high-redshift
  galaxies. Rather than short depletion times depending on the density
  of the ISM, other works have suggested for instance a top-heavy IMF
  in the early Universe as a necessary ingredient to reproduce dust
  masses locally and at high redshifts \citep{Calura2014,Gall2011}.


\subsection{Importance of the different dust formation channels}
\label{sec:disc_formationchannels}
We have implemented three different modes of dust formation in our
dust tracking model: condensation of dust in AGB ejecta, condensation
of dust in SNe ejecta, and the accretion of gas-phase metals onto
existing dust grains. Figure \ref{fig:formation} clearly shows that in
our fiducial model, the growth of dust in the ISM is the dominant
channel through which dust builds up in galaxies. Although AGB stars and SNe are necessary to
form the first grains of dust, the accretion of metals onto dust
grains rapidly takes over. This is also clear when focusing on the cosmic
formation rate of dust (Figure \ref{fig:cosmic_evol}), where the
formation rate through accretion is approximately four orders of
magnitude larger than the formation rate by stellar ejecta. These
results support the hypothesis that metal accretion is the dominant
mode of dust formation for high-redshift galaxies with large observed
dust reservoirs \citep[e.g.,][]{Dwek2007,Michalowski2015}. Only in
galaxies with stellar masses less than $10^{7}\,\rm{M}_\odot$ does the
condensation of dust in SN ejecta take over as most important channel
through which dust forms.

Besides formation, it is also important to consider the destruction of
dust in the ISM. We find that the destruction rate of dust by SNe is
much higher than the formation rate of dust by stellar
ejecta. Furthermore, Figure \ref{fig:formation} and
\ref{fig:cosmic_evol} both show that the destruction rate of dust
closely follows the formation rate of dust through accretion onto
grains. Galaxies quickly achieve a balance between these two processes. At $z
< 2$, the destruction rate of dust is approximately two thirds of the
formation rate through accretion onto grains. At higher redshifts this
number drops to approximately half of the formation rate through
metal accretion at $z\sim4$, up to almost a tenth at
$z=9$.

We can understand why the dust growth rate and destruction rate track each other so closely by examining the recipes for each of these processes.
 The rate of accretion of metals onto dust grains is set by
  the abundance of the species of interest, as well as by the volume
  density, which we derived from the SFR and molecular hydrogen surface
  density. The destruction rate is set by the number of SNe, which
  closely follows the SFR. The SFR surface density
  is set by the molecular hydrogen surface density and metallicity of
  the gas. Since the accretion timescale and destruction rate are to first
  order set by the same physical parameters (metallicity and
  molecular hydrogen/SFR surface density) it is not surprising that
  the two closely follow each other.  The variation in the ratio
  between destruction and growth is then set by the exact conditions
  under which the processes occur, depending on for instance the available
  gas-phase metals for dust growth, the available diffuse gas to be
  cleared from dust, and the effective exchange times of the molecular
clouds.

It should be noted that the `no-acc' model variant is also able to
reproduce the dust masses in galaxies without invoking any growth of
dust through accretion onto grains at all. In this model variant the
production of dust in SNe ejecta is the dominant mode of dust
formation. However, this model requires a perhaps unrealistically high
value for the condensation of dust in stellar and SN ejecta, and does
not reproduce scaling relations for DTG ratios and
DTM ratios in the local Universe as well as our fiducial
model.

\subsection{Gas-to-dust and metal-to-dust ratios}
\label{sec:disc_dust_to_gas}
 In recent years it has become clear that the DTG ratio of galaxies
in the local Universe can be described by a double power-law (and not
a single one) as a function of gas-phase metallicity
\citep{Remy-ruyer2014}. Our fiducial model successfully reproduces
such a double power-law shape. This is, however, only achieved in our fiducial model, which
suggests that it is the combination of low condensation efficiencies
and the growth of dust in the ISM which shapes the relation between
DTG ratio and gas-phase metallicity.

We found that the relation between DTG ratio and gas-phase
metallicity is constant to within $\sim 50$\% for galaxies with
gas-phase metallicities larger than 0.5 Z$_\odot$. The DTG
ratio decreases a bit with lookback time from $z=0$ to $z=4$,
but increases again at higher redshifts. In these galaxies the
reservoir of dust is fully determined by the balance between dust
growth in the ISM and the destruction of dust by SNe. We have seen in the
previous sub-section that these two processes closely follow each
other, as they depend on the same set of physical
quantities. The variations with redshift are thus the result of small
variations in the ratio between dust growth in the ISM and
destruction. These were driven by properties such as the available
metals for dust growth, the amount of diffuse gas, and the exchange
time of molecular clouds. We see the same factors at play when looking
at the evolution of the DTM ratio of galaxies as a function
of their gas-phase metallicity. Only at $z=9$ is the relation between
DTM ratio and gas-phase metallicity very different from the
$z=0$ relation.

The weak evolution between metallicity and DTG and DTM ratio is
important for observations of galaxy gas masses based on the dust
continuum. Our predictions suggest that the locally derived DTG ratio
as a function of gas-phase metallicity can be adopted within a
certainty of approximately 50 \% when estimating gas masses based on
inferred dust masses \citep[though see the high DTG ratio
    found by][in low-metallicity sub-millimeter
    galaxies]{Santini2010}.

The DTM ratio of galaxies predicted by our fiducial model cannot be
described by a single power law as a function of gas-phase
metallicity. There are three distinct regimes: one in which
condensation in ejecta dominates, one in which dust growth in the ISM
becomes increasingly important, and one in which growth in the ISM and
the destruction of dust are almost in balance. This has important
implications for modelling dust in cosmological simulations of galaxy
formation. Typically it is assumed that metals and dust scale linearly
\citep[e.g.,][]{Silva1998,Granato2000,Baugh2005,Lacey2008,Lacey2010,Fontanot2011,Niemi2012,Somerville2012,Hayward2013,Cowley2016}.
Future models need to properly account for the possible dependence of
DTM ratio on galaxy properties and redshift.

Our model predicts a clear evolutionary trend in the relation between
DTG ratio and stellar mass of galaxies. At fixed stellar mass,
high-redshift galaxies have lower DTG ratios, especially for galaxies
with stellar masses larger than $10^{10}\,\rm{M}_\odot$ and at
redshifts $z>2$. We previously saw that the relation between DTG ratio
and gas-phase metallicity evolves only weakly with time, and is
similar to the $z=0$ relation even at $z\geq5$. This means that the
evolution in the relation between DTG ratio and stellar mass must come
from the build-up of metals in galaxies in general. Indeed, we see an
evolution in the gas-phase metallicity of galaxies at fixed stellar
mass from $z=9$ to $z=3$ in Figure \ref{fig:metal_depletion}. This
implies that one cannot use a fixed number for the DTG ratio when
inferring a gas mass from massive galaxies at high-redshift
(especially at $z>2$). Knowledge of the gas-phase metallicity is
necessary to reliably estimate a galaxy gas mass. Interestingly, the
relation between stellar mass and DTM ratio is relatively constant up
to $z=6$. A close look reveals that the scatter in this relation
increases significantly with look-back time, again emphasising that
one cannot estimate the DTM ratio of a galaxy by its stellar mass and
$z=0$ relations alone.

We compared the DTM ratios predicted by our fiducial model
to the DTM ratios inferred from absorption studies up to $z=4$.
The observations suggest that the DTM ratios are similar to the
values observed in our Local Group over a large redshift and
metallicity range
\citep{Zafar2013,Sparre2014,DeCia2013,DeCia2016,Wiseman2016}. Our predictions
strongly disagree with these observational constraints and predict
much lower DTM ratios in low metallicity environments. In
our fiducial model, the growth of dust in the ISM has not yet contributed
significantly to the dust budget of the low-metallicity galaxies,
causing this disagreement. It must be noted that the nature of
high-redshift absorbers is not very clear and we have not tried to
apply similar selection criteria to make a fair
comparison. The physical conditions along an average DLA line of sight may differ
significantly from a galaxy-wide average \citep{Berry2014}.
Future models need to be able to reconcile these different
constraints in conjunction with measurements of dust in emission in
the local- and high-redshift Universe.

\subsection{The evolution of dust masses in galaxies}
In this subsection we focus on the buildup of dust in galaxies. We
found that the dust reservoirs as a function of stellar mass in
typical star-forming galaxies at $z=9$ are already larger than in
typical star-forming galaxies at $z=0$ (Figure \ref{fig:mstar_mdust}). This
suggests that the buildup of dust starts at high redshift, as
suggested by observations
\citep{Bertoldi2003,Hughes1997,Valiante2009,Venemans2012,Casey2014,Riechers2014,Watson2015}. We
saw in Section \ref{sec:disc_formationchannels} that in our fiducial model this is
mostly driven by accretion of metals onto dust grains.

The relationship between stellar mass and dust mass is approximately
constant from $z=3$ to $z=0$ \citep[see also][]{Calura2016}. At higher
redshifts the relation between stellar mass and dust mass is also
roughly constant with time, but approximately 0.4 dex
higher. In terms of galaxy global properties, the amount
of dust in the ISM of a galaxy is the balance between the amount of
gas available and the DTG ratio. At redshift $z<2$ (where the relation
between DTG ratio and stellar mass of galaxies is largely constant as
a function of time) the decrease in dust mass can be explained by the
observed and inferred decrease in the gas fraction of galaxies
\citep[e.g.,][]{Tacconi2010,Narayanan2012,Tacconi2013,
  Saintonge2013,Santini2014,Bethermin2015,Popping2015CANDELS,Scoville2016}. For
a constant DTG ratio, the dust mass of galaxies as a function of
stellar mass will therefore decrease as well. At higher redshift the
DTG ratio still increases with cosmic time. The gas mass of massive
galaxies on the other hand decreases at these redshifts \citep[e.g.,
  SPT14; ][]{Popping2015SHAM,Popping2015CANDELS}. Interestingly, there
is a balance between the decrease in gas mass and the increase in DTG
ratio, such that the relation between stellar mass and dust mass
remains constant.

The number of dusty galaxies with masses larger than
$10^9\,\rm{M}_\odot$ at the highest redshifts probed is relatively low
(see Figure \ref{fig:mass_function}). We find that the number density
of these dusty galaxies increases rapidly till $z=3$, and then keeps
growing till $z=0$.  The evolution of the dust mass function is
reflected in our predictions for the cosmic density of dust, which
grows up to a redshift of $z=0$ and remains fairly constant at lower
redshifts. This behaviour is in sharp contrast with the observed
decline in the \h2 number density of galaxies at $z<2$
\citep[PST14;][]{Walter2014,Decarli2016}.

Our results have significant consequences for studies of the redshift
distribution of sub-mm continuum selected galaxies
\citep[e.g.][]{Aravena2016,Bouwens2016,Dunlop2016}. The buildup of the dust
mass function suggests that (depending on the sensitivity) blind
surveys are most likely to pick up galaxies with redshifts lower than
three (see for example Figure 8 in \citet{Aravena2016}
  and Figure 5 in \citet{Dunlop2016}). We emphasise that our
conclusion is based only on the dust mass of galaxies. We have not
discussed how the temperature and the properties of the dust shape the
sub-mm SED and the detectability of the galaxy.

\subsection{Dust in the hot halo and ejected reservoir}
\label{sec:disc_CGM_dust}
We have presented predictions for the mass of dust in the hot halo
component in Figure \ref{fig:mstar_mdustCGM}. We found that the
reservoir of dust in the hot halo increases with host galaxy stellar
mass. At stellar masses larger than $10^{10}\,\rm{M}_\odot$, the mass
of dust in the hot halo is comparable to or larger than the mass of
dust in the cold gas. As long as dust can effectively
  escape the potential well of the dark-matter halo, the mass of dust
  in the reservoir of ejected baryons is larger than the mass of dust
  in the cold gas and hot halo. Figure \ref{fig:cosmic_evol} shows
that the density of dust in the hot halo and ejected reservoir is
always higher than the density of dust in the cold gas, especially at
redshifts $z<1.5$. These predictions suggest that a large fraction of
the dust (and metals) ever formed may be stored outside galaxies
\citep{Peeples2014,Peek2015}. If we can quantify these reservoirs
observationally (for instance through reddening studies
  around galaxies or by looking at absorbers along quasar lines of
  sight) , this will provide unique additional constraints for the
ejective feedback recipes in galaxy formation models. An
  initial comparison between observations of dust in the CGM/IGM and
  the cosmic density of dust in the hot haloes and ejected reservoirs
  of our model galaxies (Figure \ref{fig:cosmic_evol_ejected})
  suggests that our fiducial model predicts too much dust outside of
  galaxies. It may well be that a significant fraction of the ejected
  dust is destroyed in winds, or that the destruction rates of dust in
  the IGM are higher than estimated in this work. Future observations
  of absorption systems and reddening in the halos of galaxies will
  constrain these processes. At the same time, theoretical advances on
  this front will be necessary.

\citet{Peek2015} found that galaxies at $z=0$ with a luminosity of 0.1
$L_*$ have as much dust in their CGM as in the ISM. We find similar
results, where MW type galaxies have as much or even more of their
dust in the hot halo gas as in the cold gas in the galactic disk due
to the heating of dust by SNe. Our
predicted slope between host galaxy stellar mass and hot halo dust mass is
much steeper than that derived by \citet{Peek2015} for the
CGM. However, it is currently difficult to compare observations of the
CGM in detail with our model predictions, so we prefer to avoid drawing
strong conclusions. It is clear, nonetheless, that this is an important
avenue to pursue in the future in order to constrain the importance of
galactic winds in removing dust from galaxies and polluting diffuse
gas in the CGM and IGM.

\subsection{Depletion of metals}
As discussed before, a fraction of the metals in the ISM is depleted
onto dust grains, which lowers the gas-phase metallicity of
galaxies. We have shown in Figure \ref{fig:metal_depletion} that the
depletion of metals becomes relevant in galaxies with stellar masses
larger than $10^9\,\rm{M}_\odot$, where almost 40\% of the gas-phase
metals are in dust at $z=1$. This correction should always be taken
into account when comparing model predictions of the gas-phase
metallicity of galaxies to observations.  At the same time these
results also suggest that the agreement between model predictions and
observations for the gas-phase metallicity of galaxies is perhaps
worse than previously thought.  Gas-phase metallicities at $z=0$ that are too
low are a
common problem for many galaxy formation models
\citep{SomervilleDave2015}, but the depletion of metals makes the
disagreement even stronger.  To make matters worse, the
  DTM ratios predicted by our fiducial model appear to be
  too low (Figure \ref{fig:mstar_dust_to_metal}).
  More accurately reproduced
  DTM ratios may therefore result in an even lower gas-phase
  metallicity. This emphasises the need for new or modified recipes
that model the cycle of gas and metals in and out of galaxies
\citep[see for example ][]{Hirschmann2016}. These
issues are likely also linked to other problems galaxy formation
models face, such as producing too many low-mass galaxies at $z>0$ and
too-low specific star-formation rates and gas masses at $z=2$
\citep{SomervilleDave2015,Popping2015SHAM}.

\subsection{Caveats}
Here we discuss a number of caveats that should be taken into account when
interpreting our results.

\subsubsection{Dust as an ingredient of gas physics}
Dust plays an important role in the cooling and shielding of gas. We
have not included these processes in this work, but treat dust as
`normal metals' for the purposes of computing cooling rates. Future
work should self-consistently include dust as a coolant, as well as
the role of dust in the formation of molecular hydrogen. The latter is
of particular interest. Star-formation rates in our model are
calculated based on the surface density of molecular hydrogen. The
molecular hydrogen fraction of gas in our model is a function of
gas-phase metallicity, whereas this should actually be a function of
dust abundance \citep{Gnedin2011}. We have so far made the assumption
of a linear scaling between dust abundance and gas-phase metallicity,
but observations \citep{Edmunds2001,Remy-ruyer2014} and our modelling
efforts show that this assumption is incorrect. This should especially
manifest itself in low-mass and low-metallicity galaxies, where the
linear scaling between dust-abundance and metallicity clearly breaks
down. The low dust abundances (with respect to the gas-phase
metallicity) in low mass galaxies will result in lowered molecular
hydrogen fractions and consequently lower star-formation rates. It is
possible that the build up of stellar mass in low-metallicity galaxies
can thus be slowed down by adopting a dust-based molecular hydrogen
recipe (though see SPT15 for a discussion on how the self-regulating
nature of star formation implies that lowering the efficiency of star
formation does not always result in the production of fewer stars).
Preliminary tests show that (without changing any of the free
parameters) this mostly affects galaxies in haloes with masses less
than $\sim 10^{11}\,\rm{M}_\odot$. We will explore the effects of
self-consistently modelling the formation of molecular hydrogen based
on our estimated dust masses in a future work.

\subsubsection{Dust in the ejected reservoir and hot halo}
\label{sec:disc_ejected_reservoir}
In our models, when dust is ejected out of the hot halo by stellar
driven winds, it ends up in the ejected reservoir. From there it may
reaccrete back into the hot halo and ultimately the galaxy. We have
not included any physical processes acting on the dust while it is
being ejected. It may very well be that dust is destroyed while it is
driven out of the galaxy in winds. These processes could completely
alter our predictions for the mass of dust in the ejected reservoir
(and hot halo) and could also lower the dust masses in the
ISM. However, we have run tests in which all the dust in the ejected
reservoir is destroyed immediately and found that this has a minimal
effect on the dust content in the ISM of galaxies.

\citet{Elvis2002,Pipino2011} have suggested that AGN-driven winds
  can also contain dust produced in quasars. If not destroyed in the
  wind itself, this dust can provide an additional contribution to the
  dust content of galaxy halos. Similarly, we have not
explored the effects of dust-enriched winds (i.e., where the DTG ratio in
winds leaving the galaxy is higher than the galaxy mean). Such
processes have been proposed for metals \citep[i.e., metal enriched
  winds, ][]{Krumholz2011}. If dust-enriched winds exist, this would
lower the DTG ratio in galaxies and would be of particular importance
in low-mass objects. Dust-enriched winds would increase the abundance
of dust in the ejected reservoir and the hot halo.

\subsubsection{A homogeneous distribution of dust in galaxies?}
In this work we make the assumption that dust distribution is smooth
and follows the gas distribution within galaxies. In reality there may
be differences in the DTG ratio from one region of a galaxy to another
\citep{Watson2011,Smith2012,Sandstrom2013,Draine2014,Roman-Duval2014,Galametz2016}. If,
for example, thermal sputtering occurs in a region with an elevated
DTG ratio compared to the galaxy mean, a larger mass of dust will be
destroyed than predicted by our models. The opposite situation can
also occur. We expect the net effect of this to be minimal for our
model predictions. Nevertheless, we acknowledge that numerical
hydrodynamical simulations that resolve galaxy structures are
necessary to fully explore this issue.

\subsubsection{The density of the ISM}
In our model we assume that the timescale for accretion of metals onto
dust grains is a function of the gas-phase density and metallicity. As
pointed out before, SAMs do not explicitly track volume densities. In
Section \ref{sec:dust_accretion} we presented an approach to calculate
the volume density of molecular hydrogen, based on the molecular
hydrogen and SFR surface density. This approach implicitly assumes
that star formation has one fixed efficiency in molecular clouds of
varying density. Theoretical work has already shown that, depending on
the Mach number of the gas in a molecular cloud, a smaller (or larger)
fraction of the gas in a cloud may reach some critical density above
which it can collapse within a free-fall time and form
stars\citep[e.g.,][]{Krumholz2005}. Although the exact value for this
critical density may vary, this also shows that, in reality, the
star-formation efficiency does not need to be a fixed number. A higher
(lower) efficiency would result in a lower (higher) derived density of
molecular hydrogen, and therefore in a longer (shorter) timescale
for metal accretion onto dust grains. Numerical simulations of
resolved structures are necessary to properly address this issue and
derive appropriate scaling relations for one-zone models
\citep{Zhukovska2016}.

\subsubsection{The destruction of dust in the ISM}
In this work we have assumed fixed parameters for the destruction of
dust by SNe, based on theoretical work from \citet{Slavin2015}. In
their work, the authors already demonstrated that these efficiencies
may vary within a factor of two depending on the chosen values for
their modeled supernova remnant properties (density of the medium,
ionisation fraction, magnetic field strength). \citet{Temim2015} also
found that the efficiency of dust destruction may increase or decrease
depending on the density of the medium through which the SN shock
travels. \citet{Yamasawa2011} suggested that the efficiency of dust
destruction by SNe also depends on the metallicity of the cold gas
being cleared of dust. Based on observations of the LMC and the SMC,
\citet{Zhukovska2008} found that approximately 1.5 times as much
carbonaceous and silicate dust should be destroyed per SN as we chose
for our fiducial values \citep[see also the recent work
  by][]{Gioannini2017}. All these results clearly show there is still
a systematic uncertainty in how efficiently SNe can destroy dust in
the ISM. A higher (lower) destruction efficiency would result in lower
(higher) dust masses.

\section{Summary \& Conclusions}
\label{sec:summary}
We have included the tracking of dust production and destruction in a
semi-analytic model of galaxy formation and made predictions for the
dust properties of galaxies from $z=9$ to $z=0$. We present
results for different model variants for the dust production
processes. The first is our fiducial model with dust condensation
efficiencies in stellar ejecta of around 15 per cent and a density and
metallicity dependent timescale for the accretion of metals onto dust
grains. The second includes no accretion of metals onto dust grains,
the third assumes much higher dust condensation efficiencies
than in our fiducial model, whereas the fourth assumes a fixed
accretion time scale of 100 Myr. We also explored two model variants
that include the approach presented by \citet{Dwek1998} for the growth of
dust in the ISM. We summarise our main findings below.

\begin{itemize}
\item Our fiducial model successfully reproduces the trends between
  stellar mass and dust mass in the local and high-redshift
  Universe, as well as the DTG ratio of local galaxies as a
  function of their stellar mass. It furthermore accounts for a double power law relation
  between DTG ratio and gas-phase metallicity,  reproduces the dust mass function of galaxies with dust masses less than
  $10^{8.3}\,\rm{M}_\odot$, and the cosmic density of dust at $z=0$.
\item The fiducial model has problems accounting for the
  slope and exact normalization of the DTG and DTM ratio of galaxies
  as a function of their gas-phase metallicity. 
\item The dust mass of galaxies at fixed stellar mass is almost
  constant from $z=2$ to $z=0$. It decreases by $\approx 0.2$ dex from $z=3$ to $z=0$. This is mainly
  driven by a decrease in galaxy gas fractions. At higher redshift the
  relation between stellar mass and dust mass remains constant with
  time. The dust mass function of galaxies on the other hand increases rapidly from $z=9$ to
  $z=3$, after which only the number density of the galaxies with largest
  dust masses ($10^8\,\rm{M}_\odot$) keeps increasing.
\item The relation between the DTG ratio of galaxies and their
  gas-phase metallicity remains constant to within 50 \% up to
  $z=9$. There is no clear evolutionary trend in this
  relation. The DTG ratio of galaxies increases with cosmic
  time at fixed stellar mass, following the buildup of metals in a
  galaxy's ISM.
\item Our model predicts a significant reservoir of dust in the CGM
  (hot halo) of galaxies. These reservoirs can be as large or even larger than the
  reservoir of dust in the ISM of the host galaxy. Our models predict
  that even more dust is ejected from galaxies. The
    amount of dust in the ejected reservoir is significantly larger
    than observational constraints on the cosmic density of dust,
    suggesting that additional processes are needed to
    destroy this dust.
\item In our models, up to 25 \% of the gas-phase metals at redshift $z=0$ can be
  depleted onto dust. This lowers the gas-phase metallicity relation by
  $\sim0.1$ dex. This depletion should be taken into account when
  comparing model predictions to observations. Similarly, a
  significant fraction of the CGM metals may be locked up in dust.
\item Within our fiducial model the accretion of metals onto dust grains is the dominant mode of
  dust formation in galaxies. The contribution from metal accretion
  becomes increasingly important with stellar mass. Only at the lowest
  stellar masses (less than $10^7\,\rm{M}_\odot$) does the
  condensation of dust in SN ejecta become the dominant mode of
  dust formation.
\item The `high-cond' and `fix-tau' model variants cannot reproduce
  the DTG ratio of galaxies with metallicities less than 0.2
  Z$_\odot$. Furthermore the `fix-tau' model cannot reproduce the
  high dust masses observed in galaxies at $z\sim6$. A model without
  accretion of metals onto dust grains can reproduce observations
  relatively well if an unrealistically high efficiency of 100\% is
  assumed for the condensation of dust in stellar ejecta, but predicts
  too high dust masses in low-metallicity galaxies. The model variants
  that include the \citet{Dwek1998} approach for the accretion of
  metals onto dust grains predict either too high dust masses in
  low-mass galaxies ('dwek98') or too large dust masses in
  high-redshift galaxies ('dwek-evol'). We conclude that a
  model in which the rate of accretion of metals onto dust grains is set
  by the metallicity \emph{and} the density of the cold gas is
  necessary to reproduce the shape of the observed scaling relations and
  dust mass budgets.
\end{itemize}

The results presented in this paper can serve as predictions for
future surveys of the dust content of galaxies. We look forward to
observations from current and upcoming facilities such as ALMA, the
JWST, and single-dish sub-mm instruments, that will be able to
confront our predictions. In this work we have ignored the effects
dust has on the ISM physics and chemistry in galaxies, affecting the
growth rate of molecular hydrogen, and the absorption of stellar
light. In future work we will explore these effects, and their
consequences for the stellar buildup and appearance of galaxies. 

\section*{Acknowledgments}
We thank Andrew Baker, Laure Ciesla, Annalisa De Cia, Loretta Dunne, Haley Gomez, Leslie Hunt,
Suzanne Madden, Joshua Peek, Molly Peeples, Paola Santini, Raffaella
Schneider, Jason Spyromilio, Phil Wiseman, and Svitlana Zhukovska for stimulating discussions and/or
providing data. We thank the referee for useful
  suggestions. We further acknowledge the Kavli Institute for
Theoretical Physics ``Cold Universe'' workshop where some of this work
was completed. This research was supported in part by the National
Science Foundation under Grant No. NSF PHY11-25915. rss thanks the
Downsbrough family for their generous support, and acknowledges
support from the Simons Foundation through a Simons Investigator
award.

\bibliographystyle{mn2e_fix}
\bibliography{references}

\appendix
\section[]{The Dwek 1998 model}
\label{sec:appendix_dwek}
\begin{figure*}
\includegraphics[width = 0.95\hsize]{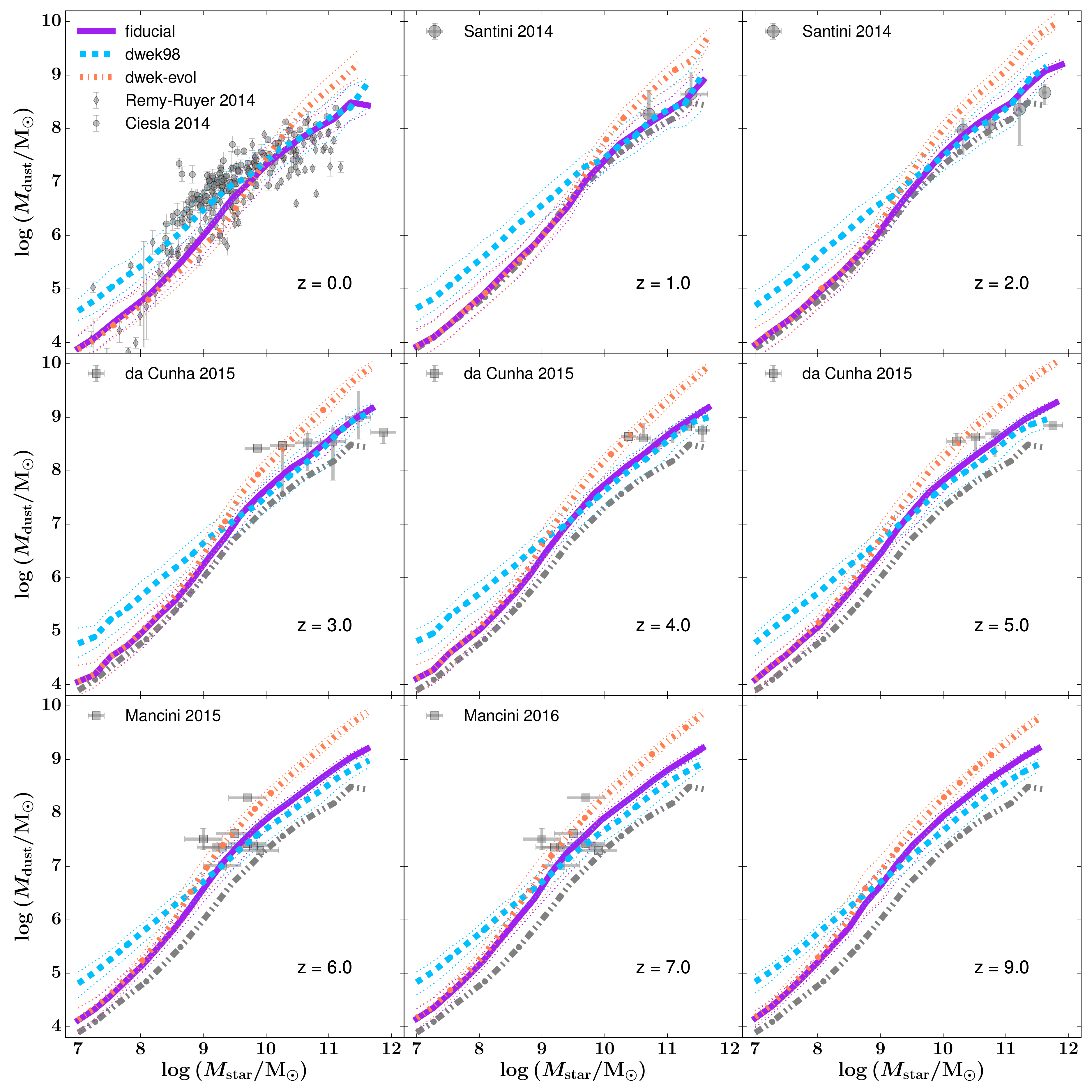}
\caption{The dust mass of galaxies as a function of their stellar mass
  and redshift for our fiducial model variant, the '`dwek98' model
  variant, and the `dwek-evol' model variant. The lines and
  observations are the same as in Figure \ref{fig:mstar_mdust}.\label{fig:mstar_mdust_dwek}}
\end{figure*}

We present the dust mass of galaxies as a function of their stellar
mass for the `dwek98', `dwek-evol', and fiducial model in Figure
\ref{fig:mstar_mdust_dwek}. We find that the predicted dust masses in
galaxies with $M_* > 10^{10.5}\,\rm{M}_\odot$ by the `dwek98' model
variant are similar to the
predictions by our fiducial model. Only at redshifts $z\geq6$ do we
see that the `dwek98' predicts dust masses approximately 0.1 dex lower
than our fiducial model. The dust masses predicted in lower-mass
galaxies by the on the other hand are up to 0.5 dex more massive for galaxies
with $M_* > 10^{8}\,\rm{M}_\odot$.  The `dwek-evol' model variant predicts dust masses similar to our
fiducial model in the mass range $M_* \leq 10^{9}\,\rm{M}_\odot$. The
predicted dust masses in galaxies with larger stellar masses is up to
$\sim 0.3$ dex larger than predicted by our fiducial model. This
especially leads to poor agreement between the predictions by the
`dwek-evol' model variant and observations at redshifts  $1\leq z\leq
5$, but also for galaxies with largest stellar masses at $z=0$.

\begin{figure*}
\includegraphics[width = 0.95\hsize]{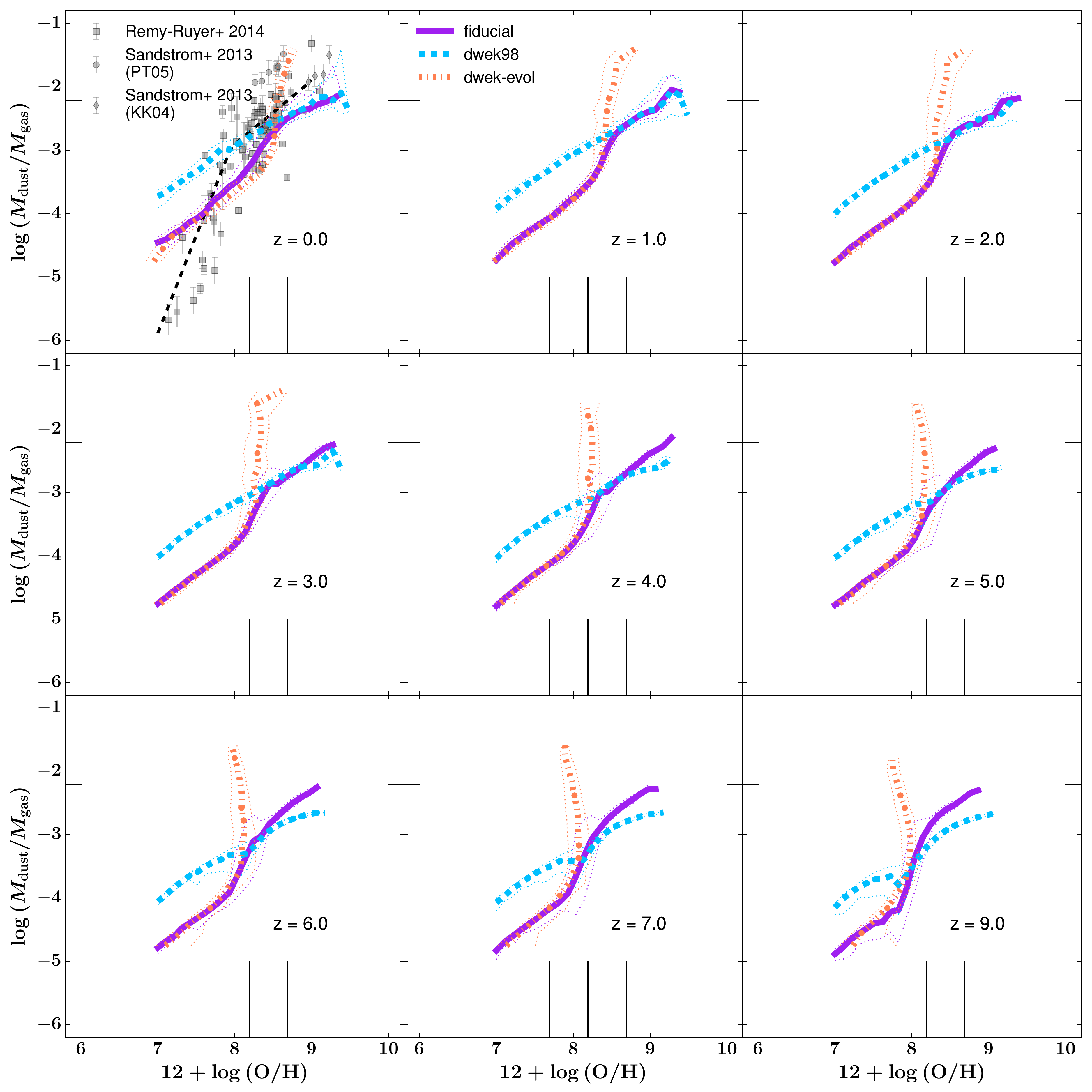}
\caption{The DTG ratio of galaxies as a function of their
  metallicity and redshift for our fiducial model variant, the
  `dwek98', and the `dwek-evol' model variant. The lines and
  observations are the same as in Figure \ref{fig:dust_to_gas}.\label{fig:dust_to_gas_dwek}}
\end{figure*}

We present the DTG ratios predicted by the `dwek98', the
`dwek-evol', and our fiducial model variants in Figure
\ref{fig:dust_to_gas_dwek}. We find that the `dwek98' variant predicts
DTG ratios at $z=0$ similar to our fiducial model for galaxies with
metallicities larger than $\sim 0.5\,\rm{Z}_\odot$. The predicted DTG
ratios are slightly below
the observational constraints. At lower metallicities the
`dwek98' model variant predicts DTG ratios that are higher
than the predictions by the fiducial model. At metallicities $<0.1\,\rm{Z}_\odot$
the `dwek98' variant predicts DTG ratios up to an order of
magnitude larger than the observational constraints.  The predicted relation between gas-phase
metallicity and DTG ratio by the `dwek98' model variant
remains fairly constant with time.

The 'dwek-evol' model variant predicts DTG ratios similar to our
fiducial model  for galaxies with a gas-phase
metallicity less than $\sim 0.5\,\rm{Z}_\odot$ at $z=0$. At $z=0$ the 'dwek-evol' variant predicts higher DTG
ratios in metal-rich galaxies, which yields better agreement with the observational constraints than we found for
the fiducial model. We see an interesting feature in the
relation between DTG ratio and gas-phase metallicity as predicted by
the 'dwek-evol' model variant for high redshifts galaxies. The DTG ratio as a function of
metallicity curves backwards at $z\geq4$. Some galaxies become so dust enriched, that their DTG ratio
keeps increasing, whereas their gas-phase metallicity decreases because large fractions of the available metals are
locked up in dust.

Putting the previous two figures together we find that the `dwek-98'
approach predicts dust masses in low-metallicity galaxies that are too
high. This is driven by the
high-condensation efficiencies assumed for stellar ejecta, similar to
the predictions by the `high-cond' model variant. \citet{RemyRuyer2014} also
explore the `dwek98' model variant in the context of the
\citet[based on the \citet{Dwek1998} approach]{Galliano2008} model for dust evolution and found similar
results. The clear discrepancy between model
predictions and observations in our local Universe at the lowest metallicities leads us to
disfavour the `dwek98' model variant. A model variant that uses lower
condensation efficiencies and the 'dwek98' approach for dust growth
with a fixed accretion timescale does better at reproducing the $z=0$
DTG ratio, but cannot reproduce the dust content of
high-redshift galaxies (similar to the `fix-tau' model variant).

The large discrepancy in the dust masses predicted by the fiducial model
and the `dwek-evol' model variant is driven by the efficient growth of dust predicted
by the \citet{Dwek1998} approach  in media
with high initial condensed fractions $f_{j,0}$ in non-solar
metallicity environments, compared to the Zhukovska approach (see
Section \ref{Sec:Dwek_approach}). The efficient growth of dust predicted by
the `dwek-evol' model variant is highlighted by the DTG ratio at high
redshifts that seems to curve backwards as a function of gas-phase
metallicity. The high dust-to-gas ratios
predicted by the 'dwek-evol' model variant result in too large
dust masses in galaxies at $1\leq z \leq 5$. \citet{Feldmann2015} imposed a maximum condensation
fraction of 70\% for dust (i.e., at most 70\% of the metals are locked
up in dust) to avoid such behaviour. We tested a model variant
that includes the ceiling in condensation fraction suggested by
\citet{Feldmann2015}. Using this ceiling, the model variant still
predicts high DTG ratios and  dust masses
at  $1\leq z \leq 5$ that are much larger than the observational
constraints. We disfavour the 'dwek-evol' variant over our fiducial model because of
the increasingly poor agreement with observations at $z>0$.

\section[]{Dust formation channels for other model variants}
\label{sec:appendix}
\begin{figure*}
\includegraphics[width = 0.95\hsize]{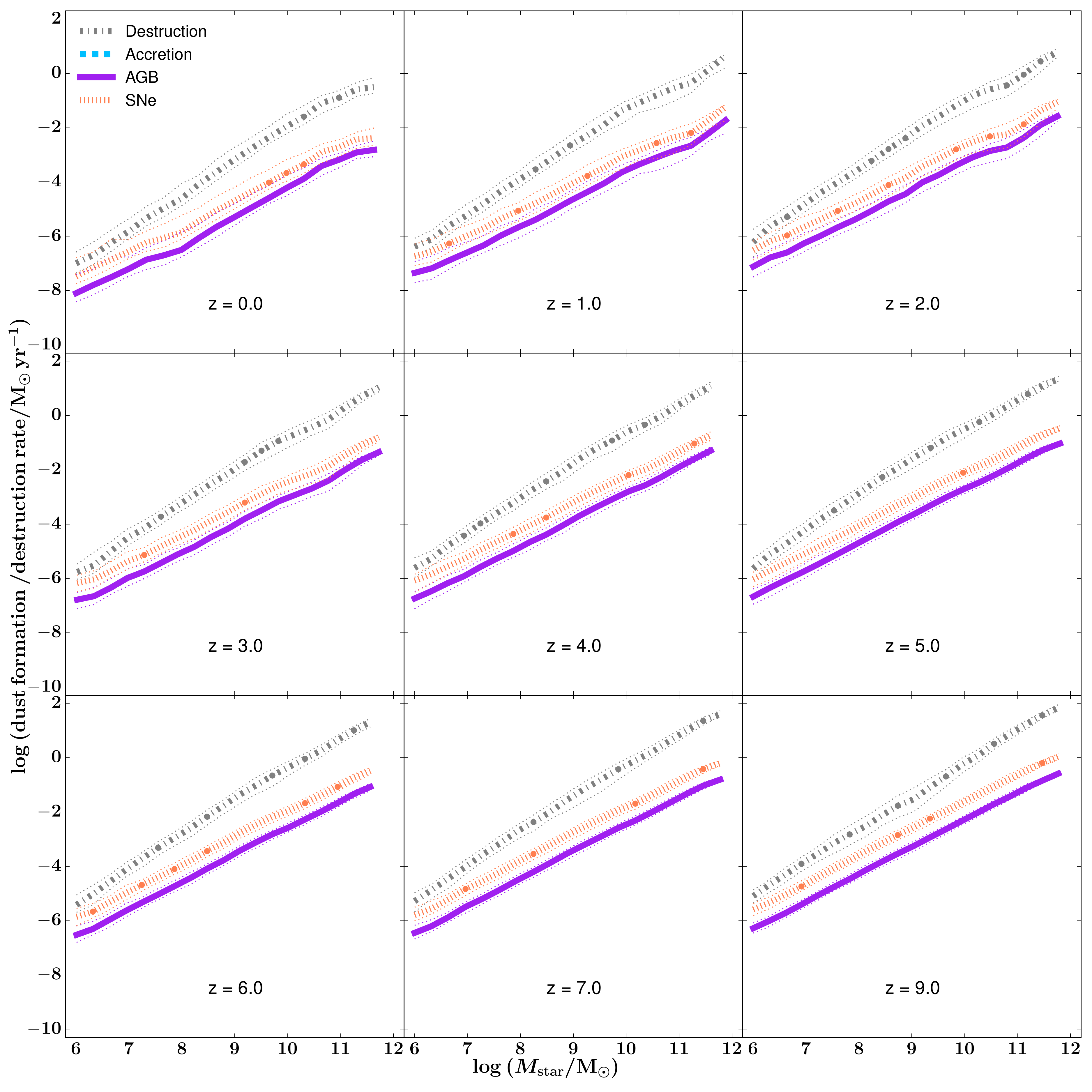}
\caption{The formation and destruction rate of dust as a function of
  stellar mass from redshift $z=9$ to $z=0$, for the `no-acc' model variant. Dust formation rates are separated into formation due to
  AGB stars, SNe, and growth of dust in the ISM. Thick lines mark the 50th percentiles, whereas the narrow
  dotted lines mark the 16th and 84th percentiles.\label{fig:formation_channel_no_acc}}
\end{figure*}

\begin{figure*}
\includegraphics[width = 0.95\hsize]{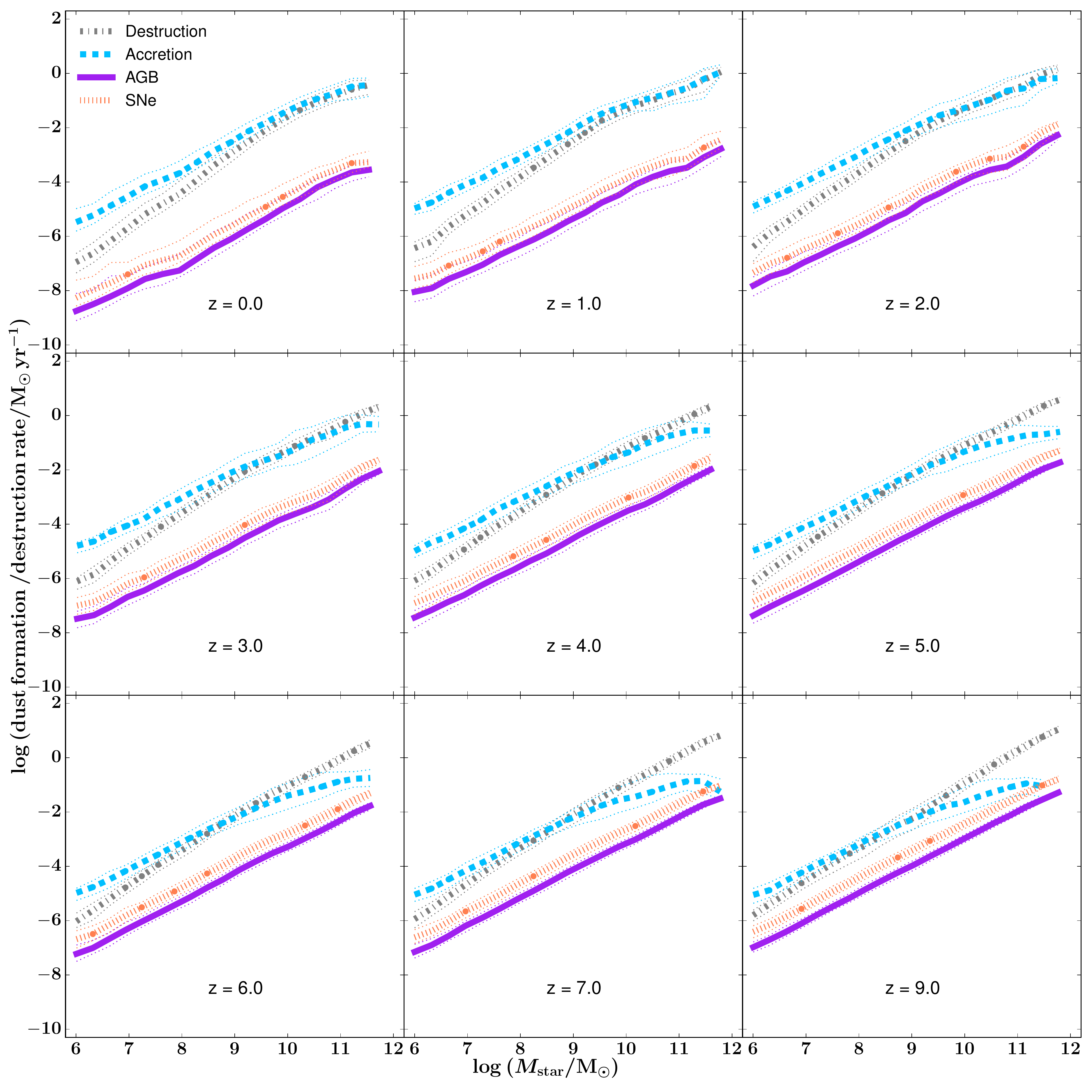}
\caption{The formation and destruction rate of dust as a function of
  stellar mass from redshift $z=9$ to $z=0$, for the `fix-tau' model variant. Dust formation rates are separated into formation due to
  AGB stars, SNe, and growth of dust in the ISM. Thick lines mark the 50th percentiles, whereas the narrow
  dotted lines mark the 16th and 84th percentiles. \label{fig:formation_channel_fix_tau}}
\end{figure*}

\begin{figure*}
\includegraphics[width = 0.95\hsize]{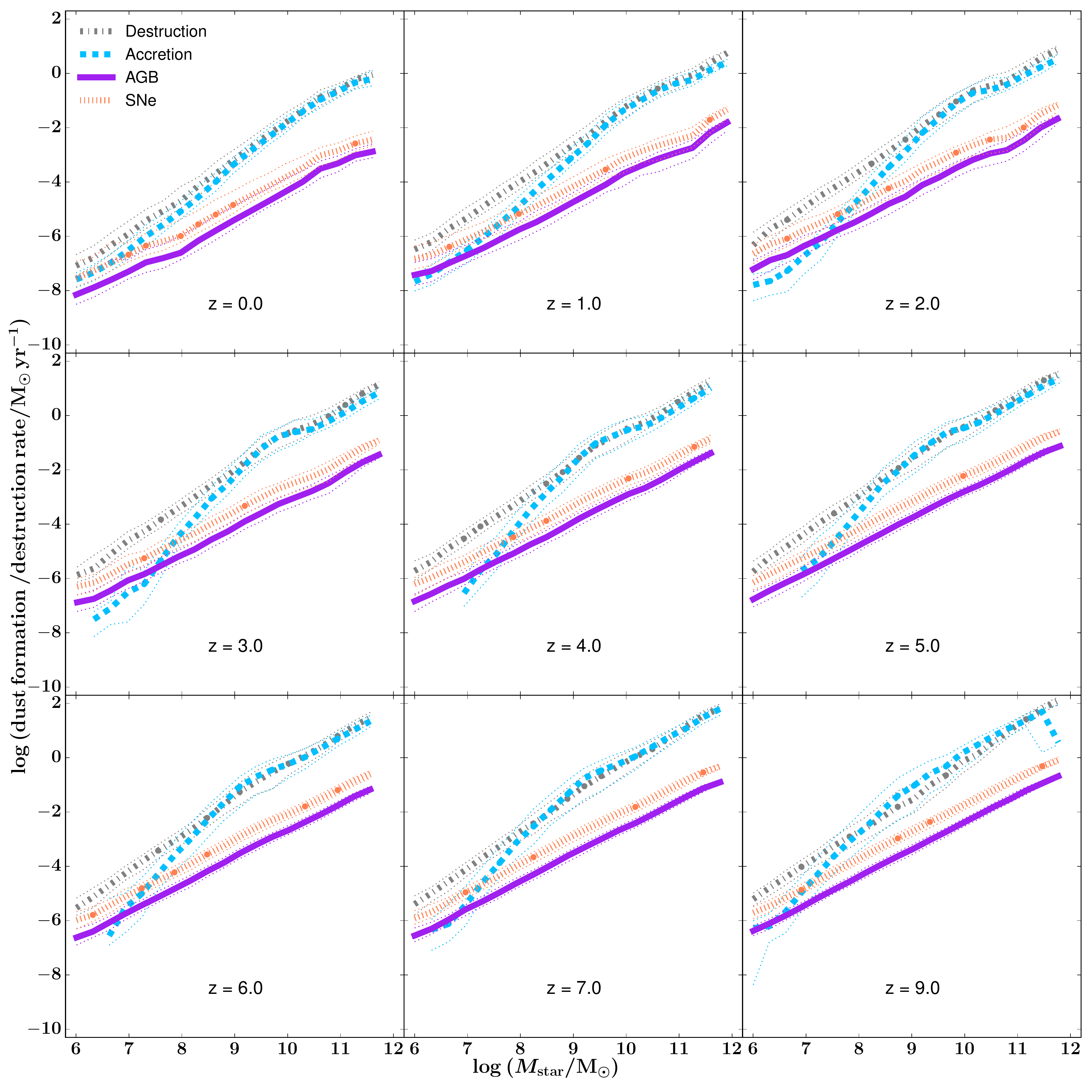}
\caption{The formation and destruction rate of dust as a function of
  stellar mass from redshift $z=9$ to $z=0$, for the `high-cond' model variant. Dust formation rates are separated into formation due to
  AGB stars, SNe, and growth of dust in the ISM. Thick lines mark the 50th percentiles, whereas the narrow
  dotted lines mark the 16th and 84th percentiles. \label{fig:formation_channel_high_cond}}
\end{figure*}

We presented the formation rate of dust in our fiducial model via the different formation
channels in Figure \ref{fig:formation}. Here we present the
same plot for the remaining model variants (Figures
\ref{fig:formation_channel_no_acc} through
\ref{fig:formation_channel_high_cond}). 

\end{document}